\documentclass[reprint,superscriptaddress,amsmath,amssymb,amsfonts,aps,prx,floatfix]{revtex4-2}
\usepackage{times}
\usepackage{graphicx}
\usepackage{dcolumn}
\usepackage{bm}
\usepackage{soul}
\usepackage{ulem}
\usepackage[colorlinks=true, citecolor=blue, linkcolor=blue, urlcolor=blue]{hyperref}
\usepackage{bbold}
\DeclareMathAlphabet\mathbfcal{OMS}{cmsy}{b}{n}
\usepackage{wasysym}
\usepackage{amssymb}
\usepackage{amsmath}
\usepackage[usenames,dvipsnames]{xcolor}

\begin{document}

\title{Parametric Instabilities of Correlated Quantum Matter}

\author{Gal Shavit}
\affiliation{Department of Physics and Institute for Quantum Information and Matter, California Institute of Technology,
Pasadena, California 91125, USA}
\affiliation{Walter Burke Institute of Theoretical Physics, California Institute of Technology, Pasadena, California 91125, USA}
\author{Gil Refael}
\affiliation{Department of Physics and Institute for Quantum Information and Matter, California Institute of Technology,
Pasadena, California 91125, USA}

\begin{abstract}

Collective bosonic excitations are a fascinating aspect of broken-symmetry correlated phases. A wealth of such phases emerged in tailored moir\'e heterostructures, where, in addition, new direct knobs of control exist.
Our work explores how the associated collective bosonic modes can be directly manipulated and amplified via parametric driving. As we show,  parametric driving is closely related to many-body quantum geometry, as it reveals information about the fidelity susceptibility of the underlying ground state with respect to changes in the control knobs of the interacting system. 
The parametric response hinges on the tunability of the hierarchy of competing ground states, which we show manifests as squeezing of the bosonic mode vacuum.
Furthermore, parametrically-driven modes can produce a high-amplitude modulation in the system that could be easily observed, and could also be used to produce a new non-equilibrium prethermal states with different symmetries.
We derive a general framework for parametric amplification of collective bosonic modes in broken-symmetry phases, and provide case-studies of current relevance. 
Elucidating the connection between the dynamics of these excitations and microscopic electronic observables is key to harnessing their immense potential.

\end{abstract}


\maketitle

\section{Introduction}\label{sec:intro}
Recent years have been marked by the discovery of correlated quantum materials, that host a variety of exotic quantum phases and offer extensive \textit{in situ} tunability.
A non-exhaustive list includes  moir\'e graphene or transition metal dichalcogenides (TMDs) materials~\cite{TBG1_CaoCorrelatedInsulator,TBG_2CaoUnconventionalSC,TBG3_EfetovAllIntegers,TBGdisplacement,TBG34goldhaber,KimTrilayerGrapheneSuperconductivity,Park2021StronglyCoupledSuperconductivityTrilayer,Park2022MATngFamily,matngYIRAN,TDBG1_Shen2020,TDBG2_Cao2020,TDBG3_Liu2020,Quasicrystal_Uri2023,TMD_kinfai0Li2021,mote2_fci_Park2023,TMD_VP_KINFAI_Zeng2023,twse2Cornell_Xia2025,twse2Dean_Guo2025},
monolayer TMDs~\cite{nbse2_Xi2016,wte2_mit,wte2_cobden},
crystalline graphene superconductors and fractional Chern insulators~\cite{ZhouYoungBLGZeeman,RTGsuperconductivityZhou2021,nadj_ISOC_BBGBLGZhang2023,BBG_electronside_li2024tunable,Nadj_zhang2024twistprogrammablesuperconductivityspinorbitcoupled,young_nadj_BBG_RTG_SC,Young_patterson2024superconductivityspincantingspinorbit,FCI_penta_longju_Lu2024,fci_hexa_chinese},
fractional quantum Hall (FQH) states in bilayer graphene~\cite{Bernalfqh1_Zibrov2017,bernalfqh2_doi:10.1126/science.aao2521,bernalfci3_doi:10.1126/science.aan8458,bernalfqh4_junzhu_PhysRevX.12.031019},
excitonic double layers~~\cite{doblelayrBLG_Dean,kinfaimak_excitonic_tmd,kinfaimak_excitonic_tmd2,kinfaicoulombdragtmd}, 
and tunable oxide interfaces~\cite{oxide_interface_Cheng2015,oxide_interface_Maniv2015,oxide_interfacePhysRevLett.119.237002}.
The combination of a rich strongly-interacting phase diagram with extraordinary tunability unlocks a myriad of unexplored opportunities.
Proposals involving the dynamical manipulations of different parameters of the system, whether by a rapid quench~\cite{quantumquanchmitra,millisquench_PhysRevX.10.021028,FQHquench,FQHbilayer_quench,ephemeral_Shavit2025,fiete_quench_3dm4-p2yq}
or by periodic driving~\cite{periodicTHz_Haque2024,periodic_photoSCEckhardt2024,periodic_metaSCChattopadhyay2025} hold promise to deepen our understanding of these materials, and to access and manipulate novel phases of matter.

Most ordered correlated phases possess \textit{collective low-energy excitations}. These modes are often at the center of unusual experimental observations, e.g., an entropic Pomeranchuk effect~\cite{pomeranchuk_Rozen2021}, reduction of ferromagnetic Curie temperatures~\cite{valley_wave_dassarme_PhysRevLett.124.046403}, and unconventional superconductivity~\cite{bultnik_kivc_modesPhysRevB.106.235157,berg_yaar_abc_PhysRevB.111.075103}.
We would like to argue that these collective excitations are also an untapped resource for out-of-equilibrium phenomena in the correlated materials that host them.

\begin{figure}
    \centering
    \includegraphics[width=9cm]{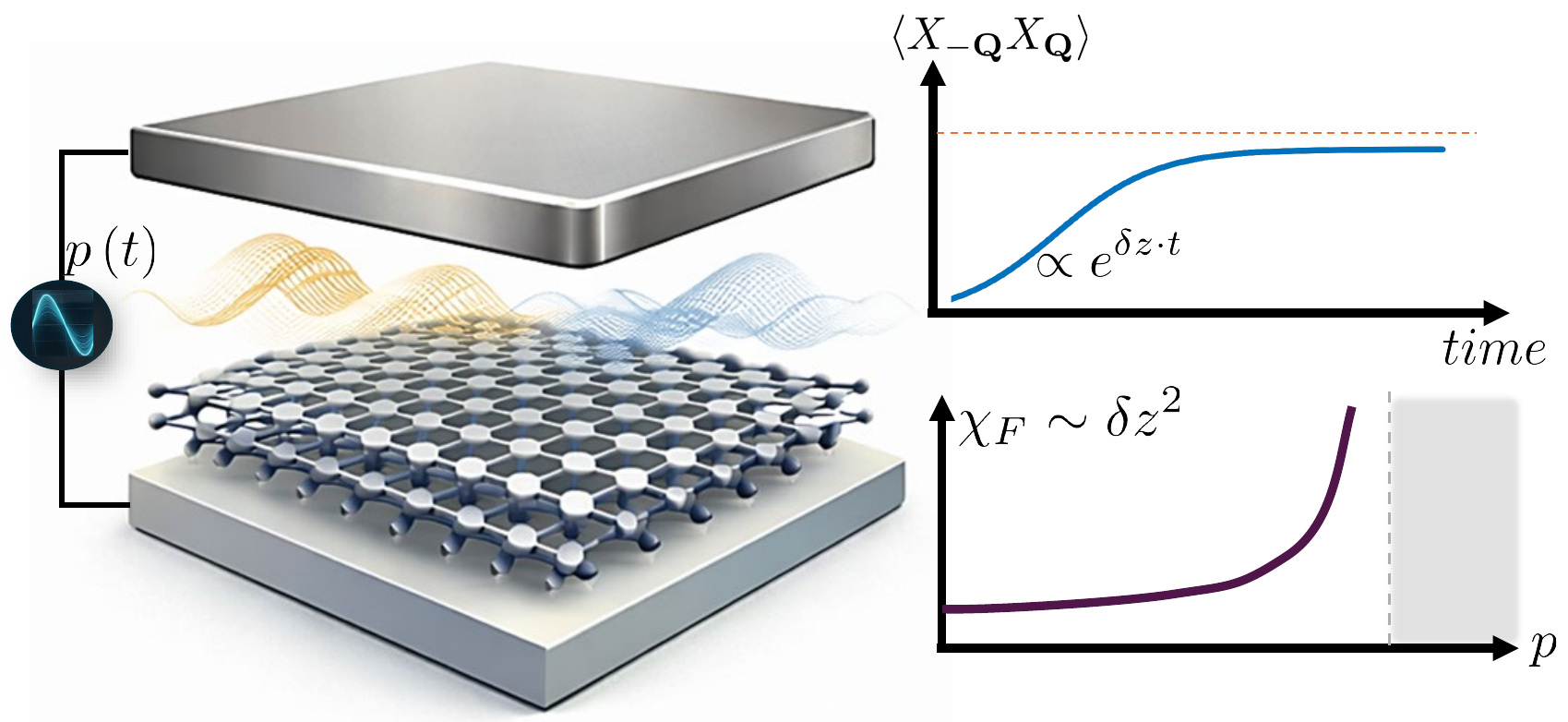}
    \caption{
    \textbf{Driving parametric instabilities of correlated electrons.}
    We consider periodic modulation of the microscopic parameters of a strongly interacting material, represented schematically by $p\left(t\right)$ (e.g. ac gate-voltage modulation).
    As a result, the effective low-energy degrees of freedom may experience a collective resonant instability, where the the fluctuations around the correlated ordered state grow exponentially.
    The latter can eventually saturate, and manifest a non-equilibrium steady state with textured excitations.
    The response of the system is intimately related to the \textit{fidelity susceptibility} $\chi_{F}$ of the correlated ground state, which tends to diverge in the vicinity of phase transitions.
    \label{fig:fig1}}
\end{figure}

\begin{figure*}
    \centering
    \includegraphics[width=18cm]{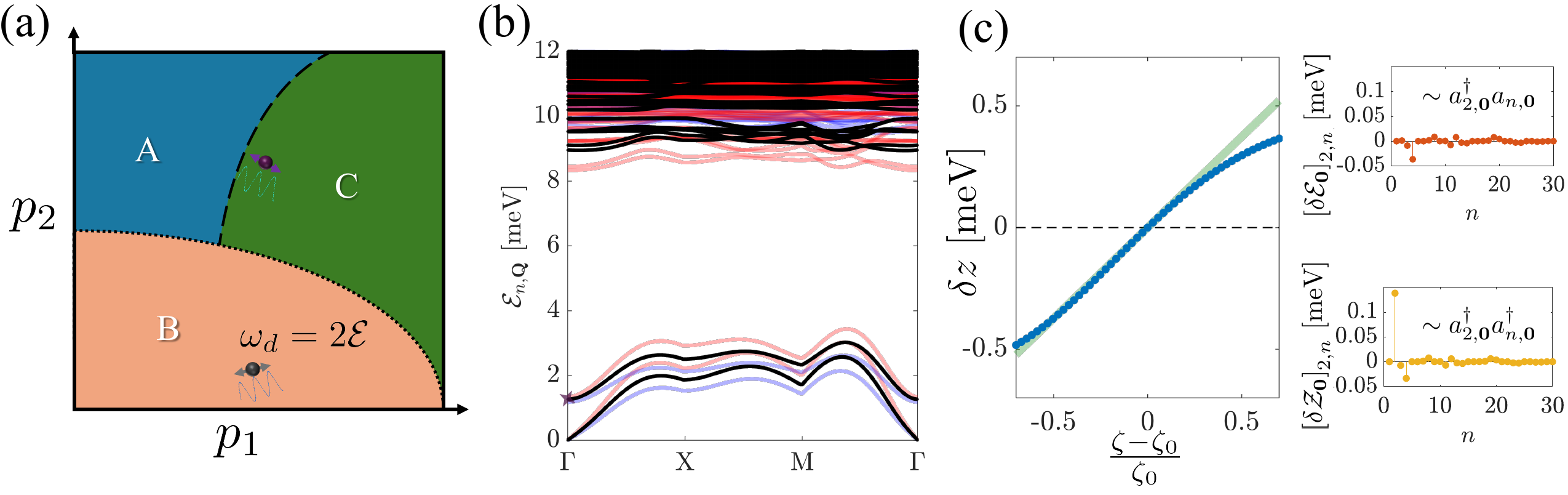}
    \caption{
    \textbf{    Parametric driving of collective bosonic excitations.}
    (a)
    A generic two-parameter ($p_1,\,p_2$) phase diagram, with distinct phases A, B, and C.
    The system is perturbed without hitting a phase boundary (e.g., gray circle in B, purple circle in C) by periodically modulating $p_1$ and $p_2$ at a frequency $\omega_d$.
    Low-lying collective excitations within a given phase, could exhibit a parametric instability and amplification, if the modulation's driving frequency coincides with twice the energy of a collective mode.
    The severity of this instability relates to the quantum fluctuations in a given phase, and thus may depend on its proximity to various phase boundaries.
    (b)
    The low-lying collective excitations for the flat-band model we study in Sec.~\ref{sec:TBGproxy} (black).
    The model has a quantum-geometry-related parameter $\zeta$, set to the value $\zeta=\zeta_0$.
    By slightly modifying this parameter, either increasing (red traces) or reducing $\zeta$ (blue traces) by 10\%, the collective mode spectrum is altered.
    (c)
    Left:
    For the $n=2$, ${\bf Q=0}$ mode, marked by a purple star in (b), we plot $\delta z$, which is the strength of the induced two-boson term (blue) as a function of deviation from $\zeta_0$.
    The light green line has the slope $\frac{\partial\delta_z}{\partial\zeta}|_{\zeta\to\zeta_0}$, representing the parametric susceptibility with respect to $\zeta$.
    Right:
    Matrix elements of the perturbed Hamiltonian, see Eq.~\eqref{eq:bosonicparametric}.
    For perturbing $\zeta_0\to1.2\zeta_0$
    we present the hybridization of this mode with other modes (top), and the respective two-boson term (bottom).
    The peaked contribution at $n=2$ is the extracted $\delta_z$ for this value of $\zeta$.
    }
    \label{fig:MAINWIDE}
\end{figure*}

In this work we study \textit{parametric driving of the low-lying bosonic collective modes} in correlated electronic systems. 
Particularly, we explore the effect of periodic modulation of the control parameters of the system at a frequency $\omega_d$, which matches the sum of the frequencies of two bosonic excitations, see Fig.~\ref{fig:MAINWIDE}a.
The control parameters we consider are typical experimental knobs such as electronic density, applied electric displacement field, or the screening of electronic interactions~\cite{tunableinteraction_jiali_doi:10.1126/science.abb8754,tunablinteraction_manchester_barrier2024coulombscreeningsuperconductivitymagicangle,tunableinteraction_ohio_gao2024doubleedgedroleinteractionssuperconducting}.
Such modulations also affect the soft-mode spectrum as demonstrated in Fig.~\ref{fig:MAINWIDE}b, in the microscopic model which we introduce in \ref{sec:casestudies}.

Parametric drives are particularly intriguing since they connect changes of underlying system parameters with an exponentially fast build-up of amplitude. 
We first discuss how an effective parametric drive is achieved (Sec.~\ref{sec:parametricsusceptibility}). 
Effective modulations must induce a time-dependent two-boson Hamiltonian term. 
Whether such a term emerges, and how strong it is, depend crucially on the microscopic details and dynamical properties of the targeted excitations as well as the suggested knob. 
In Sec.~\ref{sec:collectivederivation} we develop a general yet straightforward theoretical framework for deriving the properties of collective modes and their parametric-driving terms. 
We show how the parametric driving emerges, and connect its strength with the quantum-geometric response of the underlying many-body ground state to the driving knobs.

Our approach reveals that parametric driving can only be mediated by modulations that \textit{alter the vacuum state} of the targeted excitations. 
In fact, parametric driving strength is a direct probe of the quantum ground-state fluctuations.
Fig.~\ref{fig:MAINWIDE}c highlights how modulation of a microscopic Hamiltonian parameter translates to the effective Hamiltonian of the low-energy collective fluctuations.
It is the induced two-boson term (bottom right panel of Fig.~\ref{fig:MAINWIDE}c) that corresponds to the strength of the parametric driving.
Remarkably, the same term is intimately related to the susceptibility of the vacuum of fluctuations.
Thus, efficiently driving and probing the parametric instability of a correlated phase with respect to microscopic parameters is also a probe of the fidelity susceptibility of the ground state. This, in turn, provides an independent mapping of phase boundaries, and may reveal hidden transitions.


Triggering parametric instabilities could have dramatic signatures.
Once the parametric response of the system exceeds the inverse-lifetime of the driven mode, one expects the parent order to melt, albeit in a nonthermal way (cf. Ref.~\cite{HsiehParametricMagnons}).
If the parametric drive is counterbalanced by nonlinear dissipation terms (due to, e.g., interactions between the bosonic modes or more intricate sources of non-linearity), however, the driving can result in a non-equilibrium steady-state~\cite{parametric_magnon_bec_Demokritov2006,nonlineardamping_topologicalmagnon,parametric_afm_theory_PhysRevApplied.17.034004,kaplanprl_PhysRevLett.134.066902}.
Curiously, thoughtful targeting of specific collective excitations may drive the system towards a low-lying order just outside the reach of the equilibrium phase diagram. 

Sec.~\ref{sec:casestudies} concentrate on two examples of current interest: quantum Hall bilayers and moir\'e graphene.
We work out their relevant low-energy excitations, their susceptibility to parametric driving and relate those to the energetical hierarchy of competing many-body ground states.
These examples demonstrate the ubiquitous interplay between quantum geometry and collective modes in the presence of substantial electronic interactions (cf. Ref.~\cite{magnongeometrydassarma}).
Specifically, it appears that systems are most susceptible to changes of quantum geometric parameters when the interaction range is comparable to the length scale set by quantum geometry.
We further relate these microscopic examples to realistic systems, propose which parameters are most promising for driving, the modulation amplitudes necessary, and discuss measurable experimental consequences of the respective instabilities.

Sec.~\ref{sec:conclusions}, provides a summary of our results, and discusses further implications of this study.
Namely, the role collective excitations in correlated electronic materials play in scenarios involving a quantum quench, and the vast potential such systems hold for quantum information applications when the low-lying bosonic modes are coupled non-trivially to other systems.

\section{Parametric susceptibility}\label{sec:parametricsusceptibility}

In order to examine how can driven collective excitations result in parametric instabilities, we begin by reviewing the problem of parametric resonance in the simplest setting possible: a quantum harmonic oscillator.
Parametric resonances are well-established as a classical phenomenon. Nonetheless, to make contact with our findings in this work (and the formalism employed in Sec.~\ref{sec:collectivederivation}) it will be instructive to consider the quantum system.

\subsection{Harmonic oscillator}
Consider the Hamiltonian (setting $\hbar=1$ henceforth),
\begin{equation}
    H_{\rm harm.} = \frac{1}{2}\left(\frac{1}{m}\mathbb{p}^2+k\mathbb{x}^2 \right)
    =\omega\left(a^\dagger a+\frac{1}{2}\right),\label{eq:QHO}
\end{equation}
where $\mathbb x$ and $\mathbb p$ are canonically-conjugate operators, $\left[\mathbb{x},\mathbb{p}\right]=i$,
$\omega=\sqrt{\frac{k}{m}}$,
and
$a=\frac{1}{\sqrt{2}}\left(\sqrt{m\omega}\mathbb{x}+\frac{i}{\sqrt{m\omega}}\mathbb{p}\right)$ is a bosonic annihilation operator.

How does a small change of the bare parameters in Eq.~\eqref{eq:QHO} affect the system?
Concretely, modifying $k\to k+\delta k$, $\frac{1}{m}\to \frac{1}{m}- \frac{\delta m}{m^2}$, brings about a two-boson term $\delta z\left(a^2+a^{\dagger 2}\right)$~\footnote{Introducing a $\propto\left\{\mathbb{x},\mathbb{p}\right\}$ perturbation will change the relative phase between $a^2$ and $a^{\dagger 2}$.}, with
\begin{equation}
    \delta z= \frac{1}{2}\omega\left(\frac{\delta k}{k}+\frac{\delta m}{m}\right).\label{eq:QHOparametricterm}
\end{equation}
Consider periodically modulating this small-amplitude term with frequency $\omega_d$ and phase $\varphi_d$, i.e.,
\begin{equation}
    H_{\rm harm.}\to H_{\rm harm.} +\delta z\cos\left(\omega_dt+\varphi_d\right).
\end{equation}
On \textit{parametric resonance}, when $\omega_d=2\omega$, solving the equations of motion for $a/a^\dagger$ in the interaction picture, one finds exponential growth in the number of bosonic excitations $\hat{n}=a^\dagger a$,
\begin{equation}
    \hat{n}\left(t\right)\propto\exp\left(\left|\delta z\right|t\right).
\end{equation}
This is the essence of parametric resonance in a harmonic bosonic system.

An immediate consequence of Eq.~\eqref{eq:QHOparametricterm} is that parametric driving necessitates changing the ratio $k/\left(\frac{1}{m}\right)$.
This ratio is nothing but the oscillator ground-state fluctuation squeezing (Fig.~\ref{fig:schematicfig}a), i.e.,
\begin{equation}
    \frac{\langle\mathbb{p}^2\rangle_0}{\langle\mathbb{x}^2\rangle_0}=km,\label{eq:qhosqueez}
\end{equation}
with $\langle\cdot\rangle_0$ evaluated in the ground state.
This simple example of a quantum harmonic oscillator provides us a key insight going forward:
In order to achieve parametric resonance in a system by changing its parameters, one must \textit{alter its ground state}, or the so-called fluctuation vacuum.
Tuning parameters which effectively control the squeezing of the different fluctuation quadratures will indeed have the desired effect.
Identification of the fluctuation quadratures, the effective $\mathbb x$ and $\mathbb p$ of the collective modes, will be a key step in our analysis.

\subsection{Application of parametric driving to collective modes: Preliminaries}

Let us extend this structure for harmonic approximations of collective many-body modes. 
Our Hamiltonian $H$ will invariably be given in terms of electronic annihilation operators, 
$c_{{\bf k}\mu}$, for electrons with momentum $\bf k$, and internal index $\mu$, denoting, e.g., spin, valley, band and so on. 
In addition, we denote $|\Psi_0\rangle$ as a reasonable approximation to the ground state of $H$. 
For example, considering plasmons, magneto-rotons, or Goldstone modes as the collective excitations of interest, the appropriate reference states are the uniform electron gas, the partially filled Landau level, and the spontaneous-symmetry-broken state, respectively. 
We further assume stability of this reference state is, i.e., $|\Psi_0\rangle$ remains intact as the relevant reference state of $H\pm\delta H$ with a small perturbation $\delta H$.

Next, we would like to assess the existence and potency of parametric driving of collective modes about the reference state  $|\Psi_0\rangle$. 
For that, we must construct a quantum theory of these bosonic fluctuations (see Sec.~\ref{sec:collectivederivation}).
One then encounters two important objects of this theory relevant to the question of parametric driving (and more generally, to that of quantum quenches):
The vacuum of fluctuations, $|{\rm Vac}\rangle$, and its energy $E_{\rm Vac}$.
Notice $|{\rm Vac}\rangle$ is generally different than the original reference state $|\Psi_0\rangle$.
The former is a renormalized version of the latter, accounting for the zero-point motion of low-lying fluctuations (cf. the difference between the BCS variational state and the generalized RPA ground state in Ref.~\cite{AndersonCollective_PhysRev.112.1900}).

\begin{figure}
    \centering
    \includegraphics[width=8.8cm]{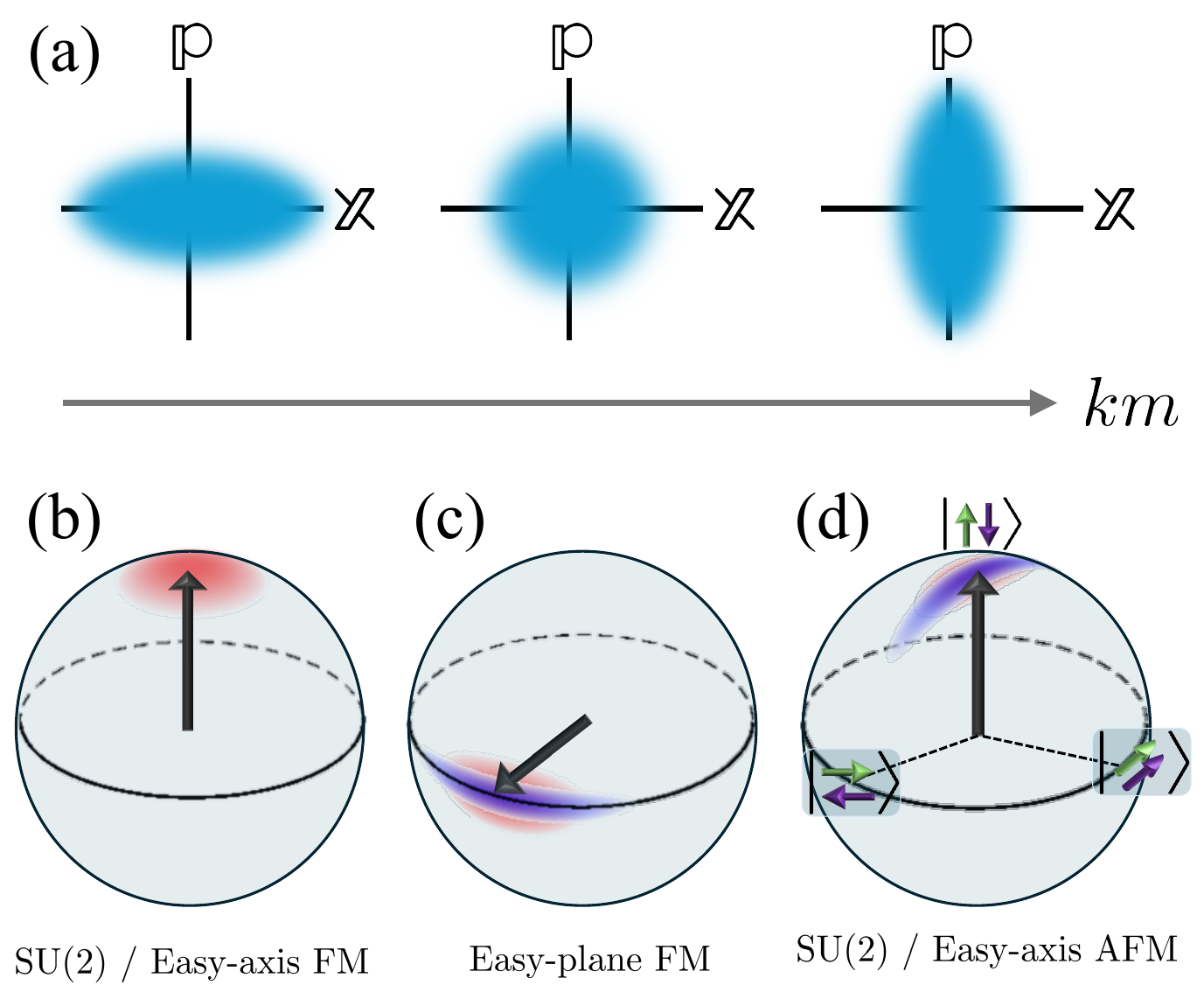}
    \caption{
   \textbf{Squeezing of vacuum fluctuations for generic ordered states.}
    (a)
    For the quantum harmonic oscillator, the ratio of uncertainty of the two quadratures in the ground state depends on parameters of the oscillator, see Eqs.~\eqref{eq:QHO},\eqref{eq:qhosqueez}.
    The squeezing in phase space is shown schematically as it evolves with the squeezing parameter $km$.
    (b)
    Schematic depiction of the Bloch sphere for an SU(2) symmetric or easy-axis (along $\hat{z}$) ferromagnet.
    The ordering direction is indicated by the black arrow.
    Small fluctuations about the ordered state are represented by the red cloud.
    A change in the parameters of the parent Hamiltonian of this phase cannot change the nature of the fluctuations without explicitly breaking the remaining U(1) symmetry.
    (c)
    Bloch sphere of an easy-plane ferromagnet.
    The out-of-plane direction is inherently harder for the vacuum fluctuations as compared to in-plane.
    The fluctuation vacuum is thus squeezed (red oval cloud).
    Modification of Hamiltonian parameters will generically alter the amount of the aforementioned squeezing, depicted by the elongated blue cloud.
    The vacuum of fluctuations is susceptible to such parameter changes.
    (d)
    Generalized Bloch sphere for an antiferromagnet.
    The north pole represents the ordered state.
    The two fluctuation quadratures correspond to either a rotation of the N\'eel vector, (an easy direction) or to ferromagnetic alignment (hard direction).
    Squeezing, sensitive to microscopic Hamiltonian parameters, is thus an unavoidable consequence.
    }
    \label{fig:schematicfig}
\end{figure}

We expect a periodic modulation of a perturbation $\delta H$ to provide the parametric driving. 
The potency of $\delta H$ is encoded into the sensitivity of the vacuum to it.
Consider the fidelity susceptibility with respect to $\delta H$~\cite{ZanardiFidelityPhysRevE.74.031123},
\begin{equation}
    \chi_{F}^{\delta H}=-\frac{\partial^2\left|\left\langle{\rm Vac}\left(H\right)|{\rm Vac}\left(H+\lambda\delta H\right)\right\rangle\right|}{\partial\lambda^2}|_{\lambda\to0},\label{eq:mainsusceptibilities}
\end{equation}
and the energy susceptibility:
\begin{equation}
    \kappa^{\delta H}=-\frac{\partial^2E_{\rm Vac}}{\partial\lambda^2}|_{\lambda\to0}.\label{eq:susceptmain}
\end{equation}
These two quantities control the two-boson terms in the Hamiltonian, necessary for parametric driving.
Consequently, \textit{a perturbation which does not alter the vacuum of quantum fluctuations cannot induce parametric instabilities}.

As an example, consider a ferromagnet with either SU(2) or easy-axis symmetry, see Fig.~\ref{fig:schematicfig}a.
Small departures from the ordered state must manifest as finite magnetization fluctuation in the directions perpendicular to the order.
On the other hand, these fluctuations are symmetric considering the remnant U(1) symmetry of the ordered state, as long as the change in the Hamiltonian does not alter the ordered state or spoil the unbroken symmetry.
Therefore, the ferromagnet will be impervious to parametric driving of this sort. 
This is the case in recently investigated moir\'e transition metal dichalcogenides (TMDs), which exhibit a valley-polarization order parameter of the easy-axis category~\cite{TMD_kinfai0Li2021,valleypolTMD_WU0_doi:10.1126/science.adg4268,valleypolTMD_WU1_Cai2023,TMD_VP_KINFAI_Zeng2023,valleypolTMD_WU2_Park2023}.
Though some properties of these ordered states are tunable by a gate voltage, they remain fundamentally outside the purview of this work, due to the restrictive nature of their symmetry breaking.

Conversely, consider an easy-plane ferromagnet, Fig.~\ref{fig:schematicfig}b.
After spontaneously ordering in some direction on the equator of the associated Bloch sphere, small out-of-plane fluctuations are greatly suppressed, while the softer in-plane collective excitation are not.
The vacuum of fluctuations is squeezed, and the degree of squeezing will depend strongly on generic Hamiltonian parameters, such as the exchange anisotropy.
Inevitably, the vacuum wavefunction is sensitive to such alterations, and parametric instabilities could then arise. 
Quantum Hall double layers are a good example of this category~\cite{QH_bilayerexp1_PhysRevLett.68.1379,QH_bilayerexp2_PhysRevLett.68.1383,QH_bilayerexp3_PhysRevLett.72.728}, and are our first case study (see Sec.~\ref{sec:QHbilayer} below).

An SU(2) (or easy-axis) \textit{antiferromagnet}, is a less trivial example of a squeezed vacuum  (Fig.~\ref{fig:schematicfig}c). 
Consider the N\'eel vector in the $\hat{z}$ direction. 
Its rotation towards the $\hat{x}$ or $\hat{y}$ directions can be considered as different modes of fluctuations.
However, unlike the ferromagnet, they are not conjugate coordinate-canonical-momentum pair~\cite{auerbach2012interacting}, see Appendix~\ref{app:magnets_goldstone_details}. Squeezing is therefore natural for an antiferromagnet as well.


Furthermore, the linear dispersion expected for the Goldstone modes in easy-plane Ferromagnets, and in antiferromagnets (as opposed to a quadratic mode in SU(2) ferromagnets) is a hallmark of squeezing (see Appendix~\ref{app:magnets_goldstone_details}).

\subsection{Parametric efficiency}

Practically speaking, one is interested in the precise strength of the perturbation-induced parametric term.
Given the expected exponential growth induced by it, it is reasonable to compare the energy of the two-boson term of interest, $\delta z$, to the energy of the driven collective mode ${\cal E}_{\bar{n},{\bf Q}}$.
This is an effective measure of the number of evolution cycles of that mode before the parametric amplification is pronounced, i.e., the exponential prefactor exceeds $O\left(1\right)$.

We define the \textit{parametric efficiency}  $\eta_{{\rm par}.}^{\left(p\right)}$ with respect to a small change $\delta p$ in the control parameter $p$ [in direct analogy to Eq.~\eqref{eq:QHOparametricterm}, where the efficiency is unity with respect to $k$ and $m^{-1}$],
\begin{equation}
    \frac{\left|\delta z\right|}{{\cal E}_{\bar{n},{\bf Q}}}\equiv\frac{1}{2}\left|\frac{\delta p}{p}\right|\eta_{{\rm par}.}^{\left(p\right)}\label{eq:parefficiency}
\end{equation}
In line with the preceding discussion, we find [see Eq.~\eqref{eq:etapargen}] that the parametric efficiency may exceed unity and diverge as the squeezing of the vacuum fluctuations becomes more pronounced.

We note that linear dissipation would compete with the driving, leading to reduced effective efficiency  $\tilde{\eta}_{\rm par.}^{\left(p\right)}=\eta_{\rm par.}^{\left(p\right)}\left(1-\frac{{\cal E}_{\bar{n},{\bf Q}}}{Q\delta z}\right)$, where $Q$ is the quality factor of the driven mode. 

\section{Collective modes: derivation and characterization}\label{sec:collectivederivation}




\subsection{Derivation of the collective mode Hamiltonian}

As we focus our attention on the properties of the lowest-lying collective modes, we can describe our system's ordered states by an electronic Slater determinant denoted $|\Psi_0\rangle$. 

The ordered state is characterized by the density matrix,
\begin{align}
    P_{\mu \nu}\left({\bf k}\right)&=\left\langle\Psi_0{\Large|}c^\dagger_{{\bf k}\mu}c_{{\bf k}\nu}{\Large|}\Psi_0\right\rangle\nonumber\\
    &\equiv\frac{1}{2}\left(\delta_{\mu\nu}+{\cal O}_{\mu\nu}\left({\bf k}\right)\right).\label{eq:densitymatrix}
\end{align}
With no loss of generality, we assume that the density matrix is diagonal in momentum $\bf k$.
Spontaneous formation of density wave orders, which defy this assumption, may be captured by working in a basis where one assigns an additional flavor index to sections of the Brillouin zone before coupling them, i.e., a folded-zone scheme.

The operator $\cal O$ can be thought of as the order parameter of $|\Psi_0\rangle$.
The Slater determinant density matrix is a projection at any $\bf k$ point, 
$\left[P\left(\bf k\right)\right]^2=P\left(\bf k\right)$ (its eigenvalues are all $0$ or $1$).
Thus, $\left[{\cal O}\left({\bf k}\right)\right]^2 =1$, and one finds $\frac{1}{\Omega}\sum_{\bf k}\left\langle{\cal O}\right({\bf k}\left)\right\rangle=n$,
where $n\equiv\frac{1}{\Omega}\sum_{\bf k}{\rm Tr}P_{\bf k}$ is the density.

Henceforth, we will assume for simplicity that $P$ (and therefore $\cal O$) are $\bf k$-independent matrices.
In some cases, this may entail applying a unitary transformation, e.g., a Bogoliubov transformation, in order to bring $P$ to this desired form.
This assumption restricts our present discussion to cases where the ordered state lacks a Fermi surface.
This restriction to insulating or semi-metallic ordered states is not necessary in principle.
It may be relaxed, within the same framework presented below, by a more complicated treatment of the fluctuations as compared to the one that follows (cf. Ref.~\cite{TBGskyrmion}).

Next, let us parametrize small fluctuations about the ordered $|\Psi_0\rangle$ state using:
\begin{equation}
    |\Psi_\phi\rangle \equiv e^{-i{\hat F[\phi]}}|\Psi_0\rangle.
\end{equation}
$\hat F$ is an hermitian {\it fluctuation generator}. Note that $|\Psi_\phi\rangle$ remains a normalized wavefunction. Also, we restrict the fluctuation subspace by restricting $\hat{F}$ to second-order in electronic creation and annihilation operators.
Following the time-dependent Hartree-Fock approach (using elements from  Refs.~\cite{TBGsoftmodes,magnongeometrydassarma,TBGskyrmion}), we define the operator ${\hat F}$ as:
\begin{equation}
    {\hat F}[\phi] = \sum_{\bf Q,p} c_{{\bf p+Q}\mu}^{\dagger}\phi_{{\bf Q}}^{\mu\nu}\left({\bf p}\right)c_{{\bf p}\nu}.\label{eq:Ffluctautiongenerator}
\end{equation}
(Summation over repeated greek indices is implied.)
Notice that one may relate the fluctuation matrix to its hermitian conjugate,
$\phi_{{\bf -Q}}\left({\bf p+Q}\right)=\phi^\dagger_{{\bf Q}}\left({\bf p}\right)$.

The matrix $\phi_{\bf Q}$ consists of the various complex amplitudes that parametrize the perturbation away from the approximate ground state. 
Crucially, pieces of the matrix $\phi_{\bf Q}$ which commute with $\cal O$ would not create a particle-hole excitation of the $|\Psi_0\rangle$ state, and lead to $|\Psi_\phi\rangle$ possessing the same density matrix as in Eq.~\eqref{eq:densitymatrix}.
Thus, $\phi_{\bf Q}$ anticommutes with the order parameter, $\left\{\phi_{\bf Q},{\cal O}\right\}=0$.

This suggests decomposing the fluctuation matrix in terms of the set $\left\{{{\cal M}^i}\right\}$, denoting the independent set of matrices in the electronic internal Hilbert space which anti-commute with $\cal O_{\mu\nu}$,
\begin{equation}
    \phi^{\mu\nu}_{{\bf Q}}\left({\bf p}\right)=\sum_{i}\left[{{\cal M}^{i}}\right]^{\mu\nu}\Phi_{{\bf Q}}^{i}\left({\bf p}\right).\label{eq:fluctuationgeneratordecomposed}
\end{equation}
$\Phi^i_{\bf Q}(\bf p)$ is now just the momentum-dependent complex amplitudes of the various anticommuting directions relative to the broken symmetry. 

To obtain the dynamics of small fluctuations $|\Psi_\phi\rangle$ we derive the Lagrangian to lowest (quadratic) order in $\Phi$ (which will be promoted from c-numbers to dynamic variables),
\begin{equation}
    {\cal L=}{\cal B}-{\cal H},
\end{equation}
where we calculate the following,
\begin{equation}
    {\cal B}= \langle\Psi_\phi| i\frac{d}{dt}|\Psi_\phi\rangle,
    \,\,\,\,\,
    {\cal H}= \langle\Psi_\phi| H|\Psi_\phi\rangle.
\end{equation}

This may be written in a compact form by using the vector $\Phi_{\bf Q}$ spanning both fluctuation-generator space and momentum space.
Specifically, the kinetic term has the form (see Appendix~\ref{app:action_details} for details),
\begin{equation}
    {\cal {B}}=\sum_{\bf Q}\frac{1}{2}\Phi_{{\bf Q}}^{\dagger}{\cal \tilde{B}}_{{\bf Q}}\frac{d}{dt}\Phi_{{\bf Q}},\label{eq:Bmatrixvectorized}
\end{equation}
\begin{equation}
    {\cal \tilde{B}}_{{\bf Q}}^{ij}\left({\bf p,p'}\right)={\rm Tr}\left\{ -\frac{i}{2}\left[{\cal M}^{i},{\cal M}^{j}\right]{\cal O}\right\} \delta_{{\bf pp'}}.\label{eq:Bmain}
\end{equation}
This expression has notable consequences.
The $\cal B$ term in the effective action determines the commutation relations between the  operators in the quantized fluctuation Hamiltonian.
Eq.~\eqref{eq:Bmain} indicates that $\cal O$ defines  pairs of conjugate canonical variables analogous to $p{\dot q}$ terms in classical Lagrangian formulations.
The operators $\cal O$, $p$, and $q$ are determined by forming a triplet product with non-vanishing (and imaginary) trace, ${\rm Tr}\left\{\left[p,q\right]{\cal O}\right\}\neq0$.

The simplest example is an order parameter encoded by a Pauli $s_z$ operator, with canonically conjugate variables $s_x,s_y$.
Indeed, in the ordered phase, $s_z$ may be approximated by its expectation value, and
$\left[s_x,s_y\right]\sim i\left\langle s_z\right\rangle\propto i$, a canonical commutation relation.
We take great pains to emphasize this identification, as it becomes crucial in determining the parametric susceptibility of any system of interest (including the harmonic oscillator example illustrated in the previous section).

Expanding once more to lowest order in fluctuations, we calculate the second part of the action,
\begin{equation}
    {\cal H}\approx E_0-\frac{1}{2}\langle\Psi_0| \left[{\hat F},\left[{\hat F},H\right]\right]|\Psi_0\rangle,\label{eq:Hexpansion}
\end{equation}
which leads to an expression similar to~\eqref{eq:Bmatrixvectorized} for the Hamiltonian part,
\begin{equation}
    {\cal {H}}=\sum_{\bf Q}\frac{1}{2}\Phi_{{\bf Q}}^{\dagger}{\cal \tilde{H}}_{{\bf Q}}\Phi_{{\bf Q}}.\label{eq:Hmatrixvectorized}
\end{equation}
The exact details ${\cal \tilde{H}}_{{\bf Q}}^{ij}\left({\bf p,p'}\right)$ follow the microscopic calculation of the expectation value of the double commutator in Eq.~\eqref{eq:Hexpansion}, see Appendix~\ref{app:action_details}.

The equation of motion extracted from the effective Lagrangian is,
\begin{equation}
    i\frac{d}{dt}\Phi_{\bf Q}=i{\cal \tilde{B}}^{-1}_{{\bf Q}}{\cal \tilde{H}}_{{\bf Q}}\Phi_{\bf Q}
    \equiv
    {\cal D}_{{\bf Q}}\Phi_{\bf Q},\label{eq:EOM}
\end{equation}
where ${\cal D}_{{\bf Q}}$ is the dynamic matrix, which has eigenvalue pairs $\pm{\cal E}_{n,{\bf Q}}$.
The spectrum of collective excitation is comprised of the positive eigenvalues of this matrix~\footnote{The collective mode spectrum is more commonly recovered by solving the characteristic equation
$\det\left\{ i\omega\tilde{{\cal B}}_{{\bf Q}}-\tilde{{\cal H}}_{{\bf Q}}\right\} _{\omega={\cal E}_{n,{\bf Q}}}=0$~\cite{auerbach2012interacting}, which is of course an equivalent method.}. The spectrum of collective modes is not our main object of interest here. Instead, we are interested in the situations where a small perturbation $\delta H$ to the equilibrium Hamiltonian $H$ is capable of driving a parametric instability.

Conceptually, our program would proceed as follows:
First, construct bosonic variables out of the canonically conjugate pairs implied by Eq.~\eqref{eq:Bmain}.
A generalized Bogoliubov transformation is then required to diagonalize $\cal H$, such that 
\begin{equation}
    {\cal H}=\sum_{n,\bf Q}{\cal E}_{n,{\bf Q}}a_{n,{\bf Q}}^\dagger a_{n,{\bf Q}}.\label{eq:Hdiagonalstatic}
\end{equation}
The fluctuation amplitudes $\Phi_{\bf Q}$ can be related to the $a_{n,{\bf Q}}$ eigenmodes by the transformation [see Eqs.~\eqref{eq:Ddiagonalization},\eqref{eq:diagonalizedH}],
\begin{equation}
    a_{n,{\bf Q}}=\left[T^{-1}_{\bf Q}\Phi_{\bf Q}\right]_n.
\end{equation}
Formally, we may now promote the $a_{n,{\bf Q}}$ variables to quantum 
bosonic annihilation operators, obtaining the quantum Hamiltonian of the low-energy degrees of freedom.

\subsection{Parametric term}

We proceed by extracting the \textit{strength of the parametric driving} introduced by a small perturbation $\lambda\delta H$ to the microscopic Hamiltonian.
This may be accomplished by projection of the perturbation into the low-energy subspace of bosonic fluctuations.
[The perturbation-free Hamiltonian of these latter degrees of freedom is given by Eq.~\eqref{eq:Hdiagonalstatic}.]
Concretely, what is the amplitude of the anomalous two-boson terms in this projected Hamiltonian?
Generically, the contribution of the microscopic perturbation 
has the general form,
\begin{equation}
    \delta{\cal H}=\frac{1}{2}\sum_{{\bf Q}}\begin{pmatrix}a_{{\bf Q}}^{\dagger} & a_{-{\bf Q}}\end{pmatrix}\delta{\Lambda}_{\bf Q}\begin{pmatrix}a_{{\bf Q}}\\
a_{-{\bf Q}}^{\dagger}
\end{pmatrix}, \label{eq:bosonicparametric}
\end{equation}
\begin{equation}
    \delta{\Lambda}_{\bf Q}
    =
    \begin{pmatrix}\delta{\cal E}_{{\bf Q}} & \delta {\cal Z}_{{\bf Q}}\\
\delta {\cal Z}_{{\bf Q}}^{\dagger} & \delta{\cal E}_{{\bf Q}}
\end{pmatrix},
\end{equation}
where $a_{\bf Q}$ is a column vector of all the different $a_{n,{\bf Q}}$ which diagonalize the unperturbed excitation Hamiltonian~\eqref{eq:Hdiagonalstatic}.
Clearly, the \textit{off-diagonal}  $\delta{\cal Z}_{\bf Q}$ part of this matrix [analogous to Eq.~\eqref{eq:QHOparametricterm} in the single oscillator case] encodes the parametric susceptibility -- our main concern in this work. 

Practically speaking, the most challenging part of the scheme described above is the Bogoliubov transformation.
This step becomes exceedingly difficult for large vectors $\Phi_{\bf Q}$ -- one has to solve a set of $O\left(N^2\right)$ coupled second-order equations, where $N$ scales with the volume of the system.
Instead, one may carry out the program above in one fell swoop, by considering the dynamic matrix ${\cal D}_{\bf Q}$~\cite{xiao2009theoryBosonsBogoliubovEOM}.
It is readily diagonalized by a similarity transformation,
\begin{equation}
    T_{\bf Q}^{-1}{\cal D}_{\bf Q}T_{\bf Q}= \begin{pmatrix}{\cal E}_{{\bf Q}} & 0\\
0 & -{\cal E}_{{\bf Q}}
\end{pmatrix}.\label{eq:Ddiagonalization}
\end{equation}
${\cal E}_{{\bf Q}}$ is a diagonal matrix whose entries are the positive ${\cal E}_{n,{\bf Q}}$.
Now, we wish to make use of this transformation to diagonalize the bosonic Hamiltonian, and arrive at Eq.~\eqref{eq:Hdiagonalstatic}.
As detailed in Ref.~\cite{xiao2009theoryBosonsBogoliubovEOM} (also see Appendix~\ref{app:remarkbogo}), the same matrices $T_{\bf Q}$ also encode the Bogoliubov transformation itself, \textit{diagonalizing the Hamiltonian} by hermitian congruence,
\begin{equation}
    T_{\bf Q}^{\dagger}{\cal \tilde{H}}_{{\bf Q}}T_{\bf Q}=
    \begin{pmatrix}{\cal E}_{{\bf Q}} & 0\\
0 & {\cal E}_{{\bf Q}}
\end{pmatrix}.\label{eq:diagonalizedH}
\end{equation}
We stress that the transformation matrix $T_{\bf Q}$ is generally not unitary.
Finally, $\delta{\cal H}$ can be read off as,
\begin{equation}
 \delta{\Lambda}_{\bf Q}
    =
    T_{\bf Q}^{\dagger}
    \left(
    {\cal \tilde{H}}_{{\bf Q}}^{\left[H+\lambda\delta H\right]}
    -
    {\cal \tilde{H}}_{{\bf Q}}^{\left[H\right]}
    \right)
    T_{\bf Q},\label{eq:partialbogoliubov}
\end{equation}
We note that $T_{\bf Q}$ above is the one corresponding to the unperturbed $H$, i.e., it diagonalizes ${\cal \tilde{H}}_{{\bf Q}}^{\left[H\right]}$.

In the context of parametric driving, we consider modulation of the perturbation $\delta H$ with frequency $\omega_d$.
The main objects of interest are thus specific elements of the matrices $\left[\delta {\cal Z}_{\bf Q}\right]_{nm}$, such that
\begin{equation}
    {\cal E}_{n,{\bf Q}} + {\cal E}_{m,{\bf Q}}=\omega_d, \label{eq:resonancecondition}
\end{equation}
which is the parametric resonance condition.

\subsection{Fidelity susceptibility}
The off diagonal parts of the matrix $\delta{\cal \tilde{H}}_{\bf Q}$ directly encode the parametric terms induced by the perturbation.
Given this anomalous perturbation, let us examine what is the residual Bogoliubov transformation that diagonalizes the full \textit{perturbed} Hamiltonian.
It is given by
\begin{equation}
    \left(T^{\left[H\right]}_{\bf Q}\right)^{-1}
    T^{\left[H+\lambda\delta H\right]}_{\bf Q}
    \equiv
    \begin{pmatrix}U_{\bf Q} & V_{\bf Q}\\
V_{\bf Q}^{*} & U_{\bf Q}^{*}
\end{pmatrix}.
\end{equation}
Notice that in the $\lambda=0$ the residual transformation is the trivial identity.

Clearly, the off-diagonal component $\delta{\cal Z}_{\bf Q}$ in the perturbed Hamiltonian is finite iff $V_{\bf Q}$ is non-zero.
In the small perturbations limit $\lambda\to 0$, the parametric susceptibility and the residual transformation are directly related to one another.
For a purely off-diagonal $\delta\Lambda_{\bf Q}$, a perturbative calculation yields
\begin{equation}
    \frac{\partial\|V_{{\bf Q}}\|}{\partial\lambda}|_{\lambda\to0}
    \approx
    \sqrt{\sum_{mn}
    \frac{\left|\left[
    \frac{\partial}{\partial \lambda}\delta{\cal Z}_{{\bf Q}}\right]_{mn}\right|^2}
    {\left({\cal E}_{m,{\bf Q}}+{\cal E}_{n,{\bf Q}}\right)^2}
    }
   ,\label{eq:proprtionality}
\end{equation}
where $\|\cdot\|$ is the Frobenius norm.


We may now consider the fidelity of the unperturbed fluctuation vacuum $|{\bf {0}}_{\bf Q}\rangle$ (Appendix~\ref{app:remarkbogo}),
\begin{equation}
    \left|\langle{ {\bf 0}_{{\bf Q}}^{\left[H\right]}}
    |
    { {\bf 0}_{{\bf Q}}^{\left[H+\lambda\delta H\right]}}\rangle
    \right|=
    \left[{\det\left(\mathbb{1}_N+V_{\bf Q}V_{\bf Q}^{\dagger}\right)}\right]^{-1/4}.
\end{equation}
The momentum resolved fidelity susceptibility,
\begin{equation}
    \chi_{F,{\bf Q}}^{\delta H}=\frac{1}{2}\left(\frac{\partial\|V_{{\bf Q}}\|}{\partial\lambda}|_{\lambda\to0}\right)^{2}.\label{eq:explicitfidelity}
\end{equation}
Taken together with Eq.~\eqref{eq:proprtionality}, we find that the magnitude of the parametric driving is roughly proportional to the square root of the Fidelity susceptibility of the fluctuation vacuum.
In the next section, we make this connection even more concrete with a generic example.

\subsection{Isolated driven mode}
It is instructive to consider a microscopic perturbation $\delta H$, that creates an efficient parametric excitation in a single mode, labeled by
$\left(\bar{n},{\bf Q}\right)$.
The periodic driving of $\delta H$ adheres to the resonant condition for this mode, i.e.,
\begin{equation}
    2{\cal E}_{\bar{n},{\bf Q}}=\omega_d.
\end{equation}
For simplicity, we assume the mode is decoupled from all other modes, $\left[\delta{\cal Z}\right]_{n\bar{n}}\approx 0 ,\,\,\,\forall n\neq\bar{n}$, and we denote the induced two-boson term \[\delta z=\left[\delta{\cal Z}\right]_{\bar{n}\bar{n}}.\]
The formalism discussed above is general for arbitrary connectivity of the modes, yet for the salient features we explore here (and in the illustrative examples below) it suffices to consider this simpler isolated-mode case.

Since the parametric driving is expected to be extremely selective due to the resonance frequency, one expects contribution mainly from the driven mode to the susceptibilities introduced in Eq.~\eqref{eq:mainsusceptibilities}.
Decomposing the vacuum zero-point energy to a sum of its components, and the vacuum ground-state to a tensor products of all the different modes,
\begin{equation}
    E_{\rm Vac}=\sum_{{\bf Q},n}E^0_{{\bf Q},n},
    \,\,\,\,\,\,\,
    |{\rm Vac}\left(H\right)\rangle
    =\bigotimes_{n,{\bf Q}}|{\bf 0}_{n,{\bf Q}}\rangle,
\end{equation}
consider the mode resolved susceptibilities,
\begin{equation}
    \tilde{\chi}_{F}^{\delta H}=-\frac{\partial^2\left|\left\langle
    {\bf 0}^{\left[H\right]}_{n,{\bf Q}}
    |
    {\bf 0}^{\left[H+\lambda\delta H\right]}_{n,{\bf Q}}
    \right\rangle\right|}{\partial\lambda^2}|_{\lambda\to0},\label{eq:resolvedsusceptibilities}
\end{equation}
\begin{equation}
    \tilde{\kappa}^{\delta H}=-\frac{\partial^2E^0_{{\bf Q},n}}{\partial\lambda^2}|_{\lambda\to0}.\label{eq:resolvedsusceptibilities2}
\end{equation}
Straightforward calculation (see Appendix~\ref{app:single_mode}) reveals the direct relation between these susceptibilities and the so-called \textit{parametric susceptibility}, $\frac{\partial \left|\delta z\right|}{\partial\lambda}$,
\begin{equation}
     \frac{\partial \left|\delta z\right|}{\partial\lambda}=
     2{\cal E}_{\bar{n},{\bf Q}}\sqrt{
     \tilde{\chi}_{F}^{\delta H}
     },
     \,\,\,\,\,\,
     \frac{\partial \left|\delta z\right|}{\partial\lambda}=
     \sqrt{
     {\cal E}_{\bar{n},{\bf Q}}\tilde{\kappa}^{\delta H}
     }
     .\label{eq:seuscuptibiltyPrincipal}
\end{equation}
Notice these results align perfectly with Eqs.~\eqref{eq:proprtionality} and~\eqref{eq:explicitfidelity}.
We have thus demonstrated that parametric instabilities of the correlated system directly probe the quantum fluctuations in the reference ordered state.
Introduction of a parametrically resonant perturbation is a sensitive measure of the response of quantum fluctuations to this perturbation.

What determines the susceptibility to a given perturbation?
Making an additional simplification, we assume the single mode above is also decoupled in the original $\Phi$ basis, which is often the case (see below).
In that case we may fully characterize a general (mode resolved) fluctuation Lagrangian with the 2-by-2 matrices,
\begin{equation}
    \tilde{\cal B}_{\bar{n},\bf Q}=\begin{pmatrix} & 1\\
-1
\end{pmatrix},
\,\,\,\,\,
\tilde{{\cal H}}_{\bar{n},{\bf Q}}=\begin{pmatrix}\omega+\left|\upsilon\right|\cos\phi & \left|\upsilon\right|\sin\phi\\
\left|\upsilon\right|\sin\phi & \omega-\left|\upsilon\right|\cos\phi
\end{pmatrix},\label{eq:Hsingle}
\end{equation}
with $\upsilon=\left|\upsilon\right|e^{i\phi}$.
Upon diagonalization, the energy associated with this mode is ${\cal E}_{\bar{n},{\bf Q}}=\sqrt{\omega^2-\left|\upsilon\right|^2}$ (physically stable Lagrangians have $\omega\geq\left|\upsilon\right|\geq0$).
It is useful to define a \textit{squeezing measure},
\begin{equation}
    r=\sqrt{\frac{\omega-\left|\upsilon\right|}{\omega+\left|\upsilon\right|}}.\label{eq:rdefinition}
\end{equation}
Notice $r\in\left[0,1\right]$, and that $r\to 0$ in the limit of maximum squeezing.

Consider now the infinitesimal perturbation,
\begin{equation}
\omega\to\omega+d\omega,\,\,\,\,\,\left|\upsilon\right|\to\left|\upsilon\right|+d\upsilon,\,\,\,\,\,\phi\to\phi+d\phi.
\end{equation}
Straightforward calculation(Appendix~\ref{app:single_mode}) then reveals
The dependence of the parametric (two-boson) term $\delta z$ on the infinitesimal change of fluctuation parameters $\omega$, $\upsilon$, and $\phi$,
\begin{equation}
   \delta z=-\frac{\omega+\left|\upsilon\right|}{2}\left[\frac{\left(1+r^{2}\right)^{2}}{2r}d\left(\frac{\left|\upsilon\right|}{\omega}\right)+i\left(1-r^{2}\right)d\phi\right].\label{eq:dzMain} 
\end{equation}
Eqs. \eqref{eq:Hsingle} and \eqref{eq:dzMain} together constitute a convenient \textit{recipe} towards extracting the parametric susceptibility of the system with respect to a given perturbation.
After detailed calculations [using Eqs.~\eqref{eq:Bmain} and \eqref{eq:Hexpansion}], we employ this recipe for several case studies in Sec.~\eqref{sec:casestudies} below.

It is instructive to consider the parametric efficiency defined in Eq.~\eqref{eq:parefficiency} with respect to the $\upsilon$ and $\omega$ variables.
It is expressed entirely in terms of the squeezing parameter,
\begin{equation}
    \eta_{{\rm par}.}^{\left(\upsilon,\omega\right)}=\frac{1-r^{4}}{2r^{2}},\label{eq:etapargen}
\end{equation}
which \textit{diverges} as the squeezing becomes more extreme, $r\to 0 $.
Thus,  we uncover a dramatic effect -- the more squeezed the equilibrium fluctuation vacuum is, the more susceptible it is to parametric instabilities. 

Why does a modest change of parameters (in the original unperturbed basis) lead to such a drastic behavior?
In that basis, the vacuum state contains a superposition of many excited mode pairs with differing amplitudes, such that the expectation value of the photon number is
$\left\langle n_{a}\right\rangle =\frac{\left(1-r\right)^{2}}{4r}$, diverging with $r\to0$.
The existence of abundant fluctuations in this quantum fluctuation ground state then facilitates a large parametric (or two-boson) response, even if the changes made to the Hamiltonian are exceptionally small.
One may re-phrase this point in terms of squeezing of these original coordinate variables, employing the metaphor of a lemon: the more squeezed the unperturbed vacuum is, the easier it is to further squeeze it.

Furthermore, Eq.~\eqref{eq:dzMain} informs us that parametric driving in such system only occurs if one of the following conditions applies to the perturbation:
Either (i) it modifies the ratio between normal and anomalous term in the fluctuation Hamiltonian, or (ii) it modifies the relative phase of the anomalous term.
We stress once again that the former condition further measures and is highly sensitive to the degree of squeezing in the equilibrium vacuum.

Solution of the equations of motion in the periodically driven case (Appendix~\ref{app:single_mode}) reveals exponential growth in time for both quadrature components of the associated $\Phi_{\bf Q}$ vector proportional to $\sim e^{\left|\delta z\right|t}$.
At long times, $t\gg\left|\delta z\right|^{-1}$, the two quadratures have a relative $\pi/2$ phase.
The ratio between their (exponentially growing) amplitudes is $\sim r$,
remnant of the squeezing in the equilibrium fluctuation vacuum.

The exponential growth signals the \textit{instability of the ordered phase} with respect to the parametric resonance of this mode.
Dissipation however, may alter the eventual fate of the system (cf., Ref.~\cite{kaplan2025opticallyinducedfaradaygoldstonewaves}).
Consider linear dissipation, independent of ``photon'' number, inducing a finite collective-mode lifetime $\tau$.
A necessary condition for the parametric instability is  $\left|\delta z\right|>\tau^{-1}$.
We note that one source of dissipation expected for the systems we discuss, namely interactions with the particle-hole excitation continuum (e.g., the so-called Landau damping), is absent for the specific scenarios considered above.

Non-linear dissipation, with a rate proportional to the photon number, may lead to even richer phenomenology.
Such dissipation allows the exponential growth at early times, before reaching a non-equilibrium steady state with counter-balancing dissipation and parametric drive~\cite{nonlineardamping_topologicalmagnon}.
In this state the $\Phi_{\bf Q}$ quadratures display persistent oscillations at a frequency ${\cal E}_{n,{\bf Q}}=\omega_d/2$, establishing a novel Floquet phase of matter~\cite{phononFloquetmatterHübener2018,TRarpes_exciton_floquetdoi:10.1073/pnas.2301957120,vivek_floquet_exp_pareek2024drivingnontrivialquantumphases}.
In the next section, where we discuss concrete microscopic examples, we will encounter the ramifications and signatures of such exotic phases.

\begin{figure*}
    \centering
    \includegraphics[width=18cm]{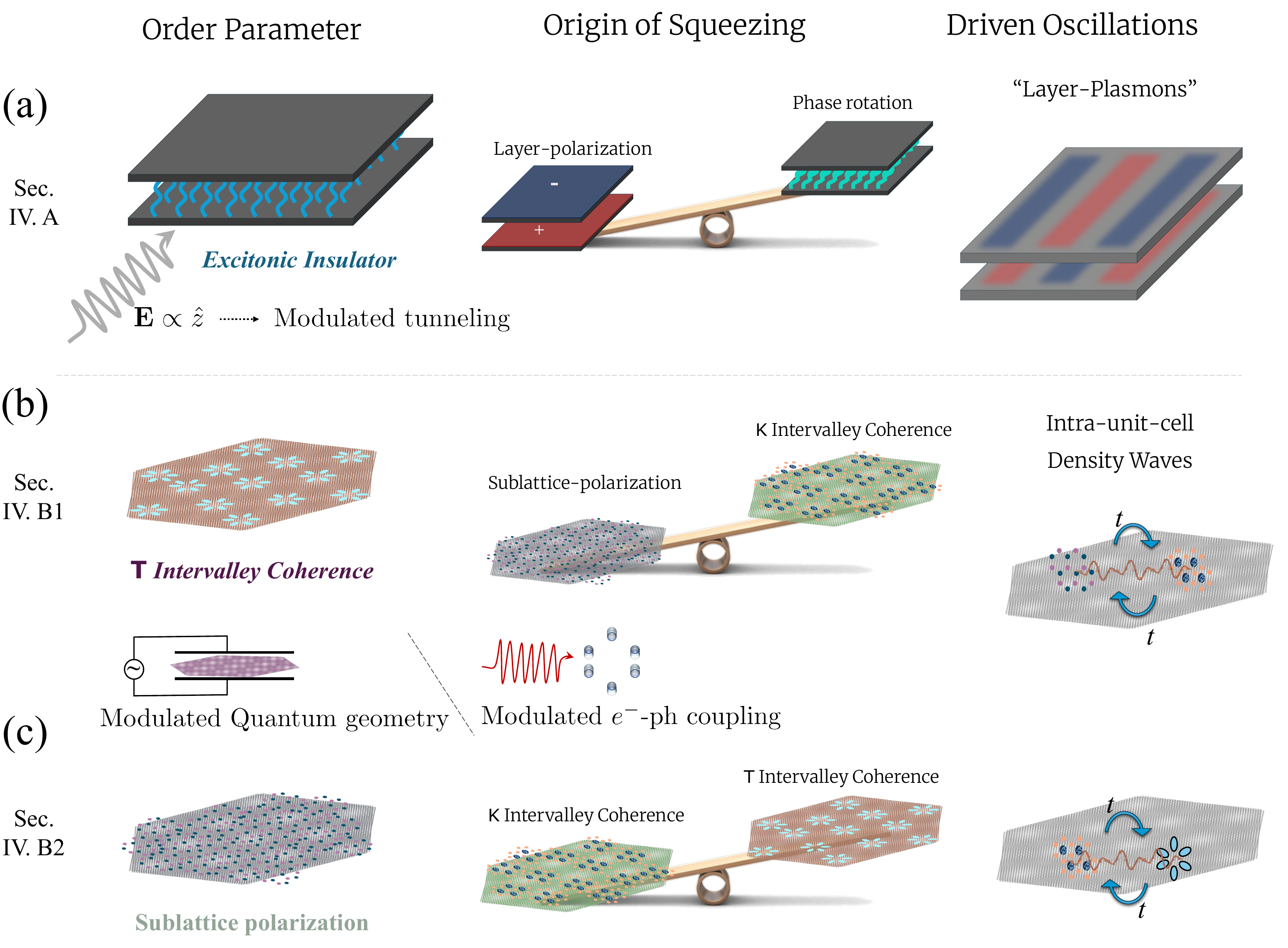}
    \caption{
    \textbf{Summary of the physical examples studied in Sec.~\ref{sec:casestudies}.}
    (a)
    In Sec.~\ref{sec:QHbilayer} we study the parametric instability of the layer-coherent excitonic insulating phase of a correlated quantum Hall bilayer.
    The instability can be driven by, e.g., Floquet modulation of the interlayer tunneling amplitude (see text).
    The apparent squeezing between the two quadratures of the low-energy collective excitation stems from the finite energy cost associated with charging the effective two-plate capacitor.
    Conversely, rotating the phase of the coherent condensate is relatively much easier for the system.
    In the driven state, the system experiences enhanced fluctuations of interlayer charge displacement, which are similar to plasmons (yet are exactly compensated between the layers).
    (b)
    In Sec.~\ref{sec:Torderflat} we examine a particular intervalley coherent ground state of moir\'e graphene.
    The graphic on the left panel illustrates a particular signature of this phase in an STM experiment.
    Focusing on one branch of the low-energy collective modes associated with this phase, we trace the squeezing in this mode to the hierarchy of competing Hartree-Fock ground states:
    One quadrature of the mode, corresponding to mixing a different intervalley coherent state into the order parameter, is lower in energy and thus easier for the system as compared to the conjugate quadrature (which involves sublattice polarization at the graphene scale).
    Upon resonant parametric driving, the inter-moir\'e unit-cell charge distribution, discernible by STM, fluctuates in time between sublattice polarization and sublattice-balanced patterns.
    (c)
    Finally, Sec.~\ref{sec:sigmaorderflat} considers sublattice-polarized order in moir\'e graphene.
    The two conjugate quadratures of the low-lying collective mode now correspond to components of two different intervalley coherent states.
    As a consequence, the driven state will display temporal fluctuation between a Kekul\'e pattern (corresponding to the T-IVC) and a trivial pattern (from the K-IVC component).
    Relevant drives for both cases, (b)--(c), are an electric displacement field (which modulates the quantum geometry), and non-linear phonon pumping (controls the electron-phonon coupling).
    }
    \label{fig:casestudysummary}
\end{figure*}

\section{Case studies}\label{sec:casestudies}
In the following we will explore several possible realizations of parametrically driven correlated materials, each highlighting different facets of our general discussion above.
For each system, we will present the relevant correlated phase of matter, and derive the corresponding bosonic fluctuations.
A summary of the phases of matter we study, the physical origin of squeezing in their collective modes, and their physical implication appears in Fig.~\ref{fig:casestudysummary}.
We discuss the susceptibility of these fluctuations to parametric driving using experimentally accessible probe, and show when it becomes anomaly strong by considering the appropriate parametric efficiency figure of merit [see Eq.~\eqref{eq:parefficiency}].

\textit{Quantum Hall double layer.--}
(Fig.~\ref{fig:casestudysummary}a)
A gate voltage controls the transition out of the layer-coherent phase of interest.
The parametric efficiency with respect to its modulation is of order unity [Eqs.~\eqref{eq:qhpareff}].
We estimate that a modulation of order $\sim$mV in the gate is sufficient to induce an appreciable effect:
Depending on non-universal details of the intrinsic dissipation in the system, the interlayer coherent order will either be (temporarily) destroyed, or a steady-state with an oscillating interlayer electric field will be established.
The oscillations will occur at half the driving frequency, and may be detected by tunneling spectroscopy, a well-established technique in the study of such systems.

\textit{Twisted graphene toy model.--}
(Fig.~\ref{fig:casestudysummary}b--c)
We explore two different ordering possibilities in such systems.
In a valley-coherent ground state, the parametric efficiency diverges as $\Delta_{\rm TK}^{-1}$ [Eq.~\eqref{eq:etautk}], where $\Delta_{\rm TK}$ is the energy difference between the ordered mean-field ground state and its closest competitor.
This parameter is intimately related to electron-phonon coupling of specific modes~\cite{TIVCphonons}, making it  susceptible to periodic driving of these modes~\cite{Floquet_parametric_Kiselev2024}.
Parametrically driving the appropriate collective mode promotes a time-periodic sublattice polarization in the system (with half the driving frequency), which could be transiently detected by a second-harmonic-generation probe.

Conversely, when the electronic order corresponds to sublattice polarization, the parametric efficiency diverges in a manner inversely proportional to the distance from the phase transition to the lowest lying competing order [Eq.~\eqref{eq:parametricsuszeta}].
We discuss how these divergences lead to detectable experimental signatures, even with small parameter modulations (e.g., $\sim$ 10 mV/nm of the applied displacement fields). 
For example, a time resolved STM probe would detect suppression of the ordered phase and oscillations in the real space charge distribution.

\subsection{Quantum Hall double layer (Fig.~\ref{fig:casestudysummary}a)}\label{sec:QHbilayer}
We first consider a double layer quantum Hall system at total odd integer filling $\nu=2n+1$~\cite{QHbilayerMacdonald,QHbilayerZhang}.
Both layers experience equal external chemical potential, such that in the absence of interactions and interlayer tunneling both are at filling factor $n+\frac{1}{2}$.
Our analysis below easily captures the salient features of many strongly-interacting (or flat-band) double layers, expected to show a similar (or generalized) excitonic ordered state, including electron-hole double layers~\cite{excitonic_eh_inasgas}, graphene or bilayer graphene double layers~\cite{Bilayergraphenesuperfluidbistritzer,doblelayrBLG_Dean}, TMD double layers~\cite{novoselovtmdproposal_Fogler2014,kinfaimak_excitonic_tmd,kinfaimak_excitonic_tmd2,kinfaicoulombdragtmd}, and moir\'e systems~\cite{doubleTBLGexcitonic,tbg_excitonic}.

Projecting the problem to the $n$th Landau level, the degrees of freedom are described by fermionic annihilation operators $c_{\bf k}=\left(c_{{\bf k}1},c_{{\bf k}2}\right)^T$ belonging to the two layers, and Pauli operators $\ell_i$, such that $\ell_z=\pm 1$ for the different layers.  
The projected microscopic Hamiltonian we consider has three parts,
\begin{equation}
    H_{\rm QH-2\ell}=H_{\rm C} + H_z+H_{\rm tun.}
\end{equation}
The interaction terms are given by
\begin{equation}
    H_{\rm C}=\frac{1}{2A}\sum_{\bf q}V_{\bf q}^0\tilde{\rho}_{\bf q}\tilde{\rho}_{\bf -q},
    \,\,\,\,\,
    H_z=\frac{1}{2A}\sum_{\bf q}V_{\bf q}^z\tilde{\rho}^z_{\bf q}\tilde{\rho}^z_{\bf -q},\label{eq:HChz_QH}
\end{equation}
where the projected density operators are
\[\tilde{\rho}_{\bf q}=\sum_{\bf k}c^\dagger_{\bf k+q}\Lambda_{{\bf k},{\bf k+q}} c_{\bf k}
\,\,\,\,\,
\tilde{\rho}^z_{\bf q}=\sum_{\bf k}c^\dagger_{\bf k+q}\ell_z\Lambda_{{\bf k},{\bf k+q}} c_{\bf k},\]
\[\Lambda_{{\bf k},{\bf k+q}}^{\alpha\beta}=\delta^{\alpha\beta}f\left({\bf q}\right)e^{-i\frac{{\Omega}}{2}{\bf k}\times\left({\bf k+q}\right)}
,\]
with
$f\left({\bf q}\right)=e^{-\frac{1}{4}{\Omega}\left|{\bf q}\right|^{2}}L_n\left({\Omega}\left|{\bf q}\right|^{2}\right)$ (${\Omega}=l_B^2$, i.e., the magnetic length squared, $L_n$ is the $n$ Laguerre polynomial).

With nominal layer separation $d$, the two interaction strengths are given by
\begin{equation}
    V_{{\bf q}}^{0/z}=\frac{V_{{\bf q}}^{C}\left(1\pm e^{-qd}\right)}{2},
\end{equation}
with $V_{{\bf q}}^{C}$ the Fourier transformed Coulomb repulsion in two dimensions.
The $H_z$ interaction therefore accounts for the difference between intra- and interlayer interactions.
The zero momentum component of $V_{{\bf q}}^{z}$ is the capacitive energy of the effective capacitor formed by the ``double-plate'' structure~\cite{QHbilayerMacdonald}.
As such, the system is expected to disfavor layer-polarization ($\ell_z$ order), instead tending towards inter-layer coherence, forming an effective easy-plane $U\left(1\right)$ ferromagnet.

The $U\left(1\right)$ symmetry may be broken \textit{explicitly} by single-particle tunneling between the layers~\footnote{The ``$\rm sas$'' label indicates that in the absence of interactions, the interlayer tunneling splits the energy of the layer-symmetric and layer-antisymmetric electron wavefunctions by an energy of $\pm\Delta_{\rm sas}$.},
\begin{equation}
    H_{{\rm tun.}}=-\Delta_{\rm sas}\sum_{{\bf k}}c_{{\bf k}}^{\dagger}\ell_x c_{{\bf k}}.
\end{equation}
One expects the appearance of this term to open a gap in the otherwise gapless Goldstone mode, demonstrated below.

Our reference ordered state may thus be characterized by the matrix
${\cal O}=\ell_x$, such that the relevant fluctuation generators are readily found to be $\left\{{{\cal M}^i}\right\}=\ell_y,\,\ell_z$.
The fluctuations correspond to a rotation of the coherent interlayer phase ($\ell_y$) and layer charge imbalance ($\ell_z$).
These identifications allow us to employ the scheme outlined in Sec.~\ref{sec:collectivederivation} in a straight forward manner, which we carry out in detail in Appendix~\ref{app:QHbilayer}.
Focusing our attention on the lowest lying Goldstone mode mentioned above, we recover the following effective single-mode parameters,
\begin{equation}
    \omega^{\rm QH-2\ell}_{\bf Q}
    =
    \frac{\rho^0_s}{2}\left|{\bf Q}\right|^2+\frac{1}{2}E_{\rm easy} + \Delta_{\rm sas},
\end{equation}
\begin{equation}
    \upsilon^{\rm QH-2\ell}_{\bf Q}
    =
    \frac{\rho^z_s}{2}\left|{\bf Q}\right|^2+\frac{1}{2}E_{\rm easy} ,
\end{equation}
where $\rho_s^0$ and $\rho_s^z$ are stiffness parameters proportional to integrals over $V^0_{\bf q}$ and $V^z_{\bf q}$, respectively, and the easy-plane anisotropy is
\begin{equation}
    E_{\rm easy}=\frac{1}{A}\sum_{{\bf q}}\left[V_{{\bf 0}}^{z}-V_{{\bf q}}^{z}f^{2}\left({\bf q}\right)\right].\label{eq:Eez}
\end{equation}
At small momenta, the spectrum of this mode is
\begin{equation}
    {\cal E}_{{\bf Q}}=\sqrt{\left({\cal E}_{{\bf 0}}^{{\rm QH-2\ell}}\right)^{2}+\left(\Delta_{{\rm sas}}\rho_{s}^{0}+E_{{\rm easy}}\frac{\rho_{s}^{0}-\rho_{s}^{z}}{2}\right)\left|{\bf Q}\right|^{2}},
\end{equation}
with the gap at zero momentum
\begin{equation}
    {\cal E}_{\bf 0}^{\rm QH-2\ell}=\sqrt{\Delta_{\rm sas}\left(E_{\rm easy}+\Delta_{\rm sas}\right)}.
\end{equation}
Notice that for $\Delta_{\rm sas}=0$, the collective mode becomes gapless and has linear dispersion, since the system spontaneously (instead of explicitly) breaks the $U\left(1\right)$ symmetry.

To demonstrate and quantify the parametric instability, let us consider driving the lowest-energy ${\bf Q}=0$ mode, i.e., modifying the parameters of the system at a frequency $\omega_d=2{\cal E}_{\bf 0}^{\rm QH-2\ell}$.
The strength of the parametric term is obtained by employing Eq.~\eqref{eq:dzMain}, 
\begin{equation}
    \delta z^{\rm QH-2\ell}=\frac{{\cal E}_{\bf 0}^{\rm QH-2\ell}}{2}\frac{\delta\left[\log\frac{\Delta_{{\rm sas}}}{E_{{\rm easy}}}\right]}{1+\frac{\Delta_{{\rm sas}}}{E_{{\rm easy}}}},\label{eq:dzQH2l}
\end{equation}
and the corresponding parametric efficiencies,
\begin{equation}
    \eta_{{\rm par}.}^{\left(\Delta_{{\rm sas}}\right)}=\eta_{{\rm par}.}^{\left(E_{{\rm easy}}\right)}=\frac{E_{{\rm easy}}}{E_{{\rm easy}}+\Delta_{{\rm sas}}}\approx1,\label{eq:qhpareff}
\end{equation}
where the last approximation is reasonable in most experimental systems, where the tunneling energy is smaller by orders of magnitude compared to the interaction anisotropy energy scale.
The upshot of Eq.~\eqref{eq:dzQH2l} is that parametric driving of the correlated double-layer system may be induced by any perturbation that modifies the energetic ratio between the interlayer tunneling and the energy associated with spontaneously creating a charge imbalance between the layers.

One direct way to achieve this is to modulate the tunneling barrier between the layers.
In a double quantum well setup~\cite{bilayertunneling_eisenstein}, this can be achieved by electrostatic modulation, effectively changing the wavefunction hybridization between the two wells making up the double layer.
The interlayer tunneling may also be controlled using light~\cite{Floquettunneling1_PhysRevB.99.201403,Floquettunneling2_PhysRevB.101.241408}.
One can modulate the amplitude of the incident light on the system at the appropriate parametric resonance frequency~\cite{LindnerPlamonkiselev2024inducingexceptionalpointsenhancing}, and thus create temporal oscillations of $\Delta_{\rm sas}$.

Alternatively, one may seek to modulate $E_{\rm easy}$, the easy-plane anisotropy.
As can be seen from the dependence of Eq.~\eqref{eq:Eez} on $f\left({\bf q}\right)$, quantum geometry plays an important role.
By modulating a perpendicular displacement field in, e.g., a double bilayer graphene system~\cite{graphene_double_layr_JIALI_DEAN}, one induces changes to the wavefunction of the active Landau levels, thus changing $E_{\rm easy}$.



Drawing from the experimental results in relevant systems~\cite{graphene_double_layr,graphene_double_layr_JIALI_DEAN}, the transition out of the layer-coherent phase has a width of order $\sim10$ mV in terms of the applied gate-voltage.
This suggests that in the vicinity of the transition a \textit{modulation of several mV} of the voltage applied are sufficient to induce a parametric instability, which becomes apparent after a few ($\lesssim 10$) driving cycles.

We note that the dependence of the collective mode Hamiltonian parameters on the quantum geometry of the projected wavefunctions (see also Appendix~\ref{app:QHbilayer} and Ref.~\cite{magnongeometrydassarma}, relating the stiffness to quantum geometry)
arises quite ubiquitously, as we demonstrate next.
The ability to temporally modulate the quantum geometric properties of correlated electrons, i.e., their Bloch wavefunctions, is thus a valuable resource in coherently driving collective excitations.

\begin{figure}
    \centering
    \includegraphics[width=9.2cm]{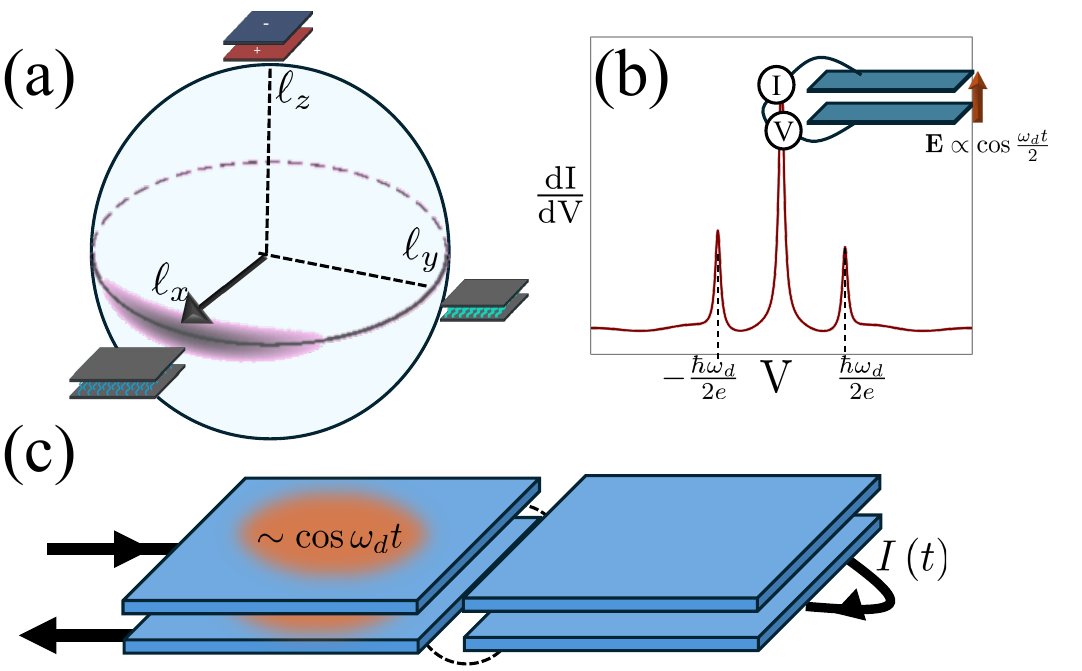}
    \caption{
    \textbf{Driving and probing quantum Hall double layer collective excitations (Sec.~\ref{sec:QHbilayer}).}
    (a)
    Schematic Bloch sphere of the interlayer-coherent ($\ell_x$) ordered phase.
    A phase-rotated coherent phase ($\ell_y$) and layer polarized phase ($\ell_z$) are schematically depicted.
    The squeezing of quantum fluctuations around the ordered state stems from the capacitive energy of the double layer system [Eq,~\eqref{eq:Eez}], manifesting an easy-plane anisotropy.
    (b)
    When the system is parametrically driven, an oscillatory perpendicular electric field is induced in the system.
    Interlayer tunneling spectroscopy is thus expected to display coherent peaks at half the driving frequency (alongside the interlayer-coherence-associated zero-bias peak).
    (c)
    In a Josephson-analogous geometry (weak links denoted by dashed black lines), the relative phase of the condensate between the driven part (driving represented by fuzzy orange oval) and the static part would result in an ac counterflow current $I\left(t\right)$ dominated by a $\omega_d/2$ harmonic.
    \label{fig:qhexample}}
\end{figure}

What are the consequences of the parametrically driven collective mode in the double layer excitonic system?
While the system is pumped, or once a (dissipation-balanced) non-equilibrium steady-state is established, two remarkable things occur in the system.
First, the effective collective isospin of the electrons acquires a finite oscillating layer polarization ($\langle\ell_z\rangle$) component.
The electric field produced by this charge oscillating between the layers is expected to manifest as a finite bias coherent interlayer tunneling peak, that should appear at a bias-voltage corresponding to half of the driving frequency~\cite{bilayertunneling_eisenstein,bilayertunneling_sipersion_eisenstein,bilayerjosephson_macdonald,AdyBilayerTheory}
$V_{\rm bias}=\hbar\omega_d/\left(2e\right)$ (see Fig.~\ref{fig:qhexample}b).
Second, the oscillating $\langle\ell_y\rangle$ (layer coherence) component modulates the coherent phase of the condensate.
A possible route to probe this sort of modulation is by employing an excitonic analog of a Josephson junction (Fig.~\ref{fig:qhexample}c): one part of the system is unperturbed, whereas  another, weakly connected  part is parametrically excited.
One thus expects an ac-current between the two parts of the junction, counterflowing on the top and bottom layers~\cite{QHbialyercounterflow1,QHbialyercounterflow2}.
Through the Josephson relations, this current will be comprised of harmonics of the $\omega_d/2$ frequency, though assuming a small modulation will lead to overwhelming dominance of the primary $\omega_d/2$ component.

\subsection{Complex orders in a flat band model}\label{sec:TBGproxy}
Flat band systems, particularly graphene moir\'e materials, have a richer landscape of collective excitations and a non-trivial complex hierarchy of symmetry breaking states ~\cite{TBG_ESLAM_PRX,tbg_globalphasediagram,TBG_AHE_mechanism,TBG_shavit_WIS,TTG_IKS,TIVCphonons}.
We apply our approach to a model which reproduces the salient features of the hierarchy of ordered states at half-filling relative to the charge neutral point in these systems. 
We will restrict our attention to a toy model which neglects the finite bandwidth of the moir\'e flat bands~\cite{TBG_BM}, the bands' fragile topological properties~\cite{TBG_fragile}, as well as the physics pertaining to the spin degrees of freedom. 
These features are important for determining the phase diagram of a material, but not so much for elucidating the issue of parametric instabilities.

The single-particle Hamiltonian of the two-dimensional (effectively spinless) system we consider here has valley $\tau_{i}$, pseudo-spin (or sub-lattice) $\sigma_{i}$, and orbital $\rho_{i}$ degrees of freedom (all Pauli matrices),
    \begin{equation}
    H_0=-t\sum_{{\bf k}}c_{{\bf k}}^{\dagger}\left[
    \hat{h}_x\left({\bf k}\right)\rho_{x}
    +\hat{h}_y\left({\bf k}\right)\rho_{y}
    -\Delta_\sigma\sigma_z
    \right]c_{{\bf k}}.\label{eq:pseudotbgsingleparticle}
\end{equation}
The matrices $\hat{h}_{x/y}\left({\bf k}\right)$ have internal structure in the $\sigma$-$\tau$ Hilbert space, and their exact form is detailed in Appendix~\ref{app:flatbandzetamodel}.
Notably, $\left[\hat{h}_x\left({\bf k}\right)\right]^2+\left[\hat{h}_y\left({\bf k}\right)\right]^2=1$ for all $\bf k$, resulting in a spectrum of completely flat bands.
The model is related to a tunable quantum metric model introduced in Refs.~\cite{tunablemetricHofmann2022,tunableMetricPRL_PhysRevLett.130.226001}.
The model is characterized by a parameter $\zeta$, corresponding to a quantum geometric length scale (in units of the unit cell) associated with the real-space spread of Wannier functions in the flat bands, referred to henceforth as the quantum metric length~\cite{quantummetriclength_ma2025universalboundarymodeslocalizationquantum}.

The Hamiltonian Eq.~\eqref{eq:pseudotbgsingleparticle} has a total of eight \textit{identically flat} bands: for $\Delta_\sigma=0$ (which we assume henceforth unless mentioned otherwise) one finds four valence bands sitting at energy $-t$, and four conduction bands at $+t$.
Our interest will lie in one set of these isolated bands, e.g., the valence bands.
Within such set, reminiscent of the decomposition of flat bands in twisted graphene, one may define a good quantum number which further classifies the bands, $\sigma_z\tau_z$~
\footnote{In TBG, this matrix would correspond to the actual Chern number associated with each of these bands, yet in our model the Berry curvature vanishes everywhere -- the model is topologically trivial, yet its quantum geometry is highly non-trivial for finite values of $\zeta$.}.

We examine interactions between the flat band electrons, projecting to the lowest lying four bands.
The simplest density-density interaction term (that will dominate the interacting physics) $H_{\rm C}$, has the same form as in Eq.~\eqref{eq:HChz_QH}, except for modified density form factors,
\begin{equation}
    \Lambda_{{\bf kk'}}={\cal F}_{{\bf kk'}}+{\cal D}_{{\bf kk'}}\sigma_z\tau_z.\label{eq:formfactors_tbglike}
\end{equation}
The matrices $\cal F$ and $\cal D$ depend on the quantum geometrical properties of the different flat bands.
Considering the half-filled scenario henceforth, completely filling two bands out of the lowest four, it is straightforward to demonstrate that on the level of Hartree-Fock, the ground state manifold is comprised of all order parameters which commute with $\sigma_z\tau_z$ (see Appendix~\ref{app:flatbandzetamodel}),
\begin{equation}
    \left[{\cal O}_{\rm g.s.},\sigma_z\tau_z\right]=0.\label{eq:gsmanifold}
\end{equation}

Next, let us consider two sub-leading interaction terms, which favor a subset of the symmetry-breaking order parameters in Eq.~\eqref{eq:gsmanifold},
\begin{equation}
    H_{\rm K}=\frac{1}{2A}\sum_{\hat{M}=\sigma_y\tau_x,\sigma_y\tau_y}\sum_{\bf q}u^{\rm K}_{\bf q}\rho^{\hat{M}}_{\bf q}\rho^{\hat{M}}_{\bf -q},\label{eq:Kinteractions}    
\end{equation}
\begin{equation}
    H_{\rm T}=\frac{1}{2A}\sum_{\hat{M}=\sigma_x\tau_x,\sigma_x\tau_y}\sum_{\bf q}u^{\rm T}_{\bf q}\rho^{\hat{M}}_{\bf q}\rho^{\hat{M}}_{\bf -q},\label{eq:Tinteractions}    
\end{equation}
with $\rho^{\hat{M}}_{\bf q}=\sum_{\bf k}c_{{\bf k+q}}^{\dagger}\Lambda_{{\bf k+q,k}}\hat{M}c_{{\bf k}}$ and $u^{\rm K/T}_{\bf q}<0$.

The $H_{\rm K}$ interaction lowers the energy of a phase with finite expectation value of the operator 
$\sigma_y e^{i\theta\tau_z}\tau_x$, which corresponds to the celebrated Kramers intervalley coherent order (KIVC) in twisted bilayer graphene (TBG).
This order has been broadly established as the ground state at half-filling in pristine TBG devices, both analytically and numerically~\cite{TBG_ESLAM_PRX,tbg_globalphasediagram,HoffmanKIVCmonte_PhysRevX.12.011061}.
This order parameter, however, is sensitive to the appearance of strain in realistic devices~\cite{KIVCstraindmrg,tbg_globalphasediagram,KIVC_straindisorder}, often leading to the emergence of a rather different intervalley coherent phase, the so-called Kekul\'e spiral~\cite{IKS_PhysRevX.11.041063}.

Still, even in nearly-pristine devices, with little to no strain, KIVC order is missing.
Instead, a third intervalley coherent phase appears in experiments~\cite{Yazdani_IKS_TIVC} -- the time-reversal invariant intervalley coherent order (TIVC).
In our model, this order is emulated by the order parameter $\sigma_x e^{i\theta\tau_z}\tau_x$, and is thus favored by the $H_{\rm T}$ interaction Hamiltonian.
It has been suggested that electron-phonon interactions involving exchange of valley momentum favor the formation of the TIVC state over KIVC~\cite{TIVCphonons}.
To capture the essence of the competition between these phases we consider the interactions in \eqref{eq:Kinteractions}--\eqref{eq:Tinteractions} short-ranged ($u^{\rm K/T}_{\bf q}$ are independent of the subscript $\bf q$, henceforth omitted), and define
\[u\equiv-\left(u^{\rm K}+u^{\rm T}\right),\]
\[\Delta_{\rm TK}\equiv u^{{\rm K}}-u^{{\rm T}},\]
\[u_\zeta\equiv\left[1-J_{0}^{4}\left(2\zeta\right)\right].\]
The TIVC phase is thus favored when $\Delta_{\rm TK}>0$.
The parameter $u_\zeta$ is related to the energy difference between the lower energy KIVC and TIVC phases and the rest of the Eq.~\eqref{eq:gsmanifold} manifold.
Its dependence on quantum geometric length scale $\zeta$ stems from the form factors in Eq.~\eqref{eq:formfactors_tbglike}, detailed in Appendix~\ref{app:flatbandzetamodel}.
We henceforth assume that $u_\zeta>\Delta_{\rm TK}$, faithfully capturing the hierarchy of competing ordered ground states (see for example Fig.~\ref{fig:TorderFigure}a).

Finally, alignment of the twisted graphene devices to the underlying hexagonal Boron Nitride (hBN) substrate breaks the inversion symmetry.
This effect is approximately captured by in our model by a single-particle sublattice polarization term in Eq.~\eqref{eq:pseudotbgsingleparticle}, $\sim\Delta_\sigma\sigma_z$~\cite{hbnmass1,hbnmass2}.
If this term is large enough, a sublattice-polarized phase, with an order parameter corresponding to $\sigma_z$, will be the energetically favored ground-state.

The following discussion is divided between two scenarios, each motivated by a distinct realization of the proposed model.
The corresponding parametric driving, and ensuing instabilities of the order parameter, will therefore have notable differences.

\subsubsection{T Intervalley Coherence (Fig.~\ref{fig:casestudysummary}b)}\label{sec:Torderflat}
\begin{figure}
    \centering
    \includegraphics[width=9cm]{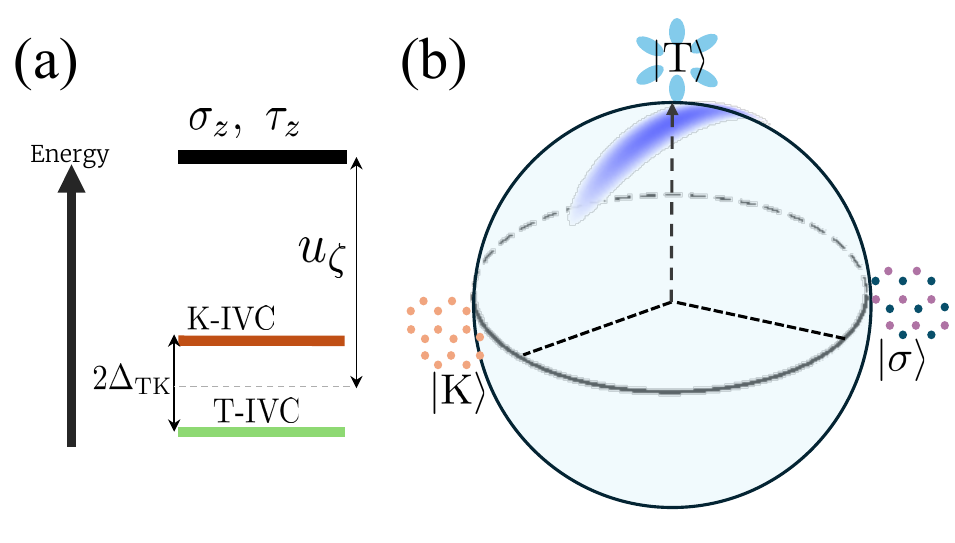}
    \caption{
    \textbf{Fluctuations in the TIVC phase of the flat band model} (Sec.~\ref{sec:Torderflat}).
    (a)
    Energy hierarchy within the low-energy manifold of the possible ordered ground states.
    The energy $u_{\zeta}$ separates the TIVC ($\sigma_x\tau_x$) and KIVC ($\sigma_y\tau_x$) orders from the rest of the manifold, and $\Delta_{\rm TK}$ further favors the TIVC.
    (b)
    The generalized Bloch sphere for the collective mode described by Eqs.~\eqref{eq:TKomega}-\eqref{eq:TKupsilon}, with the TIVC order parameter positioned at the north pole.
    The two fluctuation quadratures correspond to the easy KIVC, and the hard sublattice ($\sigma$) orders.
    The anisotropy of the schematic fluctuation cloud corresponds to the squeezing in the fluctuation vacuum.
    Next to each order we schematically draw real-space spectroscopic signatures of the analogous TBG phase: TIVC admits a Kekul\'e distortion of the graphene unit cell, $\sigma$ corresponds to finite sublattice polarization, and KIVC is not discernible in such experiments~    \cite{IVCspectroscopycalugaru}.
    Modulating the parameter $\Delta_{\rm TK}$ affects the squeezing, and thus enables parametric resonance.
    In the corresponding driven state one observes substantial sublattice oscillations at half the driving frequency.
    \label{fig:TorderFigure}}
\end{figure}

Without explicitly breaking the inversion symmetry, i.e., setting \[\Delta_{\sigma}=0,\]
the order parameter is ${\cal O}_{\rm T}=\sigma_x\tau_x$, where the $\tau_x$ part has been arbitrarily chosen, as the valley $U\left(1\right)$ symmetry is spontaneously broken.
The hierarchy of ordered ground states within the low-energy manifold we consider here is depicted in Fig.~\ref{fig:TorderFigure}a.
Restricting to collective fluctuations which leave the system in its low-energy manifold determined by Eq.~\eqref{eq:gsmanifold}~\footnote{These are operators which generate a rotated reference state $|\Psi_\phi\rangle$ whose corresponding $\cal O$ matrix lies within the low-energy manifold.}, there are four relevant generators to consider,
$\left\{{{\cal M}^i}\right\}=\tau_z,\,\sigma_x\tau_y,\,\sigma_z,\,\sigma_y\tau_x$.

Consulting again with Eq.~\eqref{eq:Bmain}, we can pair the fluctuation amplitudes associated with the operators $\tau_z$ and $\sigma_x\tau_y$ as conjugate variables.
The amplitudes corresponding to $\sigma_z$ and $\sigma_y\tau_x$ are thus a second conjugate pair.
One expects the first pair to give rise to a gapless Goldstone mode. This is because a $\tau_z$ rotation applied to the order parameter generates a very soft fluctuation, namely rotation of the order parameter in the $\tau_x$-$\tau_y$ plane.
Its partner $\sigma_x\tau_y$ acting on the order parameter generates energetically-unfavorable valley-polarization.
The difference in energetics of these two fluctuation quadratures leads to effective squeezing of this bosonic mode.

The second pair of fluctuation matrices gives rise to squeezed fluctuations as well, see Fig.~\ref{fig:TorderFigure}b.
To understand why, consider that the $\sigma_y\tau_x$ piece of $\phi_{\bf Q}$ generates finite sub-lattice polarization in $|\Psi_{\phi}\rangle$.
By our assumptions, this fluctuation is energetically disfavored compared to the KIVC component, whose generation stems from the $\sigma_z$ part of the fluctuation matrix $\phi_{\bf Q}$.
Moreover, since the KIVC order requires finite energy compared to the TIVC ${\cal O}_{\rm T}$ reference state, the collective mode spectrum associated with this second pair of generators has a \textit{finite gap}.
The dispersion of these two low-lying modes is numerically calculated and presented in Fig.~\ref{fig:MAINWIDE}b.
~\footnote{Interestingly, in this model the clear separation to two modes associated with these generator pairs is spoiled at higher momenta. Specifically, the interaction terms include finite coupling between the two modes at nonzero $\bf Q$, see Appendix~\ref{app:flatbandzetamodel}. Parametric driving of this coupling between collective modes of ordered electrons is a promising direction we leave for future work.}

At large wavelengths we can make significant analytical progress.
The gapless Goldstone mode, comprised of $\tau_z$ and $\sigma_x\tau_y$ perturbations, has effective single-mode parameters (Appendix~\ref{app:flatbandzetamodel}), 
\begin{equation}
    \omega^{\rm Goldstone}_{\bf Q}
    =
    \frac{\rho_s}{2}\left|{\bf Q}\right|^2+
    \frac{u_\zeta+\Delta_{\rm TK}}{2},
\end{equation}
\begin{equation}
    \upsilon^{\rm Goldstone}_{\bf Q}
    =
    \frac{u_\zeta+\Delta_{\rm TK}}{2}.
\end{equation}
These parameters correspond to the Hamiltonian of the form of Eq.~\eqref{eq:Hsingle}.
The linear Goldstone dispersion is approximately
${\cal E}^{\rm Goldstone}_{\bf Q}\approx\sqrt{\rho_s\left(u_\zeta+\Delta_{\rm TK}\right)}\left|{\bf Q}\right|$.

The gapped mode however, whose generators are $\sigma_z$ and $\sigma_y\tau_x$, is similarly described by
\begin{equation}
    \omega^{\rm K}_{\bf Q}
    =
    \omega^{\rm Goldstone}_{\bf Q}+\Delta_{\rm TK},\label{eq:TKomega}
\end{equation}
\begin{equation}
    \upsilon^{\rm K}_{\bf Q}
    =
    \upsilon^{\rm Goldstone}_{\bf Q}-\Delta_{\rm TK}.\label{eq:TKupsilon}
\end{equation}
Notice that in the limit where $u_\zeta\to\Delta_{\rm TK}$, the mode has no squeezing ($\upsilon^{\rm K}_{\bf Q}\to0$), confirming our expectations above -- The KIVC and sublattice-polarized orders have identical Hartree-Fock energies.
The energy gap for this mode is
\[{\cal E}^{\rm K}_{\bf 0}=\sqrt{2\Delta_{\rm TK}\left(u_\zeta+\Delta_{\rm TK}\right)}.\]

We now explore parametric driving of either of the modes above.
Driving of the uniform (${\bf Q}=0$) K mode, i.e., at a frequency $\omega_d=2{\cal E}^{\rm K}_{\bf 0}$ is particularly illuminating.
We consider small-amplitude modulations of the effective parameter 
\[\Delta_{\rm TK}\to \Delta_{\rm TK}+\delta\Delta_{\rm TK}\cos\left(2{\cal E}^{\rm K}_{\bf 0}t\right).\]
Zooming out from the specific model under consideration, this is a generic scenario of interest:
Slight modification of the system parameters, while remaining throughout the cycle within the same phase of the equilibrium phase diagram.
The parametric susceptibility, however, unlocks the possibility for intriguing out-of-equilibrium effects.

Microscopically, $\Delta_{\rm TK}$ originates in electrons interacting with specific Raman-active phonon modes~\cite{TIVCphonons,TBG_EPC_ARPES,EPCphonon_analysis}.
A promising route towards time-modulating this interaction is to optically pump an auxiliary IR-active phonon, which couples to one of these phonons non-linearly~\footnote{See recent works on non-linear phononics based parametric resonances, Refs.~\cite{kaplanprl_PhysRevLett.134.066902,kaplan2025spatiotemporalorderparametricinstabilities}}.
The \textit{amplitude} of the pump drive should be externally such that the phonon-induced electronic interaction strength oscillates with frequency $2{\cal E}^{\rm K}_{\bf 0}$.
With such experimental realization in mind, one expects that the direct effect of the driving would be to soften the interaction-inducing phonon modes (cf. Ref.~\cite{par_elph_SC}).
The induced electronic interaction are approximately inversely proportional to the phonon frequency~\cite{TIVCphonons}.
Thus, the relative amplitude $\delta\Delta_{\rm TK}/\Delta_{\rm TK}$ roughly corresponds to the relative softening of the characteristic frequency of the phonon modes $\omega_{\rm ph}$, i.e., to $\delta\omega_{\rm ph}/\omega_{\rm ph}$.


Thus, physically the modulation of $\Delta_{\rm TK}$ is coming mainly from the modification of the $u^{\rm T}$ interaction [Eq.~\eqref{eq:Tinteractions}], which favors the phonon-driven TIVC phase.
The parametric efficiency with respect to these interactions,
\begin{equation}
    \eta_{{\rm par}.}^{\left(u^{{\rm T}}\right)}=\frac{1}{1+\frac{\Delta_{{\rm TK}}}{u_{\zeta}}}\frac{u^{{\rm T}}}{\Delta_{{\rm TK}}}.\label{eq:etautk}
\end{equation}
Notice that the parametric efficiency diverges as $\Delta_{\rm TK}\to0$.
This can be appreciated from the point of view of the fidelity susceptibility, Eq.~\eqref{eq:seuscuptibiltyPrincipal}.
When $\Delta_{\rm TK}$ approaches zero, finally changing sign and crossing over, the TIVC reference state is destabilized in favor of KIVC.
The vacuum of quantum fluctuations becomes increasingly sensitive to the transition, with $\chi^F\propto\Delta_{\rm TK}^{-2}$ in its vicinity.

What are the consequences of the resonant non-equilibrium state?
The two pumped quadratures correspond to $\sigma_y\tau_x$ -- the easy direction for this mode (KIVC order), and $\sigma_z$ -- the hard direction (sublattice polarization).
The ratio of easy-to-hard quadratures is 
$\frac{1}{r}=
\frac{{\cal E}_{\bf 0}^{\rm K}}{2\Delta_{\rm TK}}$.
Remarkably, a phase which is \textit{otherwise inaccessible} in an equilibrium setting, the KIVC analog in our model, plays a key role in the dynamics of the driven state.
Moreover, this state has enhanced staggered sublattice correlations, which oscillate at half the driving frequency, $\omega_d/2$.
The resultant phase of Floquet matter, enabled by coherent manipulation of this strongly interacting system, facilitates non-linear optical responses (e.g., second-harmonic generation~\cite{SHG_Mos2,shg_graphene_hbn} due to the time-dependent inversion symmetry breaking) \textit{activated by pumping the system} at a narrow frequency window.

\subsubsection{Sublattice Polarized Order (Fig.~\ref{fig:casestudysummary}c)}\label{sec:sigmaorderflat}
\begin{figure}
    \centering
    \includegraphics[width=9cm]{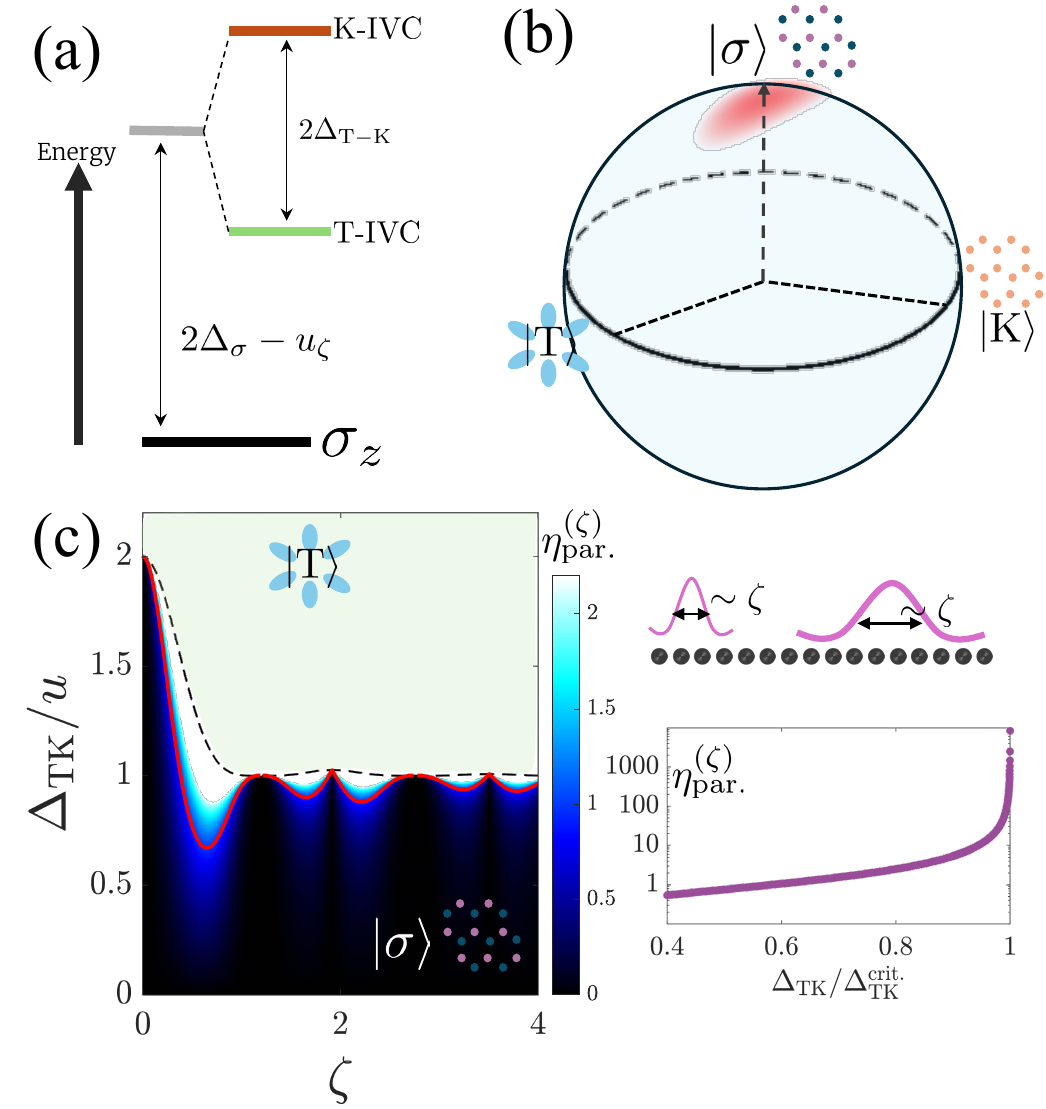}
    \caption{
    \textbf{Pseudospin (sublattice) polarized order and its parametric driving} (Sec.~\ref{sec:sigmaorderflat}).
    (a)
    Energy hierarchy with a finite external sublattice potential [$\Delta_\sigma$ in Eq.~\eqref{eq:pseudotbgsingleparticle}].
    The energy $u_{\zeta}$ brings the T/K orders down in energy, and $\Delta_{\rm TK}$ once again favors T-order over K.
    (b)
    The generalized Bloch sphere for the $\sigma$-order case, positioned the north pole.
    The two fluctuation quadratures correspond to the easy T-order, and the hard K-order, distinguished by the interaction energy $\Delta_{\rm TK}$.
    The anisotropy of the fluctuation cloud corresponds again to the vacuum.
    (c)
    Left:
    Parametric efficiency with respect to the quantum geometric parameter $\zeta$, Eq.~\eqref{eq:parametricsuszeta}.
    This parameter roughly corresponds to the Wannier spread in real-space of the wavefunctions corresponding to the active flat-bands (right schematic).
    The dashed line marks the phase boundary between the $\sigma$ phase below and the T-order stabilized above (hatched region), in accordance with the hierarchy in (a).
    Red line traces where $\eta_{{\rm par}.}^{\left(\zeta\right)}=1$. 
    The parametric response is maximal when the interaction energy strongly depends on quantum geometry.
    The susceptibility diverges as one approaches the phase boundary, even at $\zeta$ much larger than its optimal value.
    Right:
    A vertical cut of the left colormap at $\zeta=0.65$, showing the divergence as the phase boundary is approached.
    Here, $\Delta_\sigma=u$.
    }
    \label{fig:zetasusceptflat}
\end{figure}

If the substrate-induced inversion symmetry breaking $\Delta_\sigma$ term is large enough, the preferred order parameter for the system is ${\cal O}_\sigma=\sigma_z$, i.e., sublattice polarization.
The relevant ordered state hierarchy for this case is shown in Fig.~\ref{fig:zetasusceptflat}a.
The generators which keep the system in the low energy manifold of Eq.~\eqref{eq:gsmanifold} are thus
$\left\{{\cal M}_i\right\}=\sigma_x\tau_x,\,\sigma_y\tau_x,\,\sigma_x\tau_y,\,\sigma_y\tau_y$.

We may again organize then in canonically conjugate pairs, 
$\sigma_x\tau_x$ with $\sigma_y\tau_x$, and
$\sigma_x\tau_y$ with $\sigma_y\tau_y$.
The generator containing $\sigma_y$ in each pair generates a finite TIVC component in $|\Psi_\phi\rangle$, whereas the $\sigma_x$ generator promotes KIVC correlations (see Fig.~\ref{fig:zetasusceptflat}b).
It is thus natural to expect that the corresponding collective modes will be squeezed iff $\Delta_{\rm TK}\neq 0$.
The lowest-lying modes corresponding to the two generator pairs are degenerate ($\tau_x$ and $\tau_y$ are on equal footing) and gapped (no continuous symmetry of the Hamiltonian is spontaneously broken by the ordered state).
\footnote{
We note that the $\Delta_\sigma$ scenario described here can be mapped onto the antiferromagnet discussed in Fig.~\ref{fig:schematicfig}c, with $\Delta_{\rm \sigma}$ playing the role of a staggered external magnetic field.
To see that this is the case, simply perform a Pauli matrix replacement which conserves all commutation and anti commutation relations ($s^i$ and $p^i$ are Pauli matrices describing fictitious ``spin'' and ``sublattice'' degrees of freedom):
The antiferromagnetic order is now $\sigma_z\to s^zp^z$,
N\'eel vector rotations (or T-order generators) are $\sigma_y\tau_x\to s^x$, $\sigma_y\tau_y\to s^y$, and ferromagnetic ordering (or K-order generators) is given by generators $\sigma_x\tau_x\to s^xp^z$, $\sigma_x\tau_y\to s^yp^z$.}

Continuing our analysis for the degenerate two lowest-lying modes (Appendix~\ref{app:flatbandzetamodel}), the effective parameters are,
\begin{equation}
    \omega^{\left(1,2\right)}_{\bf Q}
    =
    \frac{\rho_s}{2}\left|{\bf Q}\right|^2+
    \Delta_\sigma-
    \frac{u_\zeta}{2},
\end{equation}
\begin{equation}
    \upsilon^{\left(1,2\right)}_{\bf Q}
    =
    \frac{\Delta_{\rm TK}}{2}.
\end{equation}
As expected, the squeezing of lowest-lying collective modes is given by the breaking of T/K symmetry in the model.
The modes have the expected gap at ${\bf Q}=0$, 
\[{\cal E}^{\left(1,2\right)}_{\bf 0}=\sqrt{\left(\Delta_{\sigma}-\frac{u_{\zeta}}{2}\right)^{2}-\left(\frac{\Delta_{{\rm TK}}}{2}\right)^{2}}.\]
The gap closes when the $\sigma_z$ order becomes exactly degenerate with the TIVC phase, i.e., when $\Delta_\sigma=\frac{u_\zeta+\Delta_{\rm TK}}{2}$.

The vicinity to a phase transition affects the parametric susceptibility dramatically.
Consider small changes in the Hamiltonian parameters that affect the quantum geometry of electrons in the projected bands, modifying slightly the quantum metric length $\zeta$.
In realistic devices, this can be the result of modulating gate voltages, electric displacement fields, or strain.
More intriguing possibilities include modulation of external driving in Floquet systems, where the Floquet driving subtly controls single-particle properties~\cite{Floquet_parametric_Kiselev2024}. 
The parametric efficiency to such a modification may be extracted by employing Eq.~\eqref{eq:dzMain} (differentiating with respect to $\zeta$) and plugging it into Eq.~\eqref{eq:parefficiency},
\begin{equation}
    \eta_{{\rm par}.}^{\left(\zeta\right)}=\frac{4u\Delta_{{\rm TK}}\zeta \left|J_{0}^{3}\left(2\zeta\right)J_{1}\left(2\zeta\right)\right|}{\left(\Delta_{\sigma}-\frac{u_{\zeta}+\Delta_{{\rm TK}}}{2}\right)\left(\Delta_{\sigma}-\frac{u_{\zeta}-\Delta_{{\rm TK}}}{2}\right)}.\label{eq:parametricsuszeta}
\end{equation}
Notice that the left term in the denominator vanishes at the phase transition discussed above, resulting in the divergence of the efficiency, prominently shown in Fig.~\ref{fig:zetasusceptflat}c--d.
Once more, this can be understood as a consequence of the diverging fidelity susceptibility at the transition.

The dependence on $\zeta$ can be traced back to $u_\zeta$, and is maximized for $\zeta\approx0.5$.
This particular value is not universal -- the maximizing value will typically depend on the \textit{range of interactions}.
Under our present assumptions, the $u$ interactions are maximally short-ranged, thus a change of length scale associated with the projected band electrons, i.e., $\sim\zeta$, beyond a certain value has limited effect.
Generalizing this observation, small changes in quantum geometrical parameters are expected to have the most pronounce impact when the length-scale associated with the interaction is comparable to the one derived from the quantum metric.
The interplay between these energy scales has been demonstrated to be crucial in the context of fractional Chern insulators~\cite{ShavitFCIPhysRevLett.133.156504,PelegFCI_emanuel2025unifyingframeworkfractionalchern}.
A parametric response probe may thus shed much needed light on the prospect of realizing these phases in possible candidate materials.

Let us once more estimate the physical modulation amplitudes required to observe a significant parametric amplification.
We examine mirror-symmetric twisted-trilayer graphene, which is the analog of TBG, yet has an additional tuning knob: an electric displacement field which hybridizes the TBG bands with an additional Dirac-like band.
Empirically, this hybridization roughly leaves the bandwidth intact, while non-negligibly modifying the Bloch wavefunctions.
Ref.~\cite{efetov_diracspectroscopy_Shen2023} reported a transition between two types of symmetry-breaking correlated phase as a function of the displacement field.
Quantitatively, a change of order 50 mV/nm in displacement field translated to a $\sim30\%$ increase of the order parameter.
This indicates that a \textit{modulation of order 10 mV/nm} of the displacement field (which is less than $0.5\%$ of the experimentally accessible field of view) can trigger appreciable parametric amplification of the collective modes.

The response to driving in this scenario would be an oscillatory appearance of T-order and K-order components in the transient order parameter, with the amplitude of the former being stronger by the squeezing factor of $r^{-1}=\sqrt{1+\frac{\Delta_{{\rm TK}}}{\Delta_{\sigma}-\frac{u_{\zeta}+\Delta_{{\rm TK}}}{2}}}$ compared to the latter.
We note that in the analogous twisted graphene system, the T-order, i.e., TIVC, couples to a real-space charge density with a Kekul\'e pattern~\cite{IVCspectroscopycalugaru,IVCspectroscopyZalatel,Yazdani_IKS_TIVC,TIVCphonons}.
Thus, the parametrically driven state will give rise to a charge density wave oscillating at half the drive frequency, detectable using available time-resolved STM techniques~\cite{timeresolvedstm_review,timeresolvedstm_example}.

\section{Discussion and outlook}\label{sec:conclusions}

Parametric driving of non-trivial low-energy collective excitations is a new and exciting avenue for controlling, engineering, and probing many-body correlated states. 
The driven dynamics proposed here are enabled by recent experimental advances, the large degree of tunability, and the rich phase diagram that new correlated electronic platforms allow. 
We find that modulating the right microscopic parameters at an appropriate resonant frequency, allows one to approach phenomena that are otherwise practically inaccessible in the solid-state context.

The periodic driving we envision is simple to implement in the lab. 
Typically, modulating the voltages that control density, electrical displacement fields, or strain fields would do.
In more intricate implementations (which may be more relevant for the $\sim$THz modulation range), a modulation of an optical signal may induce driving in some effective parameters.
(This could directly drive a non-equilibrium state in the material~\cite{graphen_pulse_anomalousHall_McIver2020,FloquetGraphene_ness}, or control a thoughtfully placed photoconductive switch~\cite{PCS1_doi:10.1126/science.aat8687,PCS2_Chen2025}.)
Driving the collective modes of the system on \textit{parametric} resonance, has the added benefit of requiring very low driving power.
When the parametric resonance condition is satisfied the amplitude of the driven modes grows exponentially in time.
Conversely, on-resonance driving leads to linearly increasing amplitude with time.
Recognizing this parametric advantage, it becomes imperative to identify and isolate the kind of perturbation that will effectively drive the system, which we thoroughly explore in Sec.~\ref{sec:collectivederivation}.

The measurable signature of the parametric drive may be system-specific, and depends on the detailed character of the low-energy ordered state.
In the excitonic double layer example (Sec.~\ref{sec:QHbilayer}), interlayer tunneling spectroscopy reveals clear signatures of the parametrically driven state.
Systems with richer structure, e.g., the flat-band TBG-analog model we explore in Sec.~\ref{sec:TBGproxy}, have more detection possibilities, such as time-varying charge-density ordering and modulated optical responses.
In all cases, however, the parametric response is always \textit{sub-harmonic} with respect to the driving frequency. All of our examples have $\omega_d$ corresponding to twice the frequency of the targeted collective mode, and thus the parametric response would prominently appear at a frequency (or energy corresponding to) $\omega_d/2$.

A key result addresses the \textit{strength} of the induced parametric response. 
We find that the parametric susceptibility is intimately connected to the nature of the vacuum state and the dependence of its properties on microscopic parameters. 
As such, parametric driving emerges as a novel probe of the {\it fidelity susceptibility} in the context of solid-state materials and devices. 
This susceptibility generically gets enhanced and diverges in the vicinity of quantum phase transitions, where the wavefunction of the vacuum ground-state changes dramatically.
Therefore, mapping the readily-implemented parametric driving response across swaths of phase space unlocks a new and complimentary way to examine quantum properties in correlated materials, detect quantum fluctuations, and discover hidden phase transitions, which may be otherwise hard to reach.

The non-trivial connection between the vacuum fidelity susceptibility and the induced parametric drive was revealed through a framework that included a quantum theory of the collective many-body fluctuations.
Importantly, our analysis emphasized the dynamical properties of these collective modes, and not only their energetics (i.e., their spectrum).
The emergent conjugate-canonical-pair structure of the collective coordinates [Eq.~\eqref{eq:Bmain}] plays a key role, and our approach enables straightforward identification of such dynamical-variable pairs.
This is achieved by a particular kind of \textit{partial bosonic Bogoliubov transformation}, which isolates the parametric two-boson response of a system to certain changes of its parameters.

In our analysis, \textit{squeezing} emerged as a main organizing principle in parametric driving of electronically ordered materials.
The two quadratures of the fluctuation, analogous to ``position'' and ``momentum'' (e.g., spin fluctuations in perpendicular directions), are to be treated on equal footing from the outset.
The low-energy Hamiltonian, however [Eq.~\eqref{eq:Hexpansion}], then differentiates between the quadratures, and produces squeezing of the vacuum fluctuations.
To induce parametric driving, a modulation must \textit{alter the squeezing}, which corresponds to inducing anomalous (i.e., two-boson) driving terms in the effective low-energy theory. 
Indeed, the more squeezing the reference ordered state has, the stronger is its parametric response (see, e.g., Eq.~\eqref{eq:dzMain}). 
Intuitively, a squeezed vacuum state simply hosts a wider range of bosonic fluctuations. 

The case studies presented in Sec.~\ref{sec:casestudies} showcase the role of quantum geometry as a potent driven parameter with associated parametric instabilities. 
In strongly interacting, narrow bandwidth, electronic systems, quantum geometry is crucial. 
Indeed, the competition between the geometry-associated length scale (e.g., $\zeta$) and the range of intrinsic electronic interactions is key in determining the parametric susceptibility with respect to small changes in the quantum geometry (as evident in Fig.~\ref{fig:zetasusceptflat}b).
These observations naturally motivate further studies into ways of manipulating and engineering quantum geometric quantities, e.g., Berry curvature and the quantum metric, in correlated quantum materials.

Our framework concentrated on periodic modulations, but could also be applied to quantum parameter \textit{quenches}.
In a small quench, the reference electronic order remains intact, yet the low-energy collective degrees of freedom could be generically excited. 
Quantifying this response could use the method summarized in Eq.~\eqref{eq:partialbogoliubov}.
Inevitably, if the parametric susceptibility associated with the quench is substantial, a non-negligible amount of bosonic collective excitations will emerge.
Such excitations, recently probed in correlated optically-pumped moir\'e systems, may be long-lived and have intriguing consequences in the context of characterizing the underlying electronic order~\cite{ChenhaoTBG_Xie2024,Chenhao_ivc_xiong2025propagatingneutralmodesintervalley}. 
Also, one may ask how much work does the quench produce through exciting low-lying collective modes.
Once again, the fidelity of the fluctuation vacuum is expected to play the key quantitative role. 
Interestingly, probing the statistics of the performed work probes the corresponding Loschmidt echo~\cite{Work_eco_PhysRevLett.101.120603,Loschmidt_decoherence_PhysRevLett.96.140604}, which has thus far remained out of experimental reach in the context of electronic condensed matter systems.

An exciting future direction would be harnessing the parametric driving described here in quantum sensors and quantum information processing devices, specifically in the context of quantum-limited parametric amplification~\cite{paramp_yarivPhysRev.124.1646,QuantumTheoryfParametric,quantumlimitonnoise,josephson_paramp,paramp_RMP}.
For instance, An excitonic strongly correlated double layer (see Sec.~\ref{sec:QHbilayer}) may be employed to amplify a photonic signal fed into one of its gate voltages. 
A parametric instability may be triggered, exciting two Goldstone bosons. 
This will result in an oscillating interlayer electric field, which may be extracted as the amplified signal.
This scheme provides phase-sensitive degenerate parametric amplification, where one of the quadratures of the signal is amplified and the other attenuated.
A non-degenerate amplification setup can be explored when (i) several low-lying excitation modes are available, and (ii) joint parametric terms may be driven, e.g., $\sim a_1^\dagger a_2^\dagger$.
(This is the case for our studied flat-band model, see lower right panel of Fig.~\ref{fig:MAINWIDE}c.)

Finally, we speculate that effortless manipulation of the emergent bosonic collective excitations in the correlated electronic systems may have broad implications. 
These excitations may be \textit{coupled to other bosonic modes} in various settings, such as mechanical oscillators~\cite{mecahnicaloscillator1,mecahnnicaloscillator2}, microwave resonators~\cite{mwresonatorsPhysRevA.69.062320,SCqubit_Schoelkopf2008}, superconducting qubits~\cite{SCqubit_Schoelkopf2008,SCqubit_Houck2012}, and optical cavities~\cite{opticalcavity_tomechanicalVerhagen2012,opticalcavity_atomdoi:10.1126/science.1237125}.
Integrating solid-state tunable correlated electronic materials with more established traditional quantum platforms holds great promise for applications of hybrid quantum systems~\cite{Quantumtechnologieswithhybridsystems,clerk2020hybrid},
e.g., quantum transduction~\cite{hybridtransduction_Arnold2020,hybrid_transduction_Hönl2022,hybrid_transduction_Rochman2023,optic_microwave_transduction} and information storage~\cite{hybridmemory_PhysRevLett.123.250501,SoundMemoryCaltech}.
Controllably coupling the collective excitations explored here to other bosonic modes remains a fascinating topic for future work.

\begin{acknowledgments}
We acknowledge the support of the Institute for Quantum Information and Matter, an NSF Physics Frontiers Center (PHY-2317110).
GS acknowledges support from the Walter Burke Institute for Theoretical Physics at Caltech, and from the Yad Hanadiv Foundation through the Rothschild fellowship.
This work was performed in part at Aspen Center for Physics, which is supported by National Science Foundation grant PHY-2210452.
\end{acknowledgments}

\begin{appendix}
\begin{widetext} 
\section{Collective modes in generic magnets }\label{app:magnets_goldstone_details}
Here, we provide explicit details regarding the examples considered in Fig.~\ref{fig:schematicfig} and surrounding discussion.

\textit{SU(2) / easy-axis ferromagnet.--}
The order parameter in this case is ${\cal O}=s_z$, where $s_i$ are Pauli matrices.
By virtue of Eq.~\eqref{eq:Bmain}, the effective position and conjugate momentum can be identified by the relation
\begin{equation}
    {\rm Tr}\left\{\left[p,q\right]{\cal O}\right\}\propto i,\label{eqapp:trpqO}
\end{equation}
such that one may readily assign $p\sim s_x$, $q\sim s_y$.
At long-wavelengths, assuming the ordered state retains its U(1) symmetry with regards to $s_z$, the effective collective Hamiltonian is,
\begin{equation}
    {\cal H}^{\rm FM}\approx\sum_{\bf Q}\left[\frac{1}{2}\rho \left|{\bf Q}\right|^2+E_{\rm easy-axis}\right] \left(p_{\bf Q}^2+q_{\bf Q}^2\right),\label{eqapp:HFM}
\end{equation}
where $\rho$ is the stiffness and $E_{\rm easy-axis}$ accounts for the (possible) easy-axis anisotropy.
Clearly, Eq.~\eqref{eqapp:HFM} leads to the anticipated quadratically dispersing (possibly gapped) collective mode of a spontaneous ferromagnet.

\textit{Easy-plane ferromagnet.--}
Conversely, in this instance let us identify ${\cal O}=s_x$, such that when we invoke Eq.~\eqref{eqapp:trpqO} again, we assign $p\sim s_y$, $q\sim s_z$
Given the easy-plane nature of this problem, we expect that $s_y$ rotations are ``stiffer'', since they generate a finite $s_z$ order.
Thus, $p$ and $q$ are no longer on equal footing.
The low-energy collective Hamiltonian is generically (with $\Delta$ the easy-plane anisotropy)
\begin{align}
    {\cal H}^{\rm easy-plane}&\approx\sum_{\bf Q}\frac{1}{2}\rho \left|{\bf Q}\right|^2 \left(p_{\bf Q}^2+q_{\bf Q}^2\right)+\Delta p_{\bf Q}^2\nonumber\\
    &=\sum_{{\bf Q}}\left[\frac{1}{2}\rho\left|{\bf Q}\right|^{2}+\frac{\Delta}{2}\right]\left(p_{{\bf Q}}^{2}+q_{{\bf Q}}^{2}\right)+\frac{\Delta}{2}\left(p_{{\bf Q}}^{2}-q_{{\bf Q}}^{2}\right),\label{eqapp:Heasyplane}
\end{align}
such that in the ${\bf Q}\to0$ limit the spectrum is approximately ${\cal E}_{{\bf Q}}\approx\sqrt{\frac{\Delta\rho}{2}}\left|{\bf Q}\right|$.
One recovers the expected linear Goldstone mode associated with this system.

\textit{SU(2) antiferromagnet.--}
Here, the picture is slightly richer.
For simplicity, let us incorporate the antiferromagnetic ordering via introduction of a sublattice degree of freedom, described by Pauli matrices $\sigma_i$.
In that manner, we may treat this case using the same framework by identifying the antiferromagnetic order parameter
${\cal O}=s_z\sigma_z$.
In contrast with the easy-axis ferromagnet, here $s_x$ and $s_y$ \textit{are not conjugate variables}.
Instead, by enforcing sublattice as a good quantum number in the low-energy ordered manifold, we may identify to pairs of canonical variables [by referencing Eq.~\eqref{eqapp:trpqO} again],
\begin{equation}
    p^{\left(1\right)}=s_y\sigma_z,\,q^{\left(1\right)}=s_x,
    \,\,\,\,\,
    p^{\left(2\right)}=s_x\sigma_z,\,q^{\left(2\right)}=s_y.
\end{equation}
Notice that the $q^{\left(j\right)}$ variables induce a rotation of the N\'eel vector, whereas the $p^{\left(j\right)}$ operators introduce finite magnetization.
By definition, the latter rotations should be much stiffer.
This leads to a generic low-energy Hamiltonian,
\begin{equation}
    {\cal H}^{{\rm AFM}}\approx\sum_{j,{\bf Q}}\frac{1}{2}\rho\left|{\bf Q}\right|^{2}\left[\left(p_{{\bf Q}}^{\left(j\right)}\right)^{2}+\left(q_{{\bf Q}}^{\left(j\right)}\right)^{2}\right]+J\left(p_{{\bf Q}}^{\left(j\right)}\right)^{2},
\end{equation}
where $J>0$ encodes the antiferromagnetic tendencies.
Comparing with the easy-plane case, one immediately obtains two branches of linearly-dispersing Goldstone modes, as anticipated for such a system.

\bigskip
We note that in the three examples above we did not invoke any microscopic details regarding the systems which give rise to these ordered states.
The analysis introduced in Sec.~\ref{sec:collectivederivation} is thus quite powerful in determining universal characteristics of the low-lying collective modes, before needing to perform any calculation.
Moreover, it is now clear why linear dispersion of Goldstone modes constitutes a universal signature of collective mode squeezing.
A generic squeezing term of one of the quadratures immediately converts a (gapless) mode with a quadratic dispersion into a linear one at long wavelengths.

\section{Fluctuation action calculation details}\label{app:action_details}
\subsection{Berry phase}
Expanding to second order in $\hat F$ (using $\frac{d}{dt}|\Psi_0\rangle=0$),
\begin{align}
        {\cal B} &\approx i\langle\Psi_0| {\hat F}\frac{d}{dt}{\hat F}|\Psi_0\rangle \nonumber\\
        &= \sum_{{\bf QQ',pp'}}i\phi_{{\bf Q}}^{\mu\nu}\left({\bf p}\right)\frac{d}{dt}\phi_{{\bf Q'}}^{\alpha\beta}\left({\bf p'}\right)\langle\Psi_{0}|c_{{\bf p+Q}\mu}^{\dagger}c_{{\bf p}\nu}c_{{\bf p'+Q'}\alpha}^{\dagger}c_{{\bf p'}\beta}|\Psi_{0}\rangle\nonumber\\
        &=i\sum_{{\bf p}}{\rm Tr}\left\{ \phi_{0}\left({\bf p}\right)P^T\left({\bf p}\right)\right\} \sum_{{\bf p'}}{\rm Tr}\left\{ \frac{d}{dt}\phi_{0}\left({\bf p'}\right)P^{T}\left({\bf p'}\right)\right\} \nonumber\\
        &+i\sum_{{\bf Q,p}}{\rm Tr}\left\{ \phi_{{\bf Q}}\left({\bf p}\right)\bar{P}^{T}\left({\bf p}\right)\frac{d}{dt}\phi_{{\bf -Q}}\left({\bf p+Q}\right)P^{T}\left({\bf p+Q}\right)\right\}. 
\end{align}
Using the anti-commutation between $\cal O$ and $\phi$ and the cyclic properties of the trace, we can write schematically ${\rm Tr}\left\{\phi P^T\right\}={\rm Tr}\left\{\phi \bar{P}^T\right\}={\rm Tr}\left\{\phi\right\}=0$, eliminating the first line.
Next, since $P$ is independent of $\bf k$, 
\begin{align}
    {\cal B}&= i\sum_{{\bf Q,p}}{\rm Tr}\left\{ \phi_{{\bf Q}}\left({\bf p}\right)\frac{d}{dt}\phi_{{\bf -Q}}\left({\bf p+Q}\right)P^{T}\right\} \nonumber\\
    &=\frac{i}{2}\sum_{{\bf Q,p}}{\rm Tr}\left\{ \phi_{{\bf Q}}\left({\bf p}\right)\frac{d}{dt}\phi_{{\bf -Q}}\left({\bf p+Q}\right)P^{T}\right\} +\frac{i}{2}\sum_{{\bf Q,p}}{\rm Tr}\left\{ \frac{d}{dt}\phi_{{\bf -Q}}\left({\bf p+Q}\right)\phi_{{\bf Q}}\left({\bf p}\right)\bar{P}^{T}\right\} \nonumber\\
    &=\frac{i}{4}\frac{d}{dt}\sum_{{\bf Q,p}}{\rm Tr}\left\{ \phi_{{\bf -Q}}\left({\bf p+Q}\right)\phi_{{\bf Q}}\left({\bf p}\right)\right\}\nonumber\\
    &+\frac{i}{2}\sum_{{\bf Q,p}}{\rm Tr}\left\{ \phi_{{\bf -Q}}\left({\bf p+Q}\right)\frac{d}{dt}\phi_{{\bf Q}}\left({\bf p}\right){\cal O}\right\}, 
\end{align}
where we have used the summations under $\bf p,Q$ to perform various manipulations.
Since the first term is a complete derivative, and employing the identity $\phi_{{\bf -Q}}\left({\bf p+Q}\right)=\phi^\dagger_{{\bf Q}}\left({\bf p}\right)$ we can compactly write
\begin{equation}
    {\cal B}=\frac{i}{2}\sum_{{\bf Q,p}}{\rm Tr}\left\{ \phi^\dagger_{{\bf Q}}\left({\bf p}\right)\frac{d}{dt}\phi_{{\bf Q}}\left({\bf p}\right){\cal O}\right\}. \label{eq:Bfinalapp}
\end{equation}
Decomposed,
\begin{equation}
    {\cal B}=\sum_{{\bf Q}}\frac{1}{2}\sum_{ij{\bf pp'}}\left[\Phi_{{\bf Q}}^{j}\left({\bf p'}\right)\right]^{*}{\rm Tr}\left\{ -i{\cal M}^{i}{\cal M}^{j}{\cal O}\right\} \delta_{{\bf pp'}}\frac{d}{dt}\Phi_{{\bf Q}}^{i}\left({\bf p}\right).
\end{equation}
\subsection{Fluctuation Hamiltonian}
Expanding once more to lowest order in fluctuations,
\begin{equation}
    {\cal H}\approx E_0-\frac{1}{2}\langle\Psi_0| \left[{\hat F},\left[{\hat F},H\right]\right]|\Psi_0\rangle,
\end{equation}
where the first order term vanishes identically as a consequence of $|\Psi_0\rangle$ minimizing the Hartree-Fock energy.
Our task is thus to compute the following expectation values for various Hamiltonian terms, $\left\langle\left[{\hat F},\left[{\hat F},H\right]\right]\right\rangle$, where the expectation value is with respect to $|\Psi_0\rangle$.
Throughout the calculations detailed below, one makes extensive use of the following fermionic commutation identity,
\begin{equation}
\left[c_{1}^{\dagger}c_{2},c_{3}^{\dagger}c_{4}\right]=\delta_{23}c_{1}^{\dagger}c_{4}-\delta_{14}c_{3}^{\dagger}c_{2}.
\end{equation}

\subsubsection{Quadratic-in-fermions perturbations}
Consider a quadratic perturbation independent of momentum,
\begin{equation}
    H_{{\rm per.}}=\sum_{{\bf k}}c_{{\bf k}\alpha}^{\dagger}h^{\alpha\beta}c_{{\bf k}\beta}.
\end{equation}
We first iterate,
\begin{equation}
    \left[F,H_{{\rm per.}}\right]=\sum_{{\bf Qk}}\left[\phi_{{\bf Q}}^{\alpha s}\left({\bf k}\right)h^{s\beta}-h^{\alpha s}\phi_{{\bf Q}}^{s\beta}\left({\bf k}\right)\right]c_{{\bf k+Q}\alpha}^{\dagger}c_{{\bf k}\beta}.
\end{equation}
After computing the second commutator, and taking the Hartree-Fock expectation value with respect to $|\Psi_0\rangle$,
\begin{equation}
    \left\langle \left[{\hat F},\left[{\hat F},H_{{\rm per.}}\right]\right]\right\rangle =\sum_{{\bf Qk}}{\rm Tr}\left\{ \phi_{{\bf Q}}^{\dagger}\left({\bf k}\right)\left[\phi_{{\bf Q}}\left({\bf k}\right),h\right]{\cal O}\right\} 
    =
    2\sum_{{\bf Qk}}{\rm Tr}\left\{ \phi_{{\bf Q}}^{\dagger}\left({\bf k}\right)\phi_{{\bf Q}}\left({\bf k}\right)h{\cal O}\right\} . 
\end{equation}
In terms of the decomposition of the fluctuation matrix, we may re-write this as
\begin{equation}
    \left\langle \left[{\hat{F}},\left[{\hat{F}},H_{{\rm per.}}\right]\right]\right\rangle =\sum_{{\bf Q}}\sum_{ij{\bf pp'}}\left[\Phi_{{\bf Q}}^{j}\left({\bf p'}\right)\right]^{*}\delta_{{\bf pp'}}{\rm Tr}\left[\left\{ {\cal M}^{j}{\cal M}^{i},h\right\} {\cal O}\right] \Phi_{{\bf Q}}^{i}\left({\bf p}\right).
\end{equation}

\subsubsection{Interacting terms}
We write a generalized interaction term that will capture all the relevant kinds of interactions in the models we analyze,
\begin{equation}
    H_{\rm int.}=\frac{1}{2A}\sum_{{\bf qkk'}}V_{{\bf q}}c_{{\bf k+q}\alpha}^{\dagger}\Lambda_{{\bf k},{\bf k+q}}^{\alpha\beta}c_{{\bf k}\beta}c_{{\bf k'-q\gamma}}^{\dagger}\Lambda_{{\bf k'},{\bf k'-q}}^{\gamma\delta}c_{{\bf k'}\delta}.
\end{equation}
Here, $\Lambda$ are matrices that embody both the band-projected form factor and the structure of the interaction terms in the internal index space spanned by the Greek letters.

Upon first iteration of the commutator, one finds
\begin{equation}
    \left[\hat{F},H_{\rm int.}\right]	=\frac{1}{2A}\sum_{{\bf Qq}}V_{{\bf q}}\left\{ \rho_{{\bf q,Q}}^{\phi},\rho_{{\bf -q}}\right\}, 
\end{equation}
where the following quantities and matrices were defined,
\begin{equation}
    \rho_{{\bf -q}}=\sum_{{\bf k'}}\Lambda_{{\bf k'},{\bf k'-q}}^{\gamma\delta}c_{{\bf k'-q\gamma}}^{\dagger}c_{{\bf k'}\delta},
\end{equation}
\begin{equation}
    \rho_{{\bf q,Q}}^{\phi}=\sum_{{\bf k}}M_{{\bf q,Q}}^{\alpha\beta}\left({\bf k}\right)c_{{\bf k+q+Q}\alpha}^{\dagger}c_{{\bf k}\beta},
\end{equation}
and
\begin{equation}
    M_{{\bf q,Q}}^{\alpha\beta}\left({\bf k}\right)\equiv\left[\phi_{{\bf Q}}^{\alpha s}\left({\bf k+q}\right)\Lambda_{{\bf k},{\bf k+q}}^{s\beta}-\Lambda_{{\bf k+Q},{\bf k+q+Q}}^{\alpha s}\phi_{{\bf Q}}^{s\beta}\left({\bf k}\right)\right].
\end{equation}

Now, let us split the next iteration into two parts and write,
\begin{equation}
    \left[\hat{F},\left[\hat{F},H_{\rm int.}\right]\right]=I_{1}+I_{2},
\end{equation}
with
\begin{equation}
    I_{1}=\left[\hat{F},\frac{1}{2A}\sum_{{\bf Qq}}V_{{\bf q}}\rho_{{\bf q,Q}}^{\phi}\rho_{{\bf -q}}\right],
\end{equation}
\begin{equation}
    I_{2}=\left[\hat{F},\frac{1}{2A}\sum_{{\bf Qq}}V_{{\bf q}}\rho_{{\bf -q}}\rho_{{\bf q,Q}}^{\phi}\right].
\end{equation}
Calculating the different commutators we find
\begin{align}
    I_{1}&=\frac{1}{2A}\sum_{{\bf Qq}}\sum_{{\bf Q'}}\sum_{{\bf k}}V_{{\bf q}}\left[\phi_{{\bf Q'}}^{\alpha s}\left({\bf k+q+Q}\right)M_{{\bf q,Q}}^{s\beta}\left({\bf k}\right)-M_{{\bf q,Q}}^{\alpha s}\left({\bf k+Q'}\right)\phi_{{\bf Q'}}^{s\beta}\left({\bf k}\right)\right]c_{{\bf k+q+Q+Q'}\alpha}^{\dagger}c_{{\bf k}\beta}\rho_{{\bf -q}}
    \nonumber\\
    &+\frac{1}{2A}\sum_{{\bf Qq}}\sum_{{\bf Q'}}V_{{\bf q}}\rho_{{\bf q,Q}}^{\phi}\rho_{{\bf -q,Q'}}^{\phi},
\end{align}
and
\begin{align}
    I_{2}&=\frac{1}{2A}\sum_{{\bf Q'}}\sum_{{\bf Qq}}V_{{\bf q}}\rho_{{\bf -q,Q'}}^{\phi}\rho_{{\bf q,Q}}^{\phi}
    \nonumber\\
    &+\frac{1}{2A}\sum_{{\bf Q'}}\sum_{{\bf Qq}}\sum_{{\bf k}}V_{{\bf q}}\rho_{{\bf -q}}\left[\phi_{{\bf Q'}}^{\alpha s}\left({\bf k+q+Q}\right)M_{{\bf q,Q}}^{s\beta}\left({\bf k}\right)-M_{{\bf q,Q}}^{\alpha s}\left({\bf k+Q'}\right)\phi_{{\bf Q'}}^{s\beta}\left({\bf k}\right)\right]c_{{\bf k+q+Q+Q'}\alpha}^{\dagger}c_{{\bf k}\beta}.
\end{align}
We may collect the different terms together by manipulating the momentum summations, and we find
\begin{align}
    \left[\hat{F},\left[\hat{F},H_{\rm int.}\right]\right]&=\frac{1}{2A}\sum_{{\bf Qq}}\sum_{{\bf Q'}}\sum_{{\bf k}}V_{{\bf q}}\left[\phi_{{\bf Q'}}^{\alpha s}\left({\bf k+q+Q}\right)M_{{\bf q,Q}}^{s\beta}\left({\bf k}\right)-M_{{\bf q,Q}}^{\alpha s}\left({\bf k+Q'}\right)\phi_{{\bf Q'}}^{s\beta}\left({\bf k}\right)\right]\left\{ c_{{\bf k+q+Q+Q'}\alpha}^{\dagger}c_{{\bf k}\beta},\rho_{{\bf -q}}\right\} 
    \nonumber\\
    &+\frac{1}{A}\sum_{{\bf Qq}}\sum_{{\bf Q'}}V_{{\bf q}}\rho_{{\bf q,Q}}^{\phi}\rho_{{\bf -q,Q'}}^{\phi}.
\end{align}

Finally, we evaluate the expectation value of this double commutator in the ground state, defining $\Lambda_{\bf k,k}\equiv\Lambda_0$ and exploiting the momentum independence of the density matrix, yielding the following expression,
\begin{align}
\left\langle \left[\hat{F},\left[\hat{F},H_{{\rm int.}}\right]\right]\right\rangle  & =\sum_{{\bf Q}}\sum_{{\bf kk'}}\frac{V_{{\bf k-k'}}}{A}{\rm Tr}\left\{ \phi_{{\bf Q}}^{\dagger}\left({\bf {\bf k'}}\right)\Lambda_{{\bf k+Q},{\bf k'+Q}}\phi_{{\bf Q}}\left({\bf {\bf k}}\right)\left(P^{T}\Lambda_{{\bf {\bf k'}},{\bf {\bf k}}}P^{T}+\bar{P}^{T}\Lambda_{{\bf {\bf k'}},{\bf {\bf k}}}\bar{P}^{T}\right)\right\} \nonumber\\
 & -\sum_{{\bf Q}}\sum_{{\bf kk'}}\frac{V_{{\bf k-k'}}}{A}{\rm Tr}\left\{ \Lambda_{{\bf k},{\bf k'}}\phi_{{\bf Q}}^{\dagger}\left({\bf {\bf k}}\right)\phi_{{\bf Q}}\left({\bf {\bf k}}\right)\left(P^{T}\Lambda_{{\bf {\bf k'}},{\bf {\bf k}}}P^{T}+\bar{P}^{T}\Lambda_{{\bf k'},{\bf k}}\bar{P}^{T}\right)\right\} \nonumber\\
 & ...\nonumber\\
 & +\sum_{{\bf Q}}\sum_{{\bf kk'}}\frac{V_{{\bf k-k'}}}{2A}{\rm Tr}\left\{ \left[\phi_{{\bf Q}}^{\dagger}\left({\bf k'}\right)\phi_{{\bf Q}}\left({\bf k'}\right)\Lambda_{{\bf k},{\bf k'}}+\Lambda_{{\bf {\bf k}},{\bf {\bf k}'}}\phi_{{\bf Q}}^{\dagger}\left({\bf k}\right)\phi_{{\bf Q}}\left({\bf k}\right)-2\phi_{{\bf Q}}^{\dagger}\left({\bf k'}\right)\Lambda_{{\bf k+Q},{\bf k'+Q}}\phi_{{\bf Q}}\left({\bf k}\right)\right]\bar{P}^{T}\Lambda_{{\bf k'},{\bf k}}P^{T}\right\} \nonumber\\
 & +\sum_{{\bf Q}}\sum_{{\bf kk'}}\frac{V_{{\bf k-k'}}}{2A}{\rm Tr}\left\{ \left[\phi_{{\bf Q}}^{\dagger}\left({\bf k'}\right)\phi_{{\bf Q}}\left({\bf k'}\right)\Lambda_{{\bf k},{\bf k'}}+\Lambda_{{\bf {\bf k}},{\bf {\bf k}'}}\phi_{{\bf Q}}^{\dagger}\left({\bf k}\right)\phi_{{\bf Q}}\left({\bf k}\right)-2\phi_{{\bf Q}}^{\dagger}\left({\bf k'}\right)\Lambda_{{\bf k+Q},{\bf k'+Q}}\phi_{{\bf Q}}\left({\bf k}\right)\right]P^{T}\Lambda_{{\bf k'},{\bf k}}\bar{P}^{T}\right\} \nonumber\\
 & ...\nonumber\\
 & +\sum_{{\bf Q}}\sum_{{\bf k}}\frac{V_{{\bf 0}}}{A}{\rm Tr}\left\{ \left[\left\{ \phi_{{\bf Q}}^{\dagger}\left({\bf k}\right)\phi_{{\bf Q}}\left({\bf k}\right),\Lambda_{0}\right\} -2\phi_{{\bf Q}}^{\dagger}\left({\bf k}\right)\Lambda_{0}\phi_{{\bf Q}}\left({\bf k}\right)\right]P^{T}\right\} 
 \sum_{{\bf k'}}{\rm Tr}\left\{ \Lambda_{0}P^{T}\right\} \nonumber\\
 & ...\nonumber\\
 & +\sum_{{\bf Q}}\frac{V_{{\bf Q}}}{A}\sum_{{\bf k}}{\rm Tr}\left\{ \phi_{{\bf Q}}^{\dagger}\left({\bf {\bf k}}\right)\frac{1}{2}\left[\Lambda_{{\bf k},{\bf k+Q}},{\cal O}\right]\right\} \sum_{{\bf k'}}{\rm Tr}\left\{ \phi_{{\bf Q}}\left({\bf k'}\right)\frac{1}{2}\left[\Lambda_{{\bf k'+Q},{\bf k'}},{\cal O}\right]\right\} .\label{eq:fullHfluctuation}
\end{align}
The different terms in this expression, separated by $...$ lines, contribute differently to the fluctuations, depending on the commutation relation of $P$ (or, more to the point, of $\cal O$) with the corresponding form factors of the interaction.

\subsubsection{Classifying interacting terms}
Given the matrix $\cal O$, which squares to the identity matrix, we may decompose any matrix $M$,
\begin{equation}
    M = \frac{1}{2}\left(M+{\cal O}M{\cal O}\right)
    +\frac{1}{2}\left(M-{\cal O}M{\cal O}\right).
\end{equation}
Notice that the first part commutes with $\cal O$, and the second part anti commutes with it.
This motivates us to examine two different contributions from the interacting parts, decomposing the form-factors with regards to commutation with $\cal O$:
\begin{align}
\left\langle \left[\hat{F},\left[\hat{F},H_{{\rm int.}}\right]\right]\right\rangle ^{\left[{\cal O},\Lambda\right]=0} & =\sum_{{\bf Q}}\sum_{{\bf kk'}}\frac{V_{{\bf k-k'}}}{A}{\rm Tr}\left\{ \left[\phi_{{\bf Q}}^{\dagger}\left({\bf {\bf k'}}\right)\Lambda_{{\bf k+Q},{\bf k'+Q}}-\Lambda_{{\bf k},{\bf k'}}\phi_{{\bf Q}}^{\dagger}\left({\bf {\bf k}}\right)\right]\phi_{{\bf Q}}\left({\bf {\bf k}}\right)\Lambda_{{\bf {\bf k'}},{\bf {\bf k}}}\right\} \nonumber\\
 & +\sum_{{\bf Q}}\sum_{{\bf k}}\frac{V_{{\bf 0}}}{A}{\rm Tr}\left\{ \left[\left\{ \phi_{{\bf Q}}^{\dagger}\left({\bf k}\right)\phi_{{\bf Q}}\left({\bf k}\right),\Lambda_{0}\right\} -2\phi_{{\bf Q}}^{\dagger}\left({\bf k}\right)\Lambda_{0}\phi_{{\bf Q}}\left({\bf k}\right)\right]P^{T}\right\} \sum_{{\bf k'}}{\rm Tr}\left\{ \Lambda_{0}P^{T}\right\} .
\end{align}

\begin{align}
\left\langle \left[\hat{F},\left[\hat{F},H_{{\rm int.}}\right]\right]\right\rangle ^{\left\{ {\cal O},\Lambda\right\} =0} & =\sum_{{\bf Q}}\sum_{{\bf kk'}}\frac{V_{{\bf k-k'}}}{A}{\rm Tr}\left\{ \left[\Lambda_{{\bf {\bf k}},{\bf {\bf k}'}}\phi_{{\bf Q}}^{\dagger}\left({\bf k}\right)-\phi_{{\bf Q}}^{\dagger}\left({\bf k'}\right)\Lambda_{{\bf k+Q},{\bf k'+Q}}\right]\phi_{{\bf Q}}\left({\bf k}\right)\Lambda_{{\bf k'},{\bf k}}\right\} \nonumber\\
 & +\sum_{{\bf Q}}\frac{V_{{\bf Q}}}{A}\sum_{{\bf k}}{\rm Tr}\left\{ \phi_{{\bf Q}}^{\dagger}\left({\bf {\bf k}}\right)\Lambda_{{\bf k},{\bf k+Q}}{\cal O}\right\} \sum_{{\bf k'}}{\rm Tr}\left\{ \phi_{{\bf Q}}\left({\bf k'}\right)\Lambda_{{\bf k'+Q},{\bf k'}}{\cal O}\right\}.
\end{align}
For completeness, Let us also include the expressions with the decomposed fluctuation matrices,
\begin{align}
\left\langle \left[\hat{F},\left[\hat{F},H_{{\rm int.}}\right]\right]\right\rangle ^{\left[{\cal O},\Lambda\right]=0} 
& =\sum_{{\bf Q}}\frac{1}{A}\sum_{ij{\bf kk'}}\left[\Phi_{{\bf Q}}^{j}\left({\bf k'}\right)\right]^{*}{\rm Tr}\left\{ \left[V_{{\bf k-k'}}\Lambda_{{\bf {\bf k'k}}}{\cal M}^{j}\Lambda_{{\bf k+Q},{\bf k'+Q}}-\delta_{{\bf kk'}}\sum_{{\bf p}}V_{{\bf k-p}}\Lambda_{{\bf {\bf pk}}}\Lambda_{{\bf kp}}{\cal M}^{j}\right]{\cal M}^{i}\right\} \Phi_{{\bf Q}}^{i}\left({\bf k}\right)\nonumber\\
 & +\sum_{{\bf Q}}\sum_{ij{\bf kk'}}V_{{\bf 0}}\left[\Phi_{{\bf Q}}^{j}\left({\bf k'}\right)\right]^{*}\delta_{{\bf kk'}}{\rm Tr}\left\{ \Lambda_{0}P^{T}\right\} {\rm Tr}\left\{ 2{\cal M}^{j}\left[{\cal M}^{i},\Lambda_{0}\right]P^{T}\right\} \Phi_{{\bf Q}}^{i}\left({\bf k}\right)
\end{align}

\begin{align}
\left\langle \left[\hat{F},\left[\hat{F},H_{{\rm int.}}\right]\right]\right\rangle ^{\left\{ {\cal O},\Lambda\right\} =0} 
& =\sum_{{\bf Q}}\frac{1}{A}\sum_{ij{\bf kk'}}\left[\Phi_{{\bf Q}}^{j}\left({\bf k'}\right)\right]^{*}{\rm Tr}\left\{ \left[\delta_{{\bf kk'}}\sum_{{\bf p}}V_{{\bf k-p}}\Lambda_{{\bf {\bf pk}}}\Lambda_{{\bf kp}}{\cal M}^{j}-V_{{\bf k-k'}}\Lambda_{{\bf k'k}}{\cal M}^{j}\Lambda_{{\bf k+Q},{\bf k'+Q}}\right]{\cal M}^{i}\right\} \Phi_{{\bf Q}}^{i}\left({\bf k}\right)\nonumber\\
 & +\sum_{{\bf Q}}\sum_{ij{\bf kk'}}\frac{V_{{\bf Q}}}{A}\left[\Phi_{{\bf Q}}^{j}\left({\bf k'}\right)\right]^{*}{\rm Tr}\left\{ {\cal M}^{j}\Lambda_{{\bf k'},{\bf k'+Q}}{\cal O}\right\} {\rm Tr}\left\{ {\cal M}^{i}\Lambda_{{\bf k+Q},{\bf k}}{\cal O}\right\} \Phi_{{\bf Q}}^{i}\left({\bf k}\right)
\end{align}

\section{Remarks regarding the Bogoliubov transformation}\label{app:remarkbogo}

Our use of the $T_{\bf Q}$ similarity transformation, which can be slightly modified to diagonalize the Hamiltonian itself via a Bogoliubov transformation is based on the thorough discussion in Ref.~\cite{xiao2009theoryBosonsBogoliubovEOM}.
This similarity transformation $T_{\bf Q}$ was shown to be comprised of pairs of time-reversal pairs of eigenvectors of ${\cal D}_{\bf Q}$, which can be inferred from the structure of~\eqref{eq:Ddiagonalization}.
Besides being related by complex-conjugation, the positive and negative frequency eigenvectors $|v\left(\pm{\cal E}_{n,{\bf Q}}\right)\rangle$ have opposite norm with respect to the sesquilinear form
\begin{equation}
    \langle v\left(\pm{\cal E}_{n,{\bf Q}}\right)|\left[-i\tilde{\cal B}_{{\bf Q}}\right]|v\left(\pm{\cal E}_{n,{\bf Q}}\right)\rangle=\pm 1.\label{eq:sesqulinear}
\end{equation}
A corollary to that property of the eigenvectors is that
$T_{\bf Q}^\dagger\left(i\tilde{\cal B}_{{\bf Q}}\right)T_{\bf Q}=\begin{pmatrix}1\\
 & -1
\end{pmatrix}$. 
Combining with Eqs.~\eqref{eq:EOM},~\eqref{eq:Ddiagonalization}, one finds that the same matrices $T_{\bf Q}$ must also encode the requisite Bogoliubov transformation, diagonalizing the Hamiltonian by hermitian congruence, hence the from of Eq.~\eqref{eq:diagonalizedH}.

Recall the residual Bogoliubov transformation,
\begin{equation}
    \delta T_{\bf Q}=\left(T^{\left[H\right]}_{\bf Q}\right)^{-1}
    T^{\left[H+\lambda\delta H\right]}_{\bf Q}
    \equiv
    \begin{pmatrix}U_{\bf Q} & V_{\bf Q}\\
V_{\bf Q}^{*} & U_{\bf Q}^{*}
\end{pmatrix}.
\end{equation}
The bosonic nature of this transformation implies $U_{\bf Q}U_{\bf Q}^\dagger-V_{\bf Q}V_{\bf Q}^\dagger=\mathbb{1}_N$.
This relation in turn guarantees two things.
First, if the off-diagonal $\delta{\cal Z}_{\bf Q}$ vanishes, $V_{\bf Q}$ vanishes, and $U_{\bf Q}$ simply encodes a unitary transformation.
Second, the deviation of this residual transformation form unitarity comes in $V_{\bf Q}$ to first order in the perturbation strength $\lambda$, and affects $U_{\bf Q}$ only with a second order term.
Thus, $\frac{\partial\|\delta{\cal Z}_{{\bf Q}}\|}{\partial\lambda}|_{\lambda\to0}\propto\frac{\partial\|V_{{\bf Q}}\|}{\partial\lambda}|_{\lambda\to0}$.

The perturbed bosonic Hamiltonian is diagonalized in the new basis
\begin{equation}
    \begin{pmatrix}b_{{\bf Q}}\\
b_{-{\bf Q}}^{\dagger}
\end{pmatrix}\equiv
\delta T_{\bf Q}\begin{pmatrix}a_{{\bf Q}}\\
a_{-{\bf Q}}^{\dagger}
\end{pmatrix}.
\end{equation}
At any given momentum $\bf Q$, we may write the vacuum of $b$-bosonic fluctuations $|{\bf \tilde{0}}_{\bf Q}\rangle$ in terms of $a$ operators acting on the unperturbed $a$-vacuum $|{\bf {0}}_{\bf Q}\rangle$,
\begin{equation}
    |{\bf \tilde{0}}_{\bf Q}\rangle
    =
    \frac{1}{\sqrt{\left|\det U_{\bf Q}\right|}}
    \exp \left(\frac{1}{2}a^\dagger_{\bf Q}
    \left[U_{\bf Q}^{-1}V_{\bf Q}\right]
    a^\dagger_{\bf -Q}\right)
    |{\bf {0}}_{\bf Q}\rangle.
\end{equation}
Now consider the fidelity of the unperturbed fluctuation vacuum $|{\bf {0}}_{\bf Q}\rangle$ with the perturbed vacuum $|{\bf \tilde{0}}_{\bf Q}\rangle$,
\begin{align}
    \left|\langle{\bf \tilde{0}}_{{\bf Q}}|{\bf 0}_{{\bf Q}}\rangle\right|&=\left[\det\left(U_{{\bf Q}}U_{{\bf Q}}^{\dagger}\right)\right]^{-1/4}\nonumber\\
    &=\left[\det\left(\mathbb{1}_{N}+V_{{\bf Q}}V_{{\bf Q}}^{\dagger}\right)\right]^{-1/4}\nonumber\\
    &=\exp\left[-\frac{1}{4}\log\det\left(\mathbb{1}_{N}+V_{{\bf Q}}V_{{\bf Q}}^{\dagger}\right)\right]\nonumber\\
    &=\exp\left[-\frac{1}{4}{\rm Tr}\log\left(\mathbb{1}_{N}+V_{{\bf Q}}V_{{\bf Q}}^{\dagger}\right)\right]\nonumber\\
    &\approx1-\frac{1}{4}\|V_{{\bf Q}}\|^{2}+{\cal O}\left(\|V_{{\bf Q}}\|^{4}\right),
\end{align}
from which we extract Eq.~\eqref{eq:explicitfidelity}.

\section{Single mode analysis}\label{app:single_mode}
Let us consider the general form of fluctuation terms we may get for a single $\bf Q$ mode,
\begin{equation}
    {\cal B}_{\bf Q}
    =
    \frac{1}{2}\left(
    {\mathbb P}_{-\bf Q}\partial_{t}{\mathbb X}_{\bf Q}
    -
    {\mathbb X}_{-\bf Q}\partial_{t}{\mathbb P}_{\bf Q}\right),
\end{equation}
\begin{equation}
    {\cal H}_{\bf Q}
    =
    \frac{1}{2}\omega_{\bf Q}\left(
    {\mathbb X}_{\bf -Q}{\mathbb X}_{\bf Q}
    +{\mathbb P}_{\bf -Q}{\mathbb P}_{\bf Q}
    \right)
    +
    \frac{1}{2}\Delta_{\bf Q}\left(
    {\mathbb X}_{\bf -Q}{\mathbb X}_{\bf Q}
    -
    {\mathbb P}_{\bf -Q}{\mathbb P}_{\bf Q}
    \right)
    +\frac{1}{2}\Gamma_{\bf Q}
    \left(
    {\mathbb X}_{\bf -Q}{\mathbb P}_{\bf Q}
    +{\mathbb P}_{\bf -Q}{\mathbb X}_{\bf Q}
    \right).
\end{equation}
It is useful to define $z_{\bf Q}=\Delta_{\bf Q}+i\Gamma_{\bf Q}=\left|z_{\bf Q}\right|e^{i\phi}$.
In the Pauli basis $\upsilon_i$ operating on the vector $\Phi_{\bf Q}=\begin{pmatrix}\mathbb{X}_{{\bf Q}}\\
\mathbb{P}_{{\bf Q}}
\end{pmatrix}$, we can write the appropriate matrices
\begin{equation}
    \tilde{{\cal B}}_{\bf Q}=i\upsilon_y,
\,\,\,\,\,\,
\tilde{{\cal H}}_{\bf Q}=\omega_{\bf Q}+\Delta_{\bf Q}\upsilon_z+\Gamma_{\bf Q}\upsilon_x
= \omega_{\bf Q}+\left|z_{\bf Q}\right|\upsilon_{z}e^{i\phi\upsilon_{y}},
\end{equation}
and the dynamic matrix
\begin{equation}
    {\cal D}_{\bf Q}=i\tilde{{\cal B}}^{-1}_{\bf Q}\tilde{{\cal H}}_{\bf Q}
    =
    e^{-i\frac{\phi}{2}\upsilon_{y}}\left(\omega_{{\bf Q}}\upsilon_{y}+i\left|z_{{\bf Q}}\right|\upsilon_{x}\right)e^{i\frac{\phi}{2}\upsilon_{y}}.
\end{equation}
Next, we diagonalize the dynamic matrix.
The appropriate matrices encoding the similarity transformation are
\begin{equation}
    T=\sqrt{\frac{r_{{\bf Q}}^{-1}}{2}}e^{-i\frac{\phi}{2}\upsilon_{y}}\begin{pmatrix}r_{{\bf Q}} & r_{{\bf Q}}\\
i & -i
\end{pmatrix},
\,\,\,\,\,\,
T^{-1}=\sqrt{\frac{r_{{\bf Q}}}{2}}\begin{pmatrix}r_{{\bf Q}}^{-1} & -i\\
r_{{\bf Q}}^{-1} & i
\end{pmatrix}e^{i\frac{\phi}{2}\upsilon_{y}},
\,\,\,\,\,\,
T^{\dagger}=\sqrt{\frac{r_{{\bf Q}}^{-1}}{2}}\begin{pmatrix}r_{{\bf Q}} & -i\\
r_{{\bf Q}} & i
\end{pmatrix}e^{i\frac{\phi}{2}\upsilon_{y}},
\end{equation}
where we have defined the useful parameter
\begin{equation}
    r_{{\bf Q}}=\left(\frac{\omega_{{\bf Q}}-\left|z_{{\bf Q}}\right|}{\omega_{{\bf Q}}+\left|z_{{\bf Q}}\right|}\right)^{1/2}.
\end{equation}
Notice $r_{\bf Q}\in\left[0,1\right]$, and that $r_{\bf Q}\to 0$ in the limit of maximum squeezing.
The transformation operates on the relevant matrices as,
\begin{equation}
    T^{-1}{\cal D}_{\bf Q} T=\begin{pmatrix}{\cal E}_{{\bf Q}} & 0\\
0 & -{\cal E}_{{\bf Q}}
\end{pmatrix},\,\,\,\,\,\,\,
T^{\dagger}\tilde{{\cal H}}_{\bf Q} T=\begin{pmatrix}{\cal E}_{{\bf Q}} & 0\\
0 & {\cal E}_{{\bf Q}}
\end{pmatrix},\label{eq:Tmatrices}
\end{equation}
with $ {\cal E}_{{\bf Q}}=\sqrt{\omega_{{\bf Q}}^{2}-\left|z_{{\bf Q}}\right|^{2}}$.

We now proceed to examine how the Hamiltonian changes in the diagonalized basis above when a small perturbation is introduced,
\begin{equation}
    \omega_{\bf Q}\to \omega_{\bf Q}+\delta\omega,
    \,\,\,\,\,
   \left|z_{\bf Q}\right|\to \left|z_{\bf Q}\right|+\delta z,
    \,\,\,\,\,
    \phi \to \phi+\delta\phi.
\end{equation}
The off diagonal elements in this so-called equilibrium basis will induce parametric driving.
To lowest order in these perturbations, we recover
\begin{equation}
    T^{\dagger}\left[\tilde{{\cal H}}_{{\bf Q}}+\delta\tilde{{\cal H}}_{{\bf Q}}\right]T=\begin{pmatrix}{\cal E}_{{\bf Q}} & 0\\
0 & {\cal E}_{{\bf Q}}
\end{pmatrix}+\begin{pmatrix}\delta{\cal E}_{{\bf Q}} & \delta{\cal Z}\\
\delta{\cal Z}^{*} & \delta{\cal E}_{{\bf Q}}
\end{pmatrix},\label{eq:perturbedHmatrix}
\end{equation}
\begin{equation}
    \delta{\cal Z}=\omega_{{\bf Q}}\frac{r_{{\bf Q}}+r_{{\bf Q}}^{-1}}{2}\delta\frac{\left|z_{{\bf Q}}\right|}{\omega_{{\bf Q}}}-i\left|z_{{\bf Q}}\right|\delta\phi.
\end{equation}
We can make a number of immediate observations.
First, a change of the spectrum of bosonic excitations, introduced by the perturbation, $\delta{\cal E}_{\bf Q}$, does not necessarily translate to parametric driving.
One must additionally either (i) change the ratio of anomalous to normal terms, or (ii) perturb the phase of the anomalous term.
Here, we point out an interesting effect: in the limit of strong squeezing, $r_{\bf Q}\to 0$, the parametric susceptibility to a change of the anomalous-to-normal ratio diverges as $\sim r_{\bf Q}^{-1}$.

Let us furthermore examine the energy of the bosonic vacuum.
The shift of vacuum energy compared to the unperturbed state is,
\begin{equation}
    \delta E_{{\rm Vac},\bf Q}=
    \sqrt{\left({\cal E}_{\bf Q}+\delta{\cal E}_{\bf Q}\right)^2-\left|\delta{\cal Z}\right|^2}
    -\left({\cal E}_{\bf Q}+\delta{\cal E}_{\bf Q}\right)
    \approx
    -\frac{\left|\delta{\cal Z}\right|^2}{2{\cal E}_{\bf Q}},\label{eq:vacenergysupp}
\end{equation}
where in the last step we expanded the expression to leading order in the perturbed terms.
Clearly, the susceptibility of the vacuum energy to modification introduced by $\delta \tilde{{\cal H}}_{\bf Q}$ is proportional to the magnitude of the parametric terms introduced.

Lastly, we examine the fidelity between the unperturbed vacuum wavefunction and the perturbed on,
\begin{equation}
    {\cal F}\left(\delta\tilde{{\cal H}}_{{\bf Q}}\right)=\left|\left\langle{\rm Vac}\left(\tilde{{\cal H}}_{{\bf Q}}\right)|{\rm Vac}\left(\tilde{{\cal H}}_{{\bf Q}}+\delta\tilde{{\cal H}}_{{\bf Q}}\right)\right\rangle\right|.
\end{equation}
We begin by writing the perturbed Hamiltonian in terms of the diagonalized (unperturbed) bosons [see Eq.~\eqref{eq:perturbedHmatrix}],
\begin{equation}
    H_{\bf Q}\approx
    \left(\cal{E}_{\bf Q}+\delta\cal{E}_{\bf Q}\right)
    \left(
    a^\dagger_{\bf Q}a_{\bf Q}+a^\dagger_{\bf -Q}a_{\bf -Q}
    \right)
    +\left(\delta{\cal{Z}}_{\bf Q}a_{\bf Q}a_{\bf -Q}+{\rm h.c.}\right).\label{eq:perturbedHqsupp}
\end{equation}
The ground state of this Hamiltonian may be expressed in terms of the vacuum of this $a$-bosons,
\begin{equation}
    |{\rm Vac}\left(\tilde{{\cal H}}_{{\bf Q}}+\delta\tilde{{\cal H}}_{{\bf Q}}\right)\rangle
    =\frac{1}{\cal N}
    \exp\left(-s a_{{\bf Q}}^{\dagger}a_{{\bf -Q}}^{\dagger}\right)
    |{\rm Vac}\left(\tilde{{\cal H}}_{{\bf Q}}\right)\rangle,\label{eq:vacina}
\end{equation}
where $s$ is a complex number and ${\cal N}$ is a normalization factor.
We begin by recovering the value of the latter,
\begin{align}
{\cal N} & =\langle{\rm Vac}\left(\tilde{{\cal H}}_{{\bf Q}}\right)|e^{-s^{*}a_{{\bf Q}}a_{{\bf -Q}}}e^{-sa_{{\bf Q}}^{\dagger}a_{{\bf -Q}}^{\dagger}}|{\rm Vac}\left(\tilde{{\cal H}}_{{\bf Q}}\right)\rangle\nonumber\\
 & =\sum_{m,n}\frac{\left(-s\right)^{n}\left(-s^{*}\right)^{m}}{m!n!}\langle{\rm Vac}\left(\tilde{{\cal H}}_{{\bf Q}}\right)|\left(a_{{\bf Q}}a_{{\bf -Q}}\right)^{m}\left(a_{{\bf Q}}^{\dagger}a_{{\bf -Q}}^{\dagger}\right)^{n}|{\rm Vac}\left(\tilde{{\cal H}}_{{\bf Q}}\right)\rangle\nonumber\\
 & =\sum_{n}\frac{\left|s\right|^{2n}}{n!n!}\langle{\rm Vac}\left(\tilde{{\cal H}}_{{\bf Q}}\right)|\left(a_{{\bf Q}}a_{{\bf -Q}}\right)^{n}\left(a_{{\bf Q}}^{\dagger}a_{{\bf -Q}}^{\dagger}\right)^{n}|{\rm Vac}\left(\tilde{{\cal H}}_{{\bf Q}}\right)\rangle\nonumber\\
& =\frac{1}{1-\left|s\right|^{2}}.
\end{align}
The properties of the vacuum wavefunction are thus all captured by the parameter $s$.
We note that the vacuum ``photon number'' $n_a$, i.e., the expectation value $\left\langle a^\dagger_{\bf Q}a_{\bf Q}\right\rangle$ in the perturbed vacuum is related to $s$,
\begin{equation}
    n_a=\frac{\left|s\right|^{2}}{1-\left|s\right|^{2}},
    \,\,\,\,\,
    \left|s\right|^{2}=\frac{n_{a}}{n_{a}+1}.
\end{equation}
Generally, one can readily show using~\eqref{eq:vacina} that the fidelity between vacua with different $s$-parameters,
\begin{equation}
    {\cal F}\left(s_{1},s_{2}\right)=\sqrt{\frac{\left(1-\left|s_{1}\right|^{2}\right)\left(1-\left|s_{2}\right|^{2}\right)}{1+\left|s_{1}\right|^{2}\left|s_{2}\right|^{2}-2{\rm Re}\left\{ s_{1}^{*}s_{2}\right\}}}.
\end{equation}

We must now explicitly find $s$ of the perturbed vacuum.
To do so, we demand that the annihilation operators which diagonalize the \textit{perturbed} Hamiltonian Eq.~\eqref{eq:perturbedHqsupp} annihilate the perturbed vacuum.
First, notice we may write
\begin{equation}
    T=e^{-i\frac{\phi}{2}\upsilon_{y}}\frac{1}{\sqrt{2}}\begin{pmatrix}1 & 1\\
i & -i
\end{pmatrix}
\begin{pmatrix}\frac{\sqrt{r_{{\bf Q}}}+\sqrt{r_{{\bf Q}}^{-1}}}{2} & \frac{\sqrt{r_{{\bf Q}}}-\sqrt{r_{{\bf Q}}^{-1}}}{2}\\
\frac{\sqrt{r_{{\bf Q}}}-\sqrt{r_{{\bf Q}}^{-1}}}{2} & \frac{\sqrt{r_{{\bf Q}}}+\sqrt{r_{{\bf Q}}^{-1}}}{2}
\end{pmatrix}.
\end{equation}
Here, $T$ is decomposed to three parts. The first part on the left rotates the phase of the anomalous terms, the second constructs bosonic variables out of the ``coordinate'' basis, and the last one accounts for the Bogoliubov transformation.
Taking advantage of this deconstructed form, straightforward algebra of operating on the vacuum state allows us to extract $s$ in terms of the perturbations,
\begin{equation}
    s=\frac{\tilde{r}-\tilde{r}^{-1}}{\tilde{r}+\tilde{r}^{-1}}\sqrt{\frac{\delta{\cal Z}^{*}}{\delta{\cal Z}}},
    \,\,\,\,\,
    \tilde{r}=\left(\frac{{\cal E}_{{\bf Q}}+\delta{\cal E}_{{\bf Q}}-\left|\delta{\cal Z}\right|}{{\cal E}_{{\bf Q}}+\delta{\cal E}_{{\bf Q}}+\left|\delta{\cal Z}\right|}\right)^{1/4}.
\end{equation}
Finally, we find using the fact that the unperturbed $s$ vanishes in the reference state,
\begin{align}
     {\cal F}\left(\delta\tilde{{\cal H}}_{{\bf Q}}\right)&=\sqrt{1-\left|s\right|^{2}}\nonumber\\
     &=\frac{2}{\tilde{r}+\tilde{r}^{-1}}\nonumber\\
     &\approx 1-\frac{1}{8}\frac{\left|\delta{\cal Z}\right|^2}{{\cal E}^2_{\bf Q}}.
\end{align}
Much like in Eq.~\eqref{eq:vacenergysupp}, the fidelity is sensitive to the magnitude of parametric term.

\subsection{Parametric driving}
Let us examine the equations of motion in the presence of the periodic drive with frequency and phase $\omega_d$, and $\varphi_d$, respectively.
The perturbed dynamic matrix in the almost-diagonal basis,
\begin{equation}
    \tilde{\cal D}=T^{-1}\left[{{\cal D}}_{{\bf Q}}+i{\cal \tilde{B}}^{-1}_{{\bf Q}}\delta\tilde{{\cal H}}_{{\bf Q}}\right]T=\begin{pmatrix}{\cal E}_{{\bf Q}} & 0\\
0 & -{\cal E}_{{\bf Q}}
\end{pmatrix}+\begin{pmatrix}\delta{\cal E}_{{\bf Q}} & \delta{\cal Z}\\
-\delta{\cal Z}^{*} & -\delta{\cal E}_{{\bf Q}}
\end{pmatrix}.\label{eq:perturbedDmatrix}
\end{equation}
Let us absorb the unimportant $\delta{\cal E}$ contribution into ${\cal E}_{\bf Q}$.
This will only act to modify the parametric resonance frequency.
The periodic driving is introduced by replacing
$\delta{\cal Z}\to i\varrho e^{-i\omega_d t+i\varphi_d}$, with $\varrho=\left|\delta{\cal Z}\right|$, absorbing the phase of $\delta{\cal Z}$ into that of the drive, and shifting it by $\frac{\pi}{2}$ for convenience of notation.
The equation of motion is
\begin{equation}
    i\frac{d}{dt}\tilde{\Phi}=\tilde{\cal D}\tilde{\Phi},
\end{equation}
where the vector $\tilde{\Phi}$ is related to original coordinates by $\tilde{\Phi}=T^{-1}\Phi_{\bf Q}$.
It is useful to move to a rotating frame by defining the vector
\begin{equation}
    \Psi=\begin{pmatrix}e^{i{\cal E}_{{\bf Q}}t} & 0\\
0 & e^{-i{\cal E}_{{\bf Q}}t}
\end{pmatrix}\tilde{\Phi},
\end{equation}
where we recover the equation of motion (assuming $\omega_d$ satisfies the parametric resonance condition)
\begin{equation}
    i\frac{d}{dt}\Psi=\begin{pmatrix} & i\varrho e^{i\varphi_{d}}\\
i\varrho e^{-i\varphi_{d}}
\end{pmatrix}\Psi,
\end{equation}
and its straightforward solution
\begin{equation}
    \Psi\left(t\right)=\begin{pmatrix}\cosh\varrho t & \sinh\varrho te^{i\varphi_{d}}\\
\sinh\varrho te^{-i\varphi_{d}} & \cosh\varrho t
\end{pmatrix}\Psi_{0}.
\end{equation}
Notice that the phase $\varphi_d$ determines which quadrature gets enhanced in the rotating frame and which is attenuated.
In the ``lab'' frame both oscillate with exponentially diverging amplitudes, see below.

We now manipulate to get the solution in the original coordinate basis,
\begin{align}
    \Phi\left(t\right)&=T\tilde{\Phi}\left(t\right)\nonumber\\
    &=Te^{-i{\cal E}_{{\bf Q}}t\upsilon_{z}}\Psi\left(t\right)\nonumber\\
    &=Te^{-i\left({\cal E}_{{\bf Q}}t-\frac{\varphi_{d}}{2}\right)\upsilon_{z}}e^{\varrho t\upsilon_{x}}e^{-i\frac{\varphi_{d}}{2}\upsilon_{z}}T^{-1}\Phi_{0}\nonumber\\
    &=\left[Te^{-i\left({\cal E}_{{\bf Q}}t-\frac{\varphi_{d}}{2}\right)\upsilon_{z}}T^{-1}\right]\left[Te^{\varrho t\upsilon_{x}}T^{-1}\right]\left[Te^{-i\frac{\varphi_{d}}{2}\upsilon_{z}}T^{-1}\right]\Phi_{0}\nonumber\\
    &=e^{-i\frac{\phi}{2}\upsilon_{y}}\begin{pmatrix}\cos\left({\cal E}_{{\bf Q}}t-\frac{\varphi_{d}}{2}\right) 
    & -r\sin\left({\cal E}_{{\bf Q}}t-\frac{\varphi_{d}}{2}\right)\nonumber\\
r^{-1}\sin\left({\cal E}_{{\bf Q}}t-\frac{\varphi_{d}}{2}\right) & \cos\left({\cal E}_{{\bf Q}}t-\frac{\varphi_{d}}{2}\right)
\end{pmatrix}e^{\varrho t\upsilon_{z}}\begin{pmatrix}\cos\frac{\varphi_{d}}{2} 
& -r\sin\frac{\varphi_{d}}{2}\\
r^{-1}\sin\frac{\varphi_{d}}{2} 
& \cos\frac{\varphi_{d}}{2}
\end{pmatrix}e^{i\frac{\phi}{2}\upsilon_{y}}\Phi_{0}\\
&=e^{-i\frac{\phi}{2}\upsilon_{y}}\begin{pmatrix}\cos\left({\cal E}_{{\bf Q}}t-\frac{\varphi_{d}}{2}\right) & -r\sin\left({\cal E}_{{\bf Q}}t-\frac{\varphi_{d}}{2}\right)\\
r^{-1}\sin\left({\cal E}_{{\bf Q}}t-\frac{\varphi_{d}}{2}\right) 
& \cos\left({\cal E}_{{\bf Q}}t-\frac{\varphi_{d}}{2}\right)
\end{pmatrix}\begin{pmatrix}e^{\varrho t}\cos\frac{\varphi_{d}}{2} & -e^{\varrho t}r\sin\frac{\varphi_{d}}{2}\\
e^{-\varrho t}r^{-1}\sin\frac{\varphi_{d}}{2} & e^{-\varrho t}\cos\frac{\varphi_{d}}{2}
\end{pmatrix}e^{i\frac{\phi}{2}\upsilon_{y}}\Phi_{0}\nonumber\\
&=e^{-i\frac{\phi}{2}\upsilon_{y}}\begin{pmatrix}e^{\varrho t}\cos\left({\cal E}^{*}t\right)\cos\frac{\varphi_{d}}{2}-e^{-\varrho t}\sin\left({\cal E}^{*}t\right)\sin\frac{\varphi_{d}}{2} & -r\left[e^{\varrho t}\cos\left({\cal E}^{*}t\right)\sin\frac{\varphi_{d}}{2}+e^{-\varrho t}\sin\left({\cal E}^{*}t\right)\cos\frac{\varphi_{d}}{2}\right]\\
r^{-1}\left[e^{\varrho t}\sin\left({\cal E}^{*}t\right)\cos\frac{\varphi_{d}}{2}+e^{-\varrho t}\cos\left({\cal E}^{*}t\right)\sin\frac{\varphi_{d}}{2}\right] & e^{-\varrho t}\cos\frac{\varphi_{d}}{2}\cos\left({\cal E}^{*}t\right)-e^{\varrho t}\sin\left({\cal E}^{*}t\right)\sin\frac{\varphi_{d}}{2}
\end{pmatrix}e^{i\frac{\phi}{2}\upsilon_{y}}\Phi_{0},
\end{align}
and in the last line we shifted the time $t$ for brevity,
${\cal E}^{*}t\equiv{\cal E}_{{\bf Q}}t-\frac{\varphi_{d}}{2}$.
It is useful to consider slightly different coordinates $\begin{pmatrix}\mathbb{X}\\
\mathbb{P}
\end{pmatrix}\equiv e^{i\frac{\phi}{2}\upsilon_{y}}\Phi$, and define the initial condition
$\mathbb{C}_{0}\equiv \sqrt{r^{-1}}\cos\frac{\varphi_{d}}{2}\mathbb{X}_{0}-\sqrt{r}\sin\frac{\varphi_{d}}{2}\mathbb{P}_{0}$,
and
$\mathbb{D}_{0}\equiv \sqrt{r^{-1}}\sin\frac{\varphi_{d}}{2}\mathbb{X}_{0}+\sqrt{r}\cos\frac{\varphi_{d}}{2}\mathbb{P}_{0}$,
such that
\begin{equation}
    \mathbb{X}\left(t\right)=\sqrt{r}\left[e^{\varrho t}\cos\left({\cal E}^{*}t\right)\mathbb{C}_{0}-e^{-\varrho t}\sin\left({\cal E}^{*}t\right)\mathbb{D}_{0}\right],
\end{equation}
\begin{equation}
    \mathbb{P}\left(t\right)=\sqrt{r^{-1}}\left[e^{\varrho t}\sin\left({\cal E}^{*}t\right)\mathbb{C}_{0}+e^{-\varrho t}\cos\left({\cal E}^{*}t\right)\mathbb{D}_{0}\right].
\end{equation}
Both quadratures grow exponentially in time, a consequence of the parametric resonance.
In the absence of damping, the order in the reference state is thus unstable to parametric driving.
Notice that the growth in the two quadratures is uneven, a result of the original squeezing $r$ in the unperturbed Hamiltonian.

Succinctly written, we can separate the three parts of the time evolution: the squeezing, the rotation due to the driving, and the exponential growth instability,
\begin{equation}
    \begin{pmatrix}\mathbb{X}\\
\mathbb{P}
\end{pmatrix}=\exp\left[\frac{1}{2}\upsilon_{z}\log r\right]\times\exp\left[-i\upsilon_{y}\left({\cal E}_{{\bf Q}}t-\frac{\varphi_{d}}{2}\right)\right]\times\exp\left[\upsilon_{z}\varrho t\right]\begin{pmatrix}\mathbb{C}_{0}\\
\mathbb{D}_{0}
\end{pmatrix}.
\end{equation}

\section{Quantum Hall double layer calculations}\label{app:QHbilayer}
Recall the Hamiltonian of the quantum Hall double layer problem,  
\begin{equation}
    H_{\rm QH-2\ell}=H_{\rm C} + H_z+H_{\rm tun.}
\end{equation}
The different terms are
\begin{equation}
    H_{\rm C}=\frac{1}{2A}\sum_{\bf q}V_{\bf q}^0\tilde{\rho}_{\bf q}\tilde{\rho}_{\bf -q},
    \,\,\,\,\,
    H_z=\frac{1}{2A}\sum_{\bf q}V_{\bf q}^z\tilde{\rho}^z_{\bf q}\tilde{\rho}^z_{\bf -q},
\end{equation}
\begin{equation}
    H_{{\rm tun.}}=-\Delta_{\rm sas}\sum_{{\bf k}}c_{{\bf k}}^{\dagger}\ell_x c_{{\bf k}}.
\end{equation}
The projected density operators are 
$\tilde{\rho}_{\bf q}=\sum_{\bf k}c^\dagger_{\bf k+q}\Lambda_{{\bf k},{\bf k+q}} c_{\bf k}$, $\tilde{\rho}^z_{\bf q}=\sum_{\bf k}c^\dagger_{\bf k+q}\ell_z\Lambda_{{\bf k},{\bf k+q}} c_{\bf k}$,
and the form factors 
$\Lambda_{{\bf k},{\bf k+q}}^{\alpha\beta}=\delta^{\alpha\beta}f\left({\bf q}\right)e^{-i\frac{{\Omega}}{2}{\bf k}\times\left({\bf k+q}\right)}=\delta^{\alpha\beta}f\left({\bf q}\right)e^{-i\frac{{\Omega}}{2}{\bf k}\times{\bf q}}$, 
with
$f\left({\bf q}\right)=e^{-\frac{1}{4}{\Omega}\left|{\bf q}\right|^{2}}L_n\left({\Omega}\left|{\bf q}\right|^{2}\right)$ (${\Omega}=l_B^2$, i.e., the magnetic length squared, $L_n$ is the $n$ Laguerre polynomial).

With nominal layer separation $d$, the two interaction strengths are given by
\begin{equation}
    V_{{\bf q}}^{0/z}=\frac{V_{{\bf q}}^{C}\left(1\pm e^{-qd}\right)}{2},
\end{equation}
with $V_{{\bf q}}^{C}$ the Fourier transformed Coulomb repulsion in two dimensions.

\subsection{Ground state manifold}
One may simply evaluate the Hartree-Fock energies of the various terms within the ordered state, 
\begin{align}
\frac{\left\langle H_{0}\right\rangle }{A} & =\frac{V_{{\bf 0}}^{0}}{2A^{2}}\left(\sum_{{\bf k}}{\rm Tr}\left\{ \Lambda_{{\bf k},{\bf k}}P_{\bf k}\right\} \right)^{2}+\frac{1}{2A^{2}}\sum_{{\bf kk'}}V_{{\bf k'-k}}^{0}{\rm Tr}\left\{ P^{T}_{\bf k}\Lambda_{{\bf k},{\bf k'}}\bar{P}_{\bf k'}^{T}\Lambda_{{\bf k'},{\bf k}}\right\} \nonumber\\
 & =\frac{V_{{\bf 0}}^{0}}{2}\left({\rm Tr}\left\{ P_{\bf k}\right\} \right)^{2}+\frac{1}{2A^{2}}\sum_{{\bf kk'}}V_{{\bf k'-k}}^{0}\Lambda_{{\bf k},{\bf k'}}\Lambda_{{\bf k'},{\bf k}}{\rm Tr}\left\{ P_{\bf k}^{T}\bar{P}^{T}_{\bf k'}\right\} \nonumber\\
 & =\frac{V_{{\bf 0}}^{0}}{2},
\end{align}
where in the last line we have assumed $P$ is independent of momentum -- a property of the \textit{ordered} state.
The ordered state indeed is lower in energy, because the Fock energy term is always positive unless the density matrix is constant.
We will assume this ordered state henceforth.
\begin{align}
    \frac{\left\langle H_{z}\right\rangle }{A}&=\frac{V_{{\bf 0}}^{z}}{2A^{2}}\left(\sum_{{\bf k}}{\rm Tr}\left\{ \ell_{z}\Lambda_{{\bf k},{\bf k}}P\right\} \right)^{2}+\frac{1}{2A^{2}}\sum_{{\bf kk'}}V_{{\bf k'-k}}^{z}{\rm Tr}\left\{ P^{T}\ell_{z}\Lambda_{{\bf k},{\bf k'}}\bar{P}^{T}\ell_{z}\Lambda_{{\bf k'},{\bf k}}\right\} \nonumber\\
    &=\frac{V_{{\bf 0}}^{z}}{2}\left(\frac{1}{2}{\rm Tr}\left\{ \ell_{z}{\cal O}\right\} \right)^{2}-\frac{1}{2A^{2}}\sum_{{\bf kk'}}V_{{\bf k'-k}}^{z}\left|\Lambda_{{\bf k},{\bf k'}}\right|^{2}\frac{1}{4}{\rm Tr}\left\{ {\cal O}\ell_{z}\left[{\cal O},\ell_{z}\right]\right\} , 
\end{align}
\begin{equation}
    \frac{\left\langle H_{{\rm tun.}}\right\rangle }{A}=-\frac{1}{2}\Delta_{{\rm sas}}{\rm Tr}\left\{ {\cal O}\ell_{x}\right\}. 
\end{equation}
Parameterizing ${\cal O}= X\ell_x+Y\ell_y+Z\ell_z$, with $X^2+Y^2+Z^2=1$ (so the defined $P$ is indeed a projector), the ground state energy density is
\begin{equation}
    \frac{E_{HF}}{A}=\frac{V_{{\bf 0}}^{0}+V_{{\bf 0}}^{z}}{2}-\frac{1}{2}\left(X^{2}+Y^{2}\right)\left[V_{{\bf 0}}^{z}-\frac{1}{A^{2}}\sum_{{\bf kk'}}V_{{\bf k'-k}}^{z}\left|\Lambda_{{\bf k},{\bf k'}}\right|^{2}\right]-\Delta_{{\rm sas}}X.\label{eq:QHhartreefock}
\end{equation}
The first term is the background contribution.
The layer-asymmetric interaction chooses an interlayer coherent state as the preferred interacting ground state, with arbitrary $U\left(1\right)$ phase.
The interlayer tunneling breaks this $U\left(1\right)$ symmetry explicitly, making ${\cal O}=\ell_x$ the ground state.

\subsection{Fluctuation Lagrangian}
The fluctuation operator can be decomposed into ${\cal M}_1=\ell_y$ and ${\cal M}_2=\ell_z$ components.
The Berry phase term is straightforward,
\begin{equation}
    {\cal \tilde{B}}_{{\bf Q}}^{ij}\left({\bf p,p'}\right)=2\epsilon^{ij}\delta_{{\bf pp'}}.
\end{equation}
The tunneling term produces a fluctuation Hamiltonian contribution
\begin{equation}
    {\cal \tilde{H}}_{{\rm tun.}{\bf Q}}^{ij}\left({\bf p,p'}\right)=2\Delta_{{\rm sas}}\delta_{{\bf pp'}}\delta^{ij}.
\end{equation}
The symmetric part of the Hamiltonian gives rise to
\begin{equation}
    {\cal \tilde{H}}_{{\rm 0}{\bf Q}}^{ij}\left({\bf p,p'}\right)=\left[\delta_{{\bf pp'}}\sum_{{\bf k}}\frac{V_{{\bf p-k}}^{0}}{A}\left|\Lambda_{{\bf kp}}\right|^{2}-\frac{V_{{\bf p-p'}}^{0}}{A}\Lambda_{{\bf {\bf p'p}}}\Lambda_{{\bf p+Q},{\bf p'+Q}}\right]\delta^{ij}.
\end{equation}
The layer-capacitive interaction contributes
\begin{align}
    {\cal \tilde{H}}_{{\rm z}{\bf Q}}^{ij}\left({\bf p,p'}\right)	&=-\left[\delta_{{\bf pp'}}\sum_{{\bf k}}\frac{V_{{\bf p-k}}^{z}}{A}\left|\Lambda_{{\bf kp}}\right|^{2}-\frac{V_{{\bf p-p'}}^{z}}{A}\Lambda_{{\bf {\bf p'p}}}\Lambda_{{\bf p+Q},{\bf p'+Q}}\right]\delta^{ij}\nonumber\\
	&+\left[\frac{V_{{\bf Q}}^{z}}{A}\Lambda_{{\bf p'},{\bf p'+Q}}\Lambda_{{\bf p+Q},{\bf p}}-\frac{V_{{\bf p-p'}}^{z}}{A}\Lambda_{{\bf p'p}}\Lambda_{{\bf p+Q},{\bf p'+Q}}\right]\left(1-\left(-1\right)^{i}\right)\delta^{ij}.
\end{align}
Notice that only the second line in this last contribution distinguishes between the $\ell_y$ and $\ell_z$ fluctuations -- this is the squeezing term.

\subsection{Single mode approximation}
We may get an analytical handle on the lowest lying collective mode by employing the approximation,
\begin{equation}
    \Phi_{\bf Q}^1\left({\bf p}\right)\approx
    \frac{1}{2}{\mathbb X}_{\bf Q}f\left({\bf Q}\right)e^{-i\frac{{\Omega}}{2}{\bf p}\times{\bf Q}},
    \,\,\,\,\,
    \Phi_{\bf Q}^2\left({\bf p}\right)\approx
    \frac{1}{2}{\mathbb P}_{\bf Q}f\left({\bf Q}\right)e^{-i\frac{{\Omega}}{2}{\bf p}\times{\bf Q}}.
\end{equation}
In the Hamiltonian, we find the following contributions to leading order in $\bf Q$,
\begin{equation}
    \frac{1}{2}\mathbb{X}_{{\bf -Q}}\mathbb{X}_{{\bf Q}}f^{2}\left({\bf Q}\right)
    \left[
    \frac{\rho_{s}^{0}+\rho_{s}^{z}}{2}\left|{\bf Q}\right|^{2}
    +2E_{\rm easy}
    +f^{2}\left({\bf Q}\right)V_{{\bf Q}}^{z}-V_{{\bf 0}}^{z}
    +\Delta_{{\rm sas}}
    \right],
\end{equation}
\begin{equation}
    \frac{1}{2}\mathbb{P}_{{\bf -Q}}\mathbb{P}_{{\bf Q}}f^{2}\left({\bf Q}\right)
    \left[
    \frac{\rho_{s}^{0}-\rho_{s}^{z}}{2}\left|{\bf Q}\right|^{2}
    +\Delta_{{\rm sas}}
    \right].
\end{equation}
We used the definitions,
\begin{equation}
    \rho_{s}^{0/z}=\frac{1}{A}\sum_{{\bf p}}\frac{\Omega^{2}}{4}V_{{\bf p}}^{0/z}f^{2}\left({\bf p}\right)\left|{\bf p}\right|^{2},
\end{equation}
\begin{equation}
    E_{\rm easy}=\frac{1}{2A}\sum_{{\bf p}}\left[V_{{\bf 0}}^{z}-V_{{\bf p}}^{z}f^{2}\left({\bf p}\right)\right].
\end{equation}
The $\rho_s$ quantity defines a stiffness for the fluctuations in different directions, whereas $E_{\rm easy}$ is the capacitive charging energy, or alternatively, the Hartree-Fock energy difference defining the easy-plane manifold [see Eq.~\eqref{eq:QHhartreefock}].

Since we are interested in the behavior to leading order in $\bf Q$, we may absorb into the coordinate variables the global $f^2\left({\bf Q}\right)$ factor, and renormalize the stiffness contributions to include the smooth (vanishing at ${\bf Q}\to0$) $f^{2}\left({\bf Q}\right)V_{{\bf Q}}^{z}-V_{{\bf 0}}^{z}$ interaction.
We can thus finally identify the appropriate single-mode parameters,
\begin{equation}
    \omega_{{\bf Q}}=\frac{\rho_{s}^{0}}{2}\left|{\bf Q}\right|^{2}+E_{\rm easy}+\Delta_{{\rm sas}},
\end{equation}
\begin{equation}
    z_{{\bf Q}}=\frac{\rho_{s}^{z}}{2}\left|{\bf Q}\right|^{2}+E_{\rm easy}. 
\end{equation}
The associated spectrum is
\begin{equation}
    {\cal E}_{{\bf Q}}=\sqrt{\frac{\left(\rho_{s}^{0}+\rho_{s}^{z}\right)\left(\rho_{s}^{0}-\rho_{s}^{z}\right)}{4}\left|{\bf Q}\right|^{4}+\left(\Delta_{{\rm sas}}+E_{\rm easy}\left(\frac{\rho_{s}^{0}-\rho_{s}^{z}}{\rho_{s}^{0}}\right)\right)\rho_{s}^{0}\left|{\bf Q}\right|^{2}+2\Delta_{{\rm sas}}E_{\rm easy}+\Delta_{{\rm sas}}^{2}},
\end{equation}
Let us make several observations, starting with the limit $\Delta_{\rm sas}\to0$.
In that case, for small momentum $\left|{\bf Q}\right|\ll\sqrt{\frac{8E_{\rm easy}}{\rho_{s}^{0}+\rho_{s}^{z}}}$,
one finds the expected linearly dispersing Goldstone mode,
${\cal E}_{\bf Q}\approx\sqrt{E_{\rm easy}\left(\rho_{s}^{0}-\rho_{s}^{z}\right)}\left|{\bf Q}\right|$.

In another concrete case, let us allow for finite interlayer tunneling, which is typically much smaller than $E^{\rm cap}$.
The energy of the ${\bf Q}=0$ mode, i.e., the gap of the Goldstone spectrum, is roughly the geometric mean of the $E^{\rm cap}$ and tunneling energy scales,
\begin{equation}
    {\cal E}_{{\bf 0}}=\sqrt{2E_{\rm easy}\Delta_{{\rm sas}}}\sqrt{1+\frac{\Delta_{{\rm sas}}}{2E_{\rm easy}}}.
\end{equation}
The susceptibility toward parametrically driving this mode an be easily extracted,
\begin{equation}
    \delta z=\frac{{\cal E}_{{\bf 0}}}{2\left(1+\frac{\Delta_{{\rm sas}}}{2E_{\rm easy}}\right)}d\left[\log\frac{E_{\rm easy}}{\Delta_{{\rm sas}}}\right].
\end{equation}

\section{Flat band model}\label{app:flatbandzetamodel}
As discussed in the main text, motivated by the rich and complex hierarchy of correlated ground state of twisted multilayer graphene, we consider the following simplified model.
The model is related to a tunable-metric model introduced in Refs.~\cite{tunablemetricHofmann2022,tunableMetricPRL_PhysRevLett.130.226001}.
The single-particle Hamiltonian we explore has valley $\tau_{i}$, pseudo-spin $\sigma_{i}$, and orbital $\rho_{i}$ degrees of freedom (all Pauli matrices),
\begin{equation}
    H=-t\sum_{{\bf k}}c_{{\bf k}}^{\dagger}\left[\left(\frac{1+\sigma_{z}\tau_{z}}{2}\cos\alpha_{{\bf k}}+\frac{1-\sigma_{z}\tau_{z}}{2}\cos\beta_{{\bf k}}\right)\rho_{x}+\left(\frac{1+\sigma_{z}\tau_{z}}{2}\sin\alpha_{{\bf k}}+\frac{1-\sigma_{z}\tau_{z}}{2}\sin\beta_{{\bf k}}\right)\rho_{y}+\Delta_\sigma\sigma_z\right]c_{{\bf k}}.\label{eq:pseudotbgsingleparticleapp}
\end{equation}
The momentum-dependent functions $\alpha_{\bf k}$ and $\beta_{\bf k}$ determine the quantum geometric properties of the resultant flat bands, and are related to each other by a 45-degree rotation,
\begin{equation}
    \alpha_{{\bf k}}=\zeta\left(\cos k_{x}+\cos k_{y}\right),
\end{equation}
\begin{equation}
    \beta_{{\bf k}}=\zeta\left(\cos\frac{k_{x}+k_{y}}{\sqrt{2}}+\cos\frac{k_{x}-k_{y}}{\sqrt{2}}\right).
\end{equation}
The parameter $\zeta$ controls the quantum metric of the projected flat bands.
To see this, consider without loss of generality the $\sigma_z\tau_z=1$ set of bands.
The Bloch eigenvector at momentum $\bf k$ of the valence band is given by
\begin{equation}
    {\mathsf u}_{{\bf k}}=\frac{1}{\sqrt{2}}e^{i\alpha_{{\bf k}}\frac{\rho_{z}}{2}}\left(1, i\right)=\frac{1}{\sqrt{2}}\begin{pmatrix}e^{i\frac{\alpha_{{\bf k}}}{2}}\\
 ie^{-i\frac{\alpha_{{\bf k}}}{2}}
\end{pmatrix}.
\end{equation}
Consider the Fubini-Study metric,
\[g_{\mu \nu} = {\rm Tr}\left[\partial_\mu P_{\bf k} \partial_\nu P_{\bf k}\right],\]
with the band projections $P_{\bf k}^\alpha=|{\mathsf u}_{\bf k}^\alpha\rangle\langle {\mathsf u}_{\bf k}^\alpha|$.
Straightforward calculation yields,
\begin{equation}
    g_{\mu\nu}\left({\bf k}\right) = \frac{\zeta ^2}{4} \sin\left(k_\mu \right)\sin\left(k_\nu a\right).\label{Eq:tunablemetricresult}
\end{equation}
One may define the following quantum-geometric length scale,
\begin{equation}
    \ell_{{\rm geo.}}\equiv\sqrt{\int_{{\rm BZ}}\frac{d^{2}{\bf k}}{4\pi^{2}}{\rm Trg\left({\bf k}\right)}}=\frac{\zeta}{2},
\end{equation}
which corresponds to the minimal Wannier-function spread associated with these flat bands~\cite{Vanderbiltmaximmalylocalized}.

The Hamiltonian Eq.~\eqref{eq:pseudotbgsingleparticle} has a total of eight bands: four valence bands sitting at $-t$, and four conduction bands at $+t$ (when $\Delta_\sigma=0$).
Our interest will lie in one set of these isolated bands, e.g., the valence bands.
Within such set, one may define a good quantum number which further classifies the bands, ${\cal C}=\sigma_z\tau_z$.

Let us consider the simplest form density-density interactions within the projected valence bands,
\begin{equation}
    H_{{\rm C}}=\frac{1}{2A}\sum_{{\bf qkk'}}V_{{\bf q}}c_{{\bf k+q}}^{\dagger}\Lambda_{{\bf k+q,k}}c_{{\bf k}}c_{{\bf k'-q}}^{\dagger}\Lambda_{{\bf k'-q,k'}}c_{{\bf k'}},
\end{equation}
where the form factors are
\begin{equation}
    \Lambda_{{\bf k',k}}={\cal F}_{{\bf k',k}}+{\cal D}_{{\bf k',k}}\sigma_{z}\tau_{z},
\end{equation}
with 
${\cal F}_{{\bf k',k}}=\frac{\cos\left(\alpha_{{\bf k}}-\alpha_{{\bf k'}}\right)+\cos\left(\beta_{{\bf k}}-\beta_{{\bf k'}}\right)}{2}$, and
${\cal D}_{{\bf k',k}}=\frac{\cos\left(\alpha_{{\bf k}}-\alpha_{{\bf k'}}\right)-\cos\left(\beta_{{\bf k}}-\beta_{{\bf k'}}\right)}{2}$.


What is the ground state manifold of this Hamiltonian at half-filling of the valence bands?
We employ Hartree-Fock calculations and find,
\begin{equation}
    \left\langle H_{{\rm C}}\right\rangle 
    =2A V_{0}+\frac{1}{2A}\sum_{{\bf kk'}}V_{{\bf k-k'}}\frac{{\cal D}_{{\bf k',k}}{\cal D}_{{\bf k,k'}}}{4}{\rm Tr}\left\{ \mathbb{I}-\sigma_{z}\tau_{z}{\cal O}\sigma_{z}\tau_{z}{\cal O}\right\}, 
\end{equation}
which has a very clear consequence:
The ground state manifold under the $H_{\rm C}$ interaction is the one with an order parameter which commutes with ${\cal C}$,
\begin{equation}
    \left[{\cal O}_{\rm g.s.},{\cal C}\right]=0.
\end{equation}
The ground-state energy density difference between this manifold and other generalized ferromagnets is then
$\frac{\Delta E_{{\rm C}}}{A}=\frac{1}{A^{2}}\sum_{{\bf kk'}}V_{{\bf k-k'}}\left|{\cal D}_{{\bf k',k}}\right|^{2}$.

\subsection{Symmetry breaking perturbations}
Now, let us introduce perturbations which pick out specific members of this large ${\cal O}_{\rm g.s.}$ manifold.
First, we consider an interaction term which picks out an order $\sim \sigma_y\left(\cos\theta\tau_x+i\sin\theta\tau_y\right)$ (with some arbitrary angle $\theta$),
\begin{align}
    H_{{\rm K}}	&=\frac{1}{2A}\sum_{{\bf qkk'}}u^{\rm K}_{{\bf q}}c_{{\bf k+q}}^{\dagger}\Lambda_{{\bf k+q,k}}\sigma_{y}\tau_{x}c_{{\bf k}}c_{{\bf k'-q}}^{\dagger}\Lambda_{{\bf k'-q,k'}}\sigma_{y}\tau_{x}c_{{\bf k'}}\nonumber\\
	&+\frac{1}{2A}\sum_{{\bf qkk'}}u^{\rm K}_{{\bf q}}c_{{\bf k+q}}^{\dagger}\Lambda_{{\bf k+q,k}}\sigma_{y}\tau_{y}c_{{\bf k}}c_{{\bf k'-q}}^{\dagger}\Lambda_{{\bf k'-q,k'}}\sigma_{y}\tau_{y}c_{{\bf k'}},
\end{align}
with $u^{\rm K}_{\bf q}<0$.
The ``$\rm K$'' label hints at the analogous order in TBG, which is the Kramers intervalley coherent order (KIVC).
Next, let us introduce a similar interaction which favors $\sim \sigma_x\left(\cos\theta\tau_x+i\sin\theta\tau_y\right)$ order, 
\begin{align}
    H_{{\rm T}}	&=\frac{1}{2A}\sum_{{\bf qkk'}}u^{\rm T}_{{\bf q}}c_{{\bf k+q}}^{\dagger}\Lambda_{{\bf k+q,k}}\sigma_{x}\tau_{x}c_{{\bf k}}c_{{\bf k'-q}}^{\dagger}\Lambda_{{\bf k'-q,k'}}\sigma_{x}\tau_{x}c_{{\bf k'}}\nonumber\\
	&+\frac{1}{2A}\sum_{{\bf qkk'}}u^{\rm T}_{{\bf q}}c_{{\bf k+q}}^{\dagger}\Lambda_{{\bf k+q,k}}\sigma_{x}\tau_{y}c_{{\bf k}}c_{{\bf k'-q}}^{\dagger}\Lambda_{{\bf k'-q,k'}}\sigma_{x}\tau_{y}c_{{\bf k'}},
\end{align}
with $u^{\rm T}_{\bf q}<0$.
This interaction in the case of TBG has been related to interaction between electrons and intervalley phonons, and associated with stabilization of a time-reversal invariant intervalley coherent state (TIVC) in pristine (i.e., vanishing strain) devices~\cite{TIVCphonons}.
Let us also single-out the single-particle perturbation which explicitly breaks the pseudospin symmetry at the single-particle level,
\begin{equation}
    H_{SB}=-\Delta_{\sigma}\sum_{{\bf k}}c_{{\bf k}}^{\dagger}\sigma_{z}c_{{\bf k}},
\end{equation}
akin to sublattice symmetry breaking in TBG devices.

Hartree-Fock calculations yield
\begin{align}
    \left\langle H_{{\rm K}}\right\rangle &=
    \frac{u_{0}^{{\rm K}}}{8}\left|{\rm Tr}\left\{ \sigma_{y}\left(\tau_{x}+i\tau_{y}\right){\cal O}\right\} \right|^{2}\nonumber\\
    &+\frac{1}{2A}\sum_{{\bf kk'}}u_{{\bf k-k'}}^{{\rm K}}\left(\left|{\cal F}_{{\bf k,k'}}\right|^{2}+\left|{\cal D}_{{\bf k,k'}}\right|^{2}\right){\rm Tr}\left\{ P^{T}\left(\sigma_{y}\tau_{x}\bar{P}^{T}\sigma_{y}\tau_{x}+\sigma_{x}\tau_{y}\bar{P}^{T}\sigma_{x}\tau_{y}\right)\right\} \nonumber\\
    &+\frac{1}{2A}\sum_{{\bf kk'}}u_{{\bf k-k'}}^{{\rm K}}\left({\cal F}_{{\bf k,k'}}{\cal D}_{{\bf k',k}}+{\cal D}_{{\bf k,k'}}{\cal F}_{{\bf k',k}}\right){\rm Tr}\left\{ \sigma_{z}\tau_{z}P^{T}\left(\sigma_{y}\tau_{x}\bar{P}^{T}\sigma_{y}\tau_{x}+\sigma_{y}\tau_{y}\bar{P}^{T}\sigma_{y}\tau_{y}\right)\right\}, 
\end{align}
\begin{align}
    \left\langle H_{{\rm T}}\right\rangle &=
    \frac{u_{0}^{{\rm T}}}{8}\left|{\rm Tr}\left\{ \sigma_{x}\left(\tau_{x}+i\tau_{y}\right){\cal O}\right\} \right|^{2}\nonumber\\
    &+\frac{1}{2A}\sum_{{\bf kk'}}u_{{\bf k-k'}}^{{\rm T}}\left(\left|{\cal F}_{{\bf k,k'}}\right|^{2}+\left|{\cal D}_{{\bf k,k'}}\right|^{2}\right){\rm Tr}\left\{ P^{T}\left(\sigma_{x}\tau_{x}\bar{P}^{T}\sigma_{x}\tau_{x}+\sigma_{x}\tau_{y}\bar{P}^{T}\sigma_{x}\tau_{y}\right)\right\} \nonumber\\
    &+\frac{1}{2A}\sum_{{\bf kk'}}u_{{\bf k-k'}}^{{\rm T}}\left({\cal F}_{{\bf k,k'}}{\cal D}_{{\bf k',k}}+{\cal D}_{{\bf k,k'}}{\cal F}_{{\bf k',k}}\right){\rm Tr}\left\{ \sigma_{z}\tau_{z}P^{T}\left(\sigma_{x}\tau_{x}\bar{P}^{T}\sigma_{x}\tau_{x}+\sigma_{x}\tau_{y}\bar{P}^{T}\sigma_{x}\tau_{y}\right)\right\}, 
\end{align}
\begin{equation}
    \left\langle H_{SB}\right\rangle =-A\frac{\Delta_{\sigma}}{2}{\rm Tr}\left\langle \sigma_{z}{\cal O}\right\rangle .
\end{equation}
We now calculate different ground-state energy densities relative to $\Delta E_{\rm C}$, 
\begin{equation}
    \frac{E_{{\rm T}}}{A}=2u_{0}^{{\rm T}}-\epsilon_{{\rm geo.}},
    \,\,\,\,\,\,\,
    \frac{E_{{\rm K}}}{A}=2u_{0}^{{\rm K}}-\epsilon_{{\rm geo.}},
\end{equation}
\begin{equation}
    \frac{E_{\sigma}}{A}=-2\epsilon_{{\rm geo.}}-2\Delta_{\sigma},
    \,\,\,\,\,\,\,
    \frac{E_{\tau}}{A}=-2\epsilon_{{\rm geo.}},
    \,\,\,\,\,\,\,
    \frac{E_{{\cal C}}}{A}=0.
\end{equation}
Here, $E_{\rm T/K}$ correspond to the different IVC orders, and $E_{\sigma}$, $E_\tau$, and $E_{\cal C}$ correspond to $\sigma_z$, $\tau_z$, and $\sigma_z\tau_z$ orders, respectively.
Finally,
\begin{equation}
    \epsilon_{{\rm geo.}}=-\frac{1}{A^{2}}\sum_{{\bf kk'}}\left(u_{{\bf k-k'}}^{{\rm T}}+u_{{\bf k-k'}}^{{\rm K}}\right)\left(\left|{\cal F}_{{\bf k,k'}}\right|^{2}+\left|{\cal D}_{{\bf k,k'}}\right|^{2}\right)>0,
\end{equation}
and has a quantum geometric interpretation.
Using the fact that $u_{{\bf k-k'}}^{{\rm T/K}}$ is physically only a function of $\left|{\bf k-k'}\right|$,
\begin{align}
    \epsilon_{{\rm geo.}}
    &=-\frac{1}{A^{2}}\sum_{{\bf kk'}}\left(u_{{\bf k-k'}}^{{\rm T}}+u_{{\bf k-k'}}^{{\rm K}}\right)\cos^{2}\left(\alpha_{{\bf k}}-\alpha_{{\bf k'}}\right)\nonumber\\
    &=-\frac{1}{A^{2}}\sum_{{\bf kk'}}\frac{u_{{\bf k-k'}}^{{\rm T}}+u_{{\bf k-k'}}^{{\rm K}}}{2}+\frac{1}{A^{2}}\sum_{{\bf kk'}}\frac{u_{{\bf k-k'}}^{{\rm T}}+u_{{\bf k-k'}}^{{\rm K}}}{4}\left[e^{i\left(2\alpha_{{\bf k}}-2\alpha_{{\bf k'}}\right)}+{\rm h.c.}\right]\nonumber\\
    &=-\int\frac{d{\bf k}d{\bf k'}}{\left(2\pi\right)^{4}}\frac{u_{{\bf k-k'}}^{{\rm T}}+u_{{\bf k-k'}}^{{\rm K}}}{2}+\int\frac{d{\bf k}d{\bf k'}}{\left(2\pi\right)^{4}}\frac{u_{{\bf k-k'}}^{{\rm T}}+u_{{\bf k-k'}}^{{\rm K}}}{4}\left[e^{2i\zeta\cos k_{x}}e^{2i\zeta\cos k_{y}}e^{-2i\zeta\cos k_{x}'}e^{-2i\zeta\cos k_{y}'}+{\rm h.c.}\right]\nonumber\\
    &=-\int\frac{d{\bf q}}{\left(2\pi\right)^{2}}\frac{u_{{\bf q}}^{{\rm T}}+u_{{\bf q}}^{{\rm K}}}{2}\left[1+\sum_{ab}\left[J_{a}\left(2\zeta\right)J_{b}\left(2\zeta\right)\right]^{2}e^{iaq_{x}+ibq_{y}}\right].\label{eq:egeo}
\end{align}
A comment is in order regarding the dependence of $\epsilon_{\rm geo.}$ on the parameter $\zeta$.
It is in fact related to the \textit{spatial range} of the $u^{\rm T/K}$ interactions, $R$.
The Fourier transform appearing in Eq.~\eqref{eq:egeo} clearly demonstrates that the $\zeta$-dependence is quenched if $R\gg\zeta$.
The physical interpretation is simple: $\zeta$ is related to the spatial width of a maximally local wave packet of electrons in the flat band of interest.
The sensitivity of the Hartree-Fock energy hierarchy to $\zeta$ will then be most pronounced when the range of these interactions is comparable to this width, i.e., for $\zeta\sim R$.

Next, for the sake of simplicity, we will assume that these interactions are short range, i.e., momentum independent, and define
$u_{{\bf q}}^{{\rm T}}+u_{{\bf q}}^{{\rm K}}\equiv-u<0$,
$u_{{\bf q}}^{{\rm T}}-u_{{\bf q}}^{{\rm K}}=-\Delta_{\rm TK}<0$.
Shifting all energies by a constant $-\epsilon_{\rm geo.}$, the three lowest lying manifolds have energies
\begin{equation}
    \frac{E_{{\rm T}}}{A}=-\Delta_{\rm TK},
    \,\,\,\,\,
    \frac{E_{{\rm K}}}{A}=\frac{E_{{\rm T}}}{A}+2\Delta_{\rm TK},
     \,\,\,\,\,
     \frac{E_{\sigma}}{A}=\frac{E_{{\rm T}}}{A}+\Delta_{\rm TK}+u_\zeta-2\Delta_\sigma.
\end{equation}
Here, we defined $u_\zeta\equiv u-\epsilon_{\rm geo.}=u\left[1-J_{0}^{4}\left(2\zeta\right)\right]$.
Hereafter, we assume $u_\zeta>\Delta_{\rm TK}$, such that in the absence of the explicit pseudospin symmetry breaking, the ground state is T(IVC), and its closest competitor is the K(IVC) phase.
This is presumed to be the case for pristine TBG, taking the aforementioned electron-phonon interactions into account~\cite{TIVCphonons}.
Conversely, if $\Delta_\sigma>\frac{1}{2}\left(\Delta_{\rm TK}+u_\zeta\right)$, the order parameter for the lowest-energy state is $\sigma_z$.


\subsection{Fluctuation Lagrangian}
The contribution of the primary density-density interaction is
\begin{align}
\left\langle \left[\hat{F},\left[\hat{F},H_{{\rm C}}\right]\right]\right\rangle  & =\sum_{{\bf Q}ij{\bf kk'}}\frac{4}{A}\left[\Phi_{{\bf Q}}^{j}\left({\bf k'}\right)\right]^{*}\left[V_{{\bf k-k'}}\left({\cal F}_{{\bf kk'}}{\cal F}_{{\bf k+Q,k'+Q}}+{\cal D}_{{\bf kk'}}{\cal D}_{{\bf k+Q,k'+Q}}\right)-\delta_{{\bf kk'}}\sum_{{\bf p}}V_{{\bf k-p}}\left({\cal F}_{{\bf kp}}^{2}+{\cal D}_{{\bf kp}}^{2}\right)\right]\delta^{ij}\Phi_{{\bf Q}}^{i}\left({\bf k}\right)\nonumber\\
 & +\sum_{{\bf Q}ij{\bf kk'}}\frac{4}{A}\left[\Phi_{{\bf Q}}^{j}\left({\bf k'}\right)\right]^{*}\left[V_{{\bf k-k'}}\left({\cal F}_{{\bf kk'}}{\cal D}_{{\bf k+Q,k'+Q}}+{\cal D}_{{\bf kk'}}{\cal F}_{{\bf k+Q,k'+Q}}\right)-\delta_{{\bf kk'}}\sum_{{\bf p}}V_{{\bf k-p}}\left(2{\cal F}_{{\bf kp}}{\cal D}_{{\bf kp}}\right)\right]\times\nonumber\\
 &\times\frac{{\rm Tr}\left\{ \sigma_{z}\tau_{z}{\cal M}^{j}{\cal M}^{i}\right\} }{4}\Phi_{{\bf Q}}^{i}\left({\bf k}\right).
\end{align}
By noticing the momentum sums in the second line vanish, due to the invariance of the interaction strength to 45-degree rotation in momentum space, we are left with the sole contribution
\begin{equation}
    \left\langle \left[\hat{F},\left[\hat{F},H_{{\rm C}}\right]\right]\right\rangle =\sum_{{\bf Q}ij{\bf kk'}}\frac{4}{A}\left[\Phi_{{\bf Q}}^{j}\left({\bf k'}\right)\right]^{*}\left[V_{{\bf k-k'}}\left({\cal F}_{{\bf kk'}}{\cal F}_{{\bf k+Q,k'+Q}}+{\cal D}_{{\bf kk'}}{\cal D}_{{\bf k+Q,k'+Q}}\right)-\delta_{{\bf kk'}}\sum_{{\bf p}}V_{{\bf k-p}}\left({\cal F}_{{\bf kp}}^{2}+{\cal D}_{{\bf kp}}^{2}\right)\right]\delta^{ij}\Phi_{{\bf Q}}^{i}\left({\bf k}\right).
\end{equation}

Anticipating a lowest-mode approximation in the small $\left|{\bf Q}\right|$ limit, it turns out that the leading $\bf Q$ contribution may be well captured by simply replacing,
\begin{equation}
    \Phi_{{\bf Q}}^{i}\left({\bf k}\right)\approx \frac{1}{2}{\mathbb C}^i_{\bf Q},
\end{equation}
with ${\mathbb C}^i$ the relevant coordinate for the ${\cal M}^i$ generator.
This simplifies the above expression,
\begin{equation}
    -\frac{1}{2A}\left\langle \left[\hat{F},\left[\hat{F},H_{{\rm C}}\right]\right]\right\rangle =\frac{1}{2}\sum_{{\bf Q},i}\mathbb{C}_{{\bf -Q}}^{i}\mathbb{C}_{{\bf Q}}^{i}I\left({\bf Q}\right),
\end{equation}
with
\begin{equation}
    I\left({\bf Q}\right)=\frac{1}{A^{2}}\sum_{{\bf kp}}V_{{\bf k-k'}}\left[{\cal F}_{{\bf kk'}}\left({\cal F}_{{\bf kk'}}-{\cal F}_{{\bf k+Q,k'+Q}}\right)+{\cal F}\to{\cal D}\right].
\end{equation}
Expanding in small momentum, one finds
$I\left({\bf Q}\right)\approx\frac{\rho_{s}}{2}\left|{\bf Q}\right|^{2}$,
with the stiffness given by the sum
\begin{equation}
    \rho_{s}=\zeta^{2}\frac{1}{A^{2}}\sum_{{\bf kk'}}V_{{\bf k-k'}}\left(\sin k_{x}-\sin k'_{x}\right)^{2}\cos^{2}\left(\alpha_{{\bf k}}-\alpha_{{\bf k'}}\right).
\end{equation}

\textbf{Scenario T: ${\cal O}=\sigma_x\tau_x$}

The contribution of the secondary interaction terms to the fluctuation Hamiltonian may be written as,
\begin{align}
    \left\langle \left[\hat{F},\left[\hat{F},H_{{\rm T}}\right]\right]\right\rangle 
    &=-\sum_{{\bf Q}}\frac{1}{A}\sum_{ij{\bf kk'}}\left[\Phi_{{\bf Q}}^{j}\left({\bf k'}\right)\right]^{*}u_{{\bf k-k'}}^{{\rm T}}\Lambda_{{\bf {\bf k'k}}}\Lambda_{{\bf k+Q},{\bf k'+Q}}{\rm Tr}\left\{ \left[{\cal M}^{j}-\tau_{z}{\cal M}^{j}\tau_{z}\right]{\cal M}^{i}\right\} \Phi_{{\bf Q}}^{i}\left({\bf k}\right)\nonumber\\
    &-\sum_{{\bf Q}}\sum_{ij{\bf kk'}}\frac{u_{{\bf Q}}^{{\rm T}}}{A}\left[\Phi_{{\bf Q}}^{j}\left({\bf k'}\right)\right]^{*}{\rm Tr}\left\{ {\cal M}^{j}\Lambda_{{\bf k'},{\bf k'+Q}}\tau_{z}\right\} {\rm Tr}\left\{ {\cal M}^{i}\Lambda_{{\bf k+Q},{\bf k}}\tau_{z}\right\} \Phi_{{\bf Q}}^{i}\left({\bf k}\right)\nonumber\\
    &+16\sum_{{\bf Q}}\sum_{ij{\bf kk'}}u_{{\bf 0}}^{{\rm T}}\left[\Phi_{{\bf Q}}^{j}\left({\bf k'}\right)\right]^{*}\delta_{{\bf kk'}}\delta^{ij}\Phi_{{\bf Q}}^{i}\left({\bf k}\right),
\end{align}
\begin{align}
    \left\langle \left[\hat{F},\left[\hat{F},H_{{\rm K}}\right]\right]\right\rangle 
    &=-\sum_{{\bf Q}}\frac{1}{A}\sum_{ij{\bf kk'}}\left[\Phi_{{\bf Q}}^{j}\left({\bf k'}\right)\right]^{*}u_{{\bf k-k'}}^{{\rm K}}\Lambda_{{\bf {\bf k'k}}}\Lambda_{{\bf k+Q},{\bf k'+Q}}{\rm Tr}\left\{ \left[{\cal M}^{j}-\sigma_{z}{\cal M}^{j}\sigma_{z}\right]{\cal M}^{i}\right\} \Phi_{{\bf Q}}^{i}\left({\bf k}\right)\nonumber\\
    &-\sum_{{\bf Q}}\sum_{ij{\bf kk'}}\frac{u_{{\bf Q}}^{{\rm K}}}{A}\left[\Phi_{{\bf Q}}^{j}\left({\bf k'}\right)\right]^{*}{\rm Tr}\left\{ {\cal M}^{j}\Lambda_{{\bf k'},{\bf k'+Q}}\sigma_{z}\right\} {\rm Tr}\left\{ {\cal M}^{i}\Lambda_{{\bf k+Q},{\bf k}}\sigma_{z}\right\} \Phi_{{\bf Q}}^{i}\left({\bf k}\right).
\end{align}
We can consolidate these two terms, employing the assumptions regarding the short range nature of the interaction,
\begin{align}
    \left\langle \left[\hat{F},\left[\hat{F},H_{{\rm T}}\right]\right]\right\rangle +\left\langle \left[\hat{F},\left[\hat{F},H_{{\rm K}}\right]\right]\right\rangle &=u\sum_{{\bf Q}}\frac{1}{A}\sum_{ij{\bf kk'}}\left[\Phi_{{\bf Q}}^{j}\left({\bf k'}\right)\right]^{*}\left({\cal F}_{{\bf kk'}}{\cal F}_{{\bf k+Q,k'+Q}}+{\cal D}_{{\bf kk'}}{\cal D}_{{\bf k+Q,k'+Q}}\right)\nonumber\\
    &\times{\rm Tr}\left\{ \left[{\cal M}^{j}-\tau_{z}{\cal M}^{j}\tau_{z}\right]{\cal M}^{i}\right\} \Phi_{{\bf Q}}^{i}\left({\bf k}\right)\nonumber\\
    &+\sum_{{\bf Q}}\sum_{ij{\bf kk'}}\frac{1}{A}\left[\Phi_{{\bf Q}}^{j}\left({\bf k'}\right)\right]^{*}\left(\frac{u+\Delta_{{\rm TK}}}{2}{\cal F}_{{\bf k'+Q,k'}}{\cal F}_{{\bf k+Q,k}}+\frac{u-\Delta_{{\rm TK}}}{2}{\cal D}_{{\bf k'+Q,k'}}{\cal D}_{{\bf k+Q,k}}\right)\nonumber\\
    &\times{\rm Tr}\left\{ {\cal M}^{i}\tau_{z}\right\} {\rm Tr}\left\{ {\cal M}^{i}\tau_{z}\right\} \Phi_{{\bf Q}}^{i}\left({\bf k}\right)\nonumber\\
    &+\sum_{{\bf Q}}\sum_{ij{\bf kk'}}\frac{1}{A}\left[\Phi_{{\bf Q}}^{j}\left({\bf k'}\right)\right]^{*}\left(\frac{u-\Delta_{{\rm TK}}}{2}{\cal F}_{{\bf k'+Q,k'}}{\cal F}_{{\bf k+Q,k}}+\frac{u+\Delta_{{\rm TK}}}{2}{\cal D}_{{\bf k'+Q,k'}}{\cal D}_{{\bf k+Q,k}}\right)\nonumber\\
    &\times{\rm Tr}\left\{ {\cal M}^{i}\sigma_{z}\right\} {\rm Tr}\left\{ {\cal M}^{i}\sigma_{z}\right\} \Phi_{{\bf Q}}^{i}\left({\bf k}\right)\nonumber\\
    &-16\frac{u+\Delta_{{\rm TK}}}{2}\sum_{{\bf Q}}\sum_{ij{\bf kk'}}\left[\Phi_{{\bf Q}}^{j}\left({\bf k'}\right)\right]^{*}\delta_{{\bf kk'}}\delta^{ij}\Phi_{{\bf Q}}^{i}\left({\bf k}\right)\nonumber\\
    &+u\sum_{{\bf Q}}\sum_{ij{\bf kk'}}\frac{1}{A}\left[\Phi_{{\bf Q}}^{j}\left({\bf k'}\right)\right]^{*}\left({\cal F}_{{\bf k'+Q,k'}}{\cal D}_{{\bf k+Q,k}}+{\cal F}_{{\bf k+Q,k}}{\cal D}_{{\bf k'+Q,k'}}\right)\nonumber\\
    &{\rm Tr}\left\{ {\cal M}^{i}\tau_{z}\right\} {\rm Tr}\left\{ {\cal M}^{i}\sigma_{z}\right\} \Phi_{{\bf Q}}^{i}\left({\bf k}\right)\nonumber\\
    &+u\sum_{{\bf Q}}\frac{1}{A}\sum_{ij{\bf kk'}}\left[\Phi_{{\bf Q}}^{j}\left({\bf k'}\right)\right]^{*}\left({\cal F}_{{\bf kk'}}{\cal D}_{{\bf k+Q,k'+Q}}+{\cal D}_{{\bf kk'}}{\cal F}_{{\bf k+Q,k'+Q}}\right)\nonumber\\
    &\times{\rm Tr}\left\{ \left[{\cal M}^{j}-\tau_{z}{\cal M}^{j}\tau_{z}\right]{\cal M}^{i}\sigma_{z}\tau_{z}\right\} \Phi_{{\bf Q}}^{i}\left({\bf k}\right).
\end{align}

Performing the crude lowest-mode approximation, we assign coordinates to the different generators in the following manner,
$\tau_z\to{\mathbb X_{\bf Q}^{\rm G}}$,
$\sigma_x\tau_y\to{\mathbb P_{\bf Q}^{\rm G}}$,
$\sigma_z\to{\mathbb X_{\bf Q}^{\rm K}}$,
$\sigma_y\tau_x\to{\mathbb P_{\bf Q}^{\rm K}}$.
We are going to focus below on the ${\bf Q}\to 0$ contribution of these terms.
One interesting ``casualty'' of this approximation is embodied by the last two lines in the expression above.
Specifically, this term captures a coupling at larger momenta between the gapless Goldstone mode and the so-called ``K'' mode, which may have some intriguing consequences beyond the scope of this work~\cite{DykmanPhysRevB.98.195444,DykmanPhysRevResearch.6.023162,kaplan2025opticallyinducedfaradaygoldstonewaves}.
Within this approximation, we can finally find the contribution
\begin{align}
    -\frac{\left\langle \left[\hat{F},\left[\hat{F},H_{{\rm T}}\right]\right]\right\rangle +\left\langle \left[\hat{F},\left[\hat{F},H_{{\rm K}}\right]\right]\right\rangle }{2A}&\approx\frac{u_{\zeta}+\Delta_{{\rm TK}}}{2}\sum_{{\bf Q}}\left(\mathbb{P}_{{\bf -Q}}^{{\rm G}}\mathbb{P}_{{\bf Q}}^{{\rm G}}+\mathbb{X}_{{\bf -Q}}^{{\rm G}}\mathbb{X}_{{\bf Q}}^{{\rm G}}+\mathbb{P}_{{\bf -Q}}^{{\rm G}}\mathbb{P}_{{\bf Q}}^{{\rm G}}-\mathbb{X}_{{\bf -Q}}^{{\rm G}}\mathbb{X}_{{\bf Q}}^{{\rm G}}\right)\nonumber\\
    &+\frac{u_{\zeta}+3\Delta_{{\rm TK}}}{2}\sum_{{\bf Q}}\left(\mathbb{P}_{{\bf -Q}}^{{\rm K}}\mathbb{P}_{{\bf Q}}^{{\rm K}}+\mathbb{X}_{{\bf -Q}}^{{\rm K}}\mathbb{X}_{{\bf Q}}^{{\rm K}}\right)\nonumber\\
    &+\frac{u_{\zeta}-\Delta_{{\rm TK}}}{2}\sum_{{\bf Q}}\left(\mathbb{P}_{{\bf -Q}}^{{\rm K}}\mathbb{P}_{{\bf Q}}^{{\rm K}}-\mathbb{X}_{{\bf -Q}}^{{\rm K}}\mathbb{X}_{{\bf Q}}^{{\rm K}}\right).
\end{align}

\textbf{Scenario $\sigma$: ${\cal O}=\sigma_z$}

Let us begin by considering the contribution of the additional single-particle term,
\begin{equation}
    \left\langle \left[{\hat{F}},\left[{\hat{F}},H_{\sigma}\right]\right]\right\rangle =-8\Delta_{\sigma}\sum_{{\bf Q}}\sum_{ij{\bf pp'}}\left[\Phi_{{\bf Q}}^{j}\left({\bf p'}\right)\right]^{*}\delta_{{\bf pp'}}\delta^{ij}\Phi_{{\bf Q}}^{i}\left({\bf p}\right).
\end{equation}
It is already clear that this creates a uniform gap in the spectrum of collective excitations.

The secondary interactions supply the contribution,
\begin{align}
\left\langle \left[\hat{F},\left[\hat{F},H_{{\rm K}}+H_{{\rm T}}\right]\right]\right\rangle  & =\sum_{{\bf Q}}\frac{1}{A}\sum_{ij{\bf kk'}}\left[\Phi_{{\bf Q}}^{j}\left({\bf k'}\right)\right]^{*}{\rm Tr}\left\{ 2\delta_{{\bf kk'}}\sum_{{\bf p}}\left(u_{{\bf k-p}}^{{\rm T}}+u_{{\bf k-p}}^{{\rm K}}\right)\Lambda_{{\bf {\bf pk}}}\Lambda_{{\bf kp}}{\cal M}^{j}{\cal M}^{i}\right\} \Phi_{{\bf Q}}^{i}\left({\bf k}\right)\nonumber\\
 & -\sum_{{\bf Q}}\frac{1}{A}\sum_{ij{\bf kk'}}\left[\Phi_{{\bf Q}}^{j}\left({\bf k'}\right)\right]^{*}{\rm Tr}\left\{ \left[\left(u_{{\bf k-k'}}^{{\rm T}}+u_{{\bf k-k'}}^{{\rm K}}\right)\Lambda_{{\bf k'k}}\Lambda_{{\bf k+Q},{\bf k'+Q}}\tau_{x}\left(\sigma_{x}{\cal M}^{j}\sigma_{x}+\sigma_{y}{\cal M}^{j}\sigma_{y}\right)\tau_{x}\right]{\cal M}^{i}\right\} \Phi_{{\bf Q}}^{i}\left({\bf k}\right)\nonumber\\
 & -\sum_{{\bf Q}}\sum_{ij{\bf kk'}}\frac{u_{{\bf Q}}^{{\rm K}}}{A}\left[\Phi_{{\bf Q}}^{j}\left({\bf k'}\right)\right]^{*}{\rm Tr}\left\{ {\cal M}^{j}\Lambda_{{\bf k'},{\bf k'+Q}}\sigma_{x}\tau_{y}\right\} {\rm Tr}\left\{ {\cal M}^{i}\Lambda_{{\bf k+Q},{\bf k}}\sigma_{x}\tau_{y}\right\} \Phi_{{\bf Q}}^{i}\nonumber\\
 & -\sum_{{\bf Q}}\sum_{ij{\bf kk'}}\frac{u_{{\bf Q}}^{{\rm K}}}{A}\left[\Phi_{{\bf Q}}^{j}\left({\bf k'}\right)\right]^{*}{\rm Tr}\left\{ {\cal M}^{j}\Lambda_{{\bf k'},{\bf k'+Q}}\sigma_{x}\tau_{x}\right\} {\rm Tr}\left\{ {\cal M}^{i}\Lambda_{{\bf k+Q},{\bf k}}\sigma_{x}\tau_{x}\right\} \Phi_{{\bf Q}}^{i}\nonumber\\
 & -\sum_{{\bf Q}}\sum_{ij{\bf kk'}}\frac{u_{{\bf Q}}^{{\rm T}}}{A}\left[\Phi_{{\bf Q}}^{j}\left({\bf k'}\right)\right]^{*}{\rm Tr}\left\{ {\cal M}^{j}\Lambda_{{\bf k'},{\bf k'+Q}}\sigma_{y}\tau_{y}\right\} {\rm Tr}\left\{ {\cal M}^{i}\Lambda_{{\bf k+Q},{\bf k}}\sigma_{y}\tau_{y}\right\} \Phi_{{\bf Q}}^{i}\nonumber\\
 & -\sum_{{\bf Q}}\sum_{ij{\bf kk'}}\frac{u_{{\bf Q}}^{T}}{A}\left[\Phi_{{\bf Q}}^{j}\left({\bf k'}\right)\right]^{*}{\rm Tr}\left\{ {\cal M}^{j}\Lambda_{{\bf k'},{\bf k'+Q}}\sigma_{y}\tau_{x}\right\} {\rm Tr}\left\{ {\cal M}^{i}\Lambda_{{\bf k+Q},{\bf k}}\sigma_{y}\tau_{x}\right\} \Phi_{{\bf Q}}^{i}
\end{align}
Actually, for our generators here, the second line is identically zero.

Performing the lowest-mode approximation here as well, we assign coordinates,
$\sigma_x\tau_x\to{\mathbb X_{\bf Q}^{\left(1\right)}}$,
$\sigma_y\tau_x\to{\mathbb P_{\bf Q}^{\left(1\right)}}$,
$\sigma_x\tau_y\to{\mathbb X_{\bf Q}^{\left(2\right)}}$,
$\sigma_y\tau_y\to{\mathbb P_{\bf Q}^{\left(2\right)}}$.
Focusing on the ${\bf Q}\to 0$ contributions, we write
\begin{equation}
    -\frac{\left\langle \left[{\hat{F}},\left[{\hat{F}},H_{\sigma}\right]\right]\right\rangle }{2A}=\Delta_{\sigma}\sum_{{\bf Q}}\left(\mathbb{X}_{{\bf -Q}}^{\left(1\right)}\mathbb{X}_{{\bf Q}}^{\left(1\right)}+\mathbb{P}_{{\bf -Q}}^{\left(1\right)}\mathbb{P}_{{\bf Q}}^{\left(1\right)}+\mathbb{X}_{{\bf -Q}}^{\left(2\right)}\mathbb{X}_{{\bf Q}}^{\left(2\right)}+\mathbb{P}_{{\bf -Q}}^{\left(2\right)}\mathbb{P}_{{\bf Q}}^{\left(2\right)}\right),
\end{equation}
\begin{align}
    -\frac{\left\langle \left[\hat{F},\left[\hat{F},H_{{\rm K}}+H_{{\rm T}}\right]\right]\right\rangle }{2A}&=u\left[\frac{1}{A}\sum_{{\bf p}}\left({\cal F}_{{\bf {\bf pk}}}^{2}+{\cal D}_{{\bf {\bf pk}}}^{2}\right)-1\right]\sum_{{\bf Q}}\left(\mathbb{X}_{{\bf -Q}}^{\left(1\right)}\mathbb{X}_{{\bf Q}}^{\left(1\right)}+\mathbb{P}_{{\bf -Q}}^{\left(1\right)}\mathbb{P}_{{\bf Q}}^{\left(1\right)}+\mathbb{X}_{{\bf -Q}}^{\left(2\right)}\mathbb{X}_{{\bf Q}}^{\left(2\right)}+\mathbb{P}_{{\bf -Q}}^{\left(2\right)}\mathbb{P}_{{\bf Q}}^{\left(2\right)}\right)\nonumber\\
    &+\sum_{{\bf Q}}\Delta_{{\rm TK}}\left(\mathbb{X}_{{\bf -Q}}^{\left(2\right)}\mathbb{X}_{{\bf Q}}^{\left(2\right)}+\mathbb{X}_{{\bf -Q}}^{\left(1\right)}\mathbb{X}_{{\bf Q}}^{\left(1\right)}\right)\nonumber\\
    &-\sum_{{\bf Q}}\Delta_{{\rm TK}}\left(\mathbb{P}_{{\bf -Q}}^{\left(2\right)}\mathbb{X}_{{\bf Q}}^{\left(2\right)}+\mathbb{P}_{{\bf -Q}}^{\left(1\right)}\mathbb{X}_{{\bf Q}}^{\left(1\right)}\right).
\end{align}

\end{widetext} 
\end{appendix}

\bibliographystyle{aapmrev4-2}
\bibliography{refCollective}

\end{document}